\documentclass{jfm}

\usepackage{graphicx}
\usepackage{newtxtext}
\usepackage{newtxmath}
\usepackage{natbib}
\usepackage{graphicx}
\usepackage{epstopdf, epsfig}
\usepackage{subfigure}
\usepackage{hyperref}
\usepackage[normalem]{ulem}

\hypersetup{
    colorlinks = true,
    urlcolor   = blue,
    citecolor  = black,
}
\newcommand{\Chi}{\mathrm{X}}          

\newcommand{\RomanNumeralCaps}[1]
\linenumbers

\title{Orientation dynamics of a settling spheroid in simple shear flow: bifurcations and stochastic alignment}

\author{Himanshu Mishra \and
  Anubhab Roy \corresp{\email{anubhab@iitm.ac.in}}
 }

\affiliation{Department of Applied Mechanics and Biomedical Engineering, Indian Institute of Technology Madras, Chennai, India
}

\begin{document}
\maketitle

\begin{abstract}
We investigate the orientation dynamics of a settling spheroid in simple shear
flow, combining a deterministic dynamical-systems analysis with a stochastic
Fokker-Planck treatment. The dynamics is governed by competition between the
Jeffery torque from the shear and the inertial torque from settling. When
gravity lies in the shear plane the equatorial great circle is invariant, and
the azimuthal motion on it is overdamped motion in a tilted periodic potential
controlled by a single parameter $\mathcal{R}$ combining shape anisotropy and
settling strength. That reduced system undergoes a saddle-node bifurcation on
an invariant circle at $\mathcal{R}=1$, with the period diverging as
$(1-\mathcal{R})^{-1/2}$. For oblate spheroids the
equator attracts and the bifurcation converts an attracting periodic orbit into
an attracting in-plane equilibrium. For prolate spheroids the equatorial
equilibria are transversely unstable and the vorticity-aligned director remains
the attractor at every settling strength; rapid settling draws it onto an
attracting broadside slow manifold. When gravity is parallel to the vorticity axis, a particle exhibits either log-rolling or tumbling motion, depending on its shape. Noise acts differently in the two cases. The
oblate problem supports azimuthal winding, for which we obtain an exact reduced
current and its Kramers limit, exponentially sensitive to the P\'{e}clet
number; the prolate problem shows broadside excursions about a single nematic
attractor, whose rate is set by a two-dimensional quasipotential not computed
here. Asymptotics at small and large $\mathrm{Pe}$, with numerical
Fokker-Planck solutions at arbitrary $\mathrm{Pe}$, quantify the orientation
moments across all regimes.
\end{abstract}



\section{Introduction}

A small spheroid in simple shear flow rotates along one of an
infinite family of periodic trajectories known as Jeffery orbits
\citep{jeffery1922motion}; which orbit is selected depends on
the initial orientation, rendering the long-time dynamics
indeterminate. This degeneracy complicates the prediction of
orientation-dependent quantities in suspensions, from the drag
and collision rates of ice crystals in clouds
\citep{breon2004horizontally,pinsky1998some} to the effective
viscosity of fibre suspensions in industrial flows
\citep{lundell2011fluid}. Several mechanisms can
break the degeneracy: fluid inertia, external fields, viscoelasticity or Brownian diffusion. Here we focus on some of these mechanisms and
specifically on the inertial torque generated by the
translational motion of a settling particle.

The settling-induced inertial torque, of the form $\boldsymbol{T}_{I}\propto(\boldsymbol{W}_s\cdot\boldsymbol{p})(\boldsymbol{W}_s\times\boldsymbol{p})$, was first obtained by \citet{cox1965steady}, who computed the couple
on a nearly spherical particle in a uniform stream as a perturbation in
particle Reynolds number. \citet{khayat1989inertia} obtained the complementary
slender-body limit by matched asymptotic expansions, and
\citet{dabade2015effects} bridged the two for spheroids of arbitrary aspect
ratio. Here, $\boldsymbol{W}_s$ denotes the slip velocity, and $\boldsymbol{p}$ is the unit orientation vector along the symmetry axis of a spheroid. The slip velocity $\boldsymbol{W}_s$ is defined as the difference between the particle and fluid velocities. In
a quiescent fluid, it drives the particle toward broadside-on
alignment, as confirmed in experimental studies
\citep{jayaweera1965behaviour,lopez2017inertial,cabrera2022experimental}. \citet{Joshi2026} recently introduced an inertial slender-body theory to compute forces and torques up to diameter-based Reynolds numbers of $O(10)$.
In a shear flow, the dynamics results from the competition
between the torque induced by the effects of fluid inertia and the Jeffery torque from the background shear. Prior
work has examined this competition for nearly spherical particles
\citep{subramanian2006inertial}, for neutrally buoyant spheroids
where fluid inertia arises from the background shear alone
\citep{yu2007rotation,huang2012rotation,rosen2014effect,di2024influence,wang2025inertial},
and for combined shear- and slip-induced fluid inertia
\citep{cui2025stability}. What has remained
unexplored is a systematic analysis of how the relative
orientation of the gravity and vorticity vectors controls the
preferred rotational state when the translation-induced inertial
torque is the dominant correction to the Jeffery dynamics. 

In this paper, we show that the orientation dynamics of a
settling spheroid in simple shear is governed by a single
effective parameter
$\mathcal{R}=\sqrt{\mathcal{B}^{2}
+\mathcal{K}^{2}\mathcal{F}_{p}^{2}}$,
where $\mathcal{B}$ is the Bretherton constant (measuring
particle anisotropy), $\mathcal{K}$ the settling parameter
(comparing convective and settling time scales), and
$\mathcal{F}_{p}$ is a shape factor. When gravity lies in the shear plane, the equatorial great circle
$\theta=\pi/2$ is invariant and the azimuthal motion on it reduces to
overdamped motion in a tilted periodic potential, so the reduced system
undergoes a saddle-node bifurcation on an invariant circle (SNIC) at
$\mathcal{R}=1$: below this threshold the spheroid rotates continuously within
the equator with a period that diverges as $(1-\mathcal{R})^{-1/2}$, and above
it the azimuthal rotation arrests. Whether that bifurcation
changes the observed alignment is decided by particle shape. The eigenvalue
transverse to the equator carries the sign of $-\mathcal{K}_{p}$, so the
arrested state attracts only for oblate spheroids; for prolate spheroids it is
a saddle on $S^{2}$ and the vorticity axis attracts at every $\mathcal{K}$. When gravity is parallel to the vorticity axis, the settling parameter does
not enter the azimuthal equation and acts on the polar angle alone,
selecting a tumbling limit cycle for a prolate spheroid and a
steady log-rolling orientation for an oblate one, at all $\mathcal{K}$. Closed-form
analytical solutions are obtained for all three canonical
configurations, defined by the relative orientation between the gravity and vorticity vectors.

We next examine how stochastic perturbations, arising from Brownian diffusion or turbulent fluctuations, modify these deterministic structures. The resulting orientation dynamics is governed by a Fokker--Planck equation, which is solved using a spherical harmonics expansion at arbitrary P\'{e}clet number, $\mathrm{Pe}$---defined as the ratio of diffusive to convective time scales---supplemented by asymptotic analyses in the limits $\mathrm{Pe}\to 0$ and $\mathrm{Pe}\to\infty$. The key finding is that noise
plays a qualitatively different role depending on whether the
settling-induced potential barriers are present. In the classical
Jeffery problem without settling, noise acts as a regularizing
perturbation: it diffuses over the continuum of orbit constants
on a slow $O(\mathrm{Pe})$ time scale, as analysed by
\citet{leal1971effect} and \citet{hinch1972effect}. When
settling creates isolated equilibria separated by saddle points
($\mathcal{R}>1$), noise instead drives Kramers-type escape
events: sudden phase slips in polar angle whose rate is exponentially
sensitive to $\mathrm{Pe}$ and whose statistics are governed by
large deviation theory \citep{kramers1940brownian,freidlin1998random}.
This transition from smooth orbit diffusion to rare
barrier-crossing events is the central new result of the
stochastic analysis, and connects directly to the SNIC
bifurcation structure of the deterministic problem. The
framework is closely analogous to the Fokker-Planck analysis of
polymer tumbling in turbulent shear
\citep{chertkov2005polymer,turitsyn2007polymer}. Further, \citet{vincenzi2013orientation} has obtained the exact stationary orientation distribution for a non-spherical particle in an axisymmetric Gaussian random flow, identifying four dynamical regimes that depend on the flow anisotropy and particle shape. In the polymer context, the
competition between a mean shear drift and random velocity
gradient fluctuations produces algebraic tails in the orientation
PDF and aperiodic tumbling statistics. We further examine the orientation distribution function for two flow configurations and compute the corresponding orientation moments, which provide a measure of the bulk rheological properties of the suspension.

The paper is organized as follows. In \S\ref{sec:eqm}, we
formulate the governing equations for arbitrary gravity-vorticity
alignment. Section~\ref{sec:theory} presents the deterministic
dynamical-systems analysis for the three canonical configurations
(\S\ref{sec:gz}--\S\ref{sec:gx}). Section~\ref{sec:sod}
develops the stochastic theory, including low-$\mathrm{Pe}$
asymptotics,  large-$\mathrm{Pe}$
asymptotics, numerical solutions, and the Kramers escape analysis. Conclusions are
given in \S\ref{sec:conc}.

\section{Problem formulation}\label{sec:eqm}

\begin{figure}
    \hspace{-1cm}
    \includegraphics[scale=0.45]{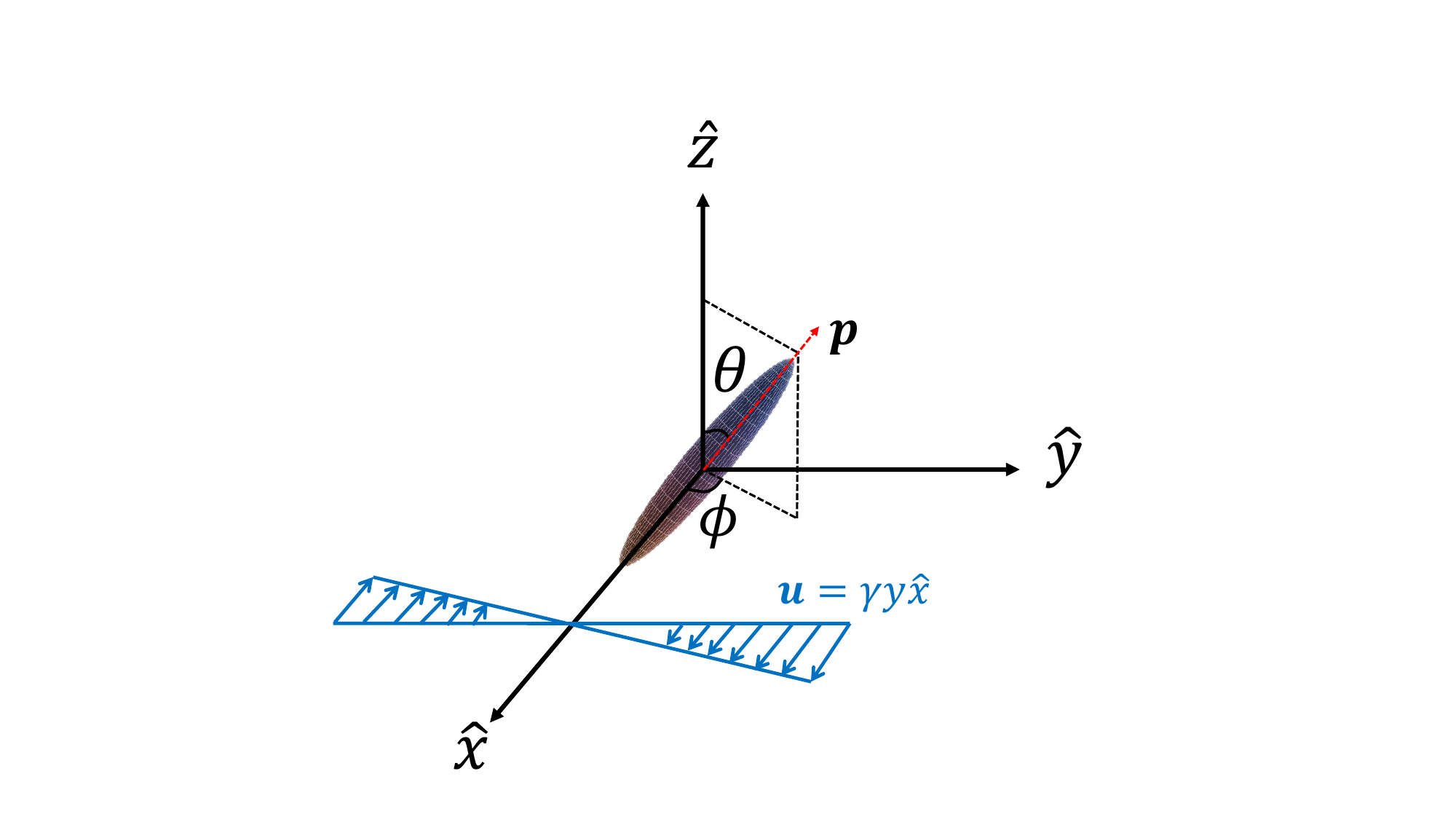}
    \caption{Schematic of a spheroidal particle in simple shear
    flow. The orientation vector $\boldsymbol{p}$ is aligned
    along the symmetry axis of a spheroid, $\theta$ and $\phi$ are the polar
    and azimuthal angles, $\gamma$ is the shear rate, and
    $\boldsymbol{u}=\gamma y\hat{\boldsymbol{x}}$ is the
    background velocity. The flow, velocity-gradient, and
    vorticity directions are $\hat{\boldsymbol{x}}$,
    $\hat{\boldsymbol{y}}$, and $\hat{\boldsymbol{z}}$,
    respectively.}
    \label{fig:sd}
\end{figure}

We consider a spheroid with orientation vector $\boldsymbol{p}$
along its symmetry axis, settling under gravity through a simple
shear flow $\boldsymbol{u}=\gamma y\hat{\boldsymbol{x}}$ with
constant shear rate $\gamma$. The coordinate system is shown in
figure~\ref{fig:sd}: $\hat{\boldsymbol{x}}$ is the flow
direction, $\hat{\boldsymbol{y}}$ is the velocity-gradient
direction, and $\hat{\boldsymbol{z}}$ is the vorticity direction.
The particle has semi-axis length $a$ along the symmetry direction and
transverse semi-axis length $b$, with an aspect ratio $r=a/b$ and
Bretherton constant
$\mathcal{B}=(r^{2}-1)/(r^{2}+1)$, so that
$0<\mathcal{B}<1$ for prolate and $-1<\mathcal{B}<0$ for oblate
spheroids. Further, length of a major dimension is denoted by $\ell$ ($\ell=a$ for a prolate and $\ell=b$ for an
oblate spheroid).

\subsection{Orientation equation and inertial corrections}

In the viscous regime, the orientation dynamics of a settling
spheroid in simple shear receives inertial corrections from two
distinct sources. The first arises from the background shear and
enters the rotation rate at $O(\mathrm{Re}_{s})$, where
$\mathrm{Re}_{s}=\gamma \ell^{2}/\nu$ is the shear Reynolds
number. This correction was derived for slender fibres by
\citet{subramanian2005inertial} and extended to spheroids of
arbitrary aspect ratio by \citet{dabade2016effect} and
\citet{marath2017effect}. The second arises from the
translational motion of a settling particle and enters at
$O(\mathrm{Re}_{p}^{2}/\mathrm{Re}_{s})$, where
$\mathrm{Re}_{p}=W\ell/\nu$ is the particle Reynolds number based
on the settling speed $W$; this is the inertial torque
derived by \citet{khayat1989inertia} and
\citet{dabade2015effects}. The general orientation equation, rendered dimensionless using $\gamma^{-1}$ as the
time scale, including both corrections takes the form
\begin{multline}\label{eq:full_pdot}
    \dot{p}_{i} =
    \underbrace{R_{ij}p_{j}
    + \mathcal{B}\left[S_{ij}p_{j}
    - p_{i}p_{j}p_{k}S_{jk}\right]}_{O(1):\;
    \text{Jeffery torque}}
    + \underbrace{\mathrm{Re}_{s}\,
    \mathcal{H}_{i}(\boldsymbol{p},\boldsymbol{S},
    \boldsymbol{R},\mathcal{B})}_{O(\mathrm{Re}_{s}):\;
    \text{shear-induced inertia}}
    - \underbrace{\mathcal{K}\mathcal{F}_{p}\,p_{k}\hat{g}_{k}
    \left(p_{i}p_{j}\hat{g}_{j}
    - \hat{g}_{i}\right)}_{O(\mathcal{K}):\;
    \text{settling-induced inertia}},
\end{multline}
where $\boldsymbol{R}$ and $\boldsymbol{S}$ are the rotation
and strain-rate tensors of the background flow,
$\hat{\boldsymbol{g}}=\boldsymbol{g}/|\boldsymbol{g}|$ is the
unit gravity vector, and
$\mathcal{H}_{i}$ is a quadratic function of
$\boldsymbol{S}$ and $\boldsymbol{R}$ whose explicit form is
given by \citet{subramanian2005inertial} and
\citet{dabade2016effect}. The settling parameter
$\mathcal{K}=\rho_{f}\tau_{p}^{2}g^{2}/(8\pi\mu\gamma)$
quantifies the importance of settling-induced inertia relative to the
imposed shear, and the modified shape factor
$\mathcal{F}_{p}=\mathcal{F}/(X_{A}Y_{A}Y_{C})$
accounts for the particle geometry through the shape factor
$\mathcal{F}$ and the resistance functions $X_{A}$, $Y_{A}$,
and $Y_{C}$
\citep{kim2013microhydrodynamics,dabade2015effects}. Here $\mathcal{F}$, $X_{A}$, $Y_{A}$ and $Y_{C}$ are normalised
by the semi-major axis $\ell$, with $\mathcal{F}<0$ for prolate ($r>1$) and
$\mathcal{F}>0$ for oblate ($r<1$) spheroids. Here, $\rho_{f}$ and $\mu$ denote the fluid density and dynamic viscosity, respectively, while $\tau_{p}=m/(6\pi\mu \ell)$ is the particle relaxation time. Here $m$ is the buoyant excess mass, $m=(\rho_{p}-\rho_{f})V$ with $V$ is the particle
volume, so that $W=\tau_{p}g$ is the isotropic Stokes settling speed. Using the total
mass instead would overstate $W$ by $\rho_{p}/(\rho_{p}-\rho_{f})$ and
$\mathcal{K}$ by the square of that factor, which is two to three orders of
magnitude for a nearly neutrally buoyant particle.

\subsection{Settling parameter: physical interpretation and turbulence regimes}

Several studies have examined the rotational dynamics of spheroidal particles in turbulence \citep{parsa2012rotation,parsa2014inertial,ni2015measurements,pujara2021shape,zaza2026angular}. These studies exploit the coupling between particle rotation and small-scale turbulent velocity gradients, allowing anisotropic particles to probe local flow structures. For example, \citet{zaza2026angular} inferred the local turbulent dissipation rate from particle angular dynamics.
In contrast, external forces such as gravity introduce preferred orientations that compete with turbulent fluctuations. Beyond a critical regime, a spheroid can exhibit a preferred alignment largely unaffected by background turbulence \citep{roy2023orientation}. Thus, while previous studies use particle orientation to probe turbulence, the present work focuses on orientation dynamics under external forcing.

In studies of non-spherical particles settling in turbulence,
the dynamics is parameterised by a settling number
$\mathrm{Sv}=W/u_{K}$, where
$u_{K}=(\nu\varepsilon)^{1/4}$ is the Kolmogorov velocity
\citep{menon2017theoretical,gustavsson2019effect,anand2020orientation}.
For simple shear with dissipation rate
$\varepsilon=\nu\gamma^{2}$, the Kolmogorov velocity reduces
to $u_{K}=\sqrt{\nu\gamma}=\gamma\ell_{\gamma}$, so that
$\mathrm{Sv}=W/\sqrt{\nu\gamma}$ and
$\mathcal{K}\sim\mathrm{Sv}^{2}$. The bifurcation threshold
$\mathcal{R}=1$ identified in \S\ref{sec:theory} thus
corresponds to a critical settling number
$\mathrm{Sv}_{c}$, defined by $\mathcal{R}=1$, above which the
azimuthal tumbling ceases. The regimes $\mathcal{K}\ll 1$ (tumbling),
$\mathcal{K}\sim O(1)$ (bifurcation), and $\mathcal{K}\gg 1$
(strong alignment) map directly onto the
small-$\mathrm{Sv}$, transitional, and
large-$\mathrm{Sv}$ regimes identified by
\cite{menon2017theoretical} and \cite{roy2023orientation} for spheroids settling in
turbulence.



The two inertial corrections in (\ref{eq:full_pdot})
originate at different outer length scales in the matched asymptotic expansion:
the shear Oseen length $\ell_{\gamma}=\sqrt{\nu/\gamma}$ for the
$O(\mathrm{Re}_{s})$ term and the settling Oseen length $\ell_{s}=\nu/W$ for
the $O(\mathcal{K})$ term. The settling parameter is the squared ratio of the
two, and follows from the definitions of $\mathrm{Re}_{p}$ and
$\mathrm{Re}_{s}$ as
\begin{equation}\label{eq:K_oseen}
    \mathcal{K}
    =\dfrac{1}{8\pi}\frac{\mathrm{Re}_{p}^{2}}
    {\mathrm{Re}_{s}}
    \sim \left(\frac{\ell_{\gamma}}{\ell_{s}}\right)^{2}
    \sim \frac{|\boldsymbol{T}_{I}|}{|\boldsymbol{T}_{J}|},
\end{equation}
the last proportionality identifying $\mathcal{K}$ with the
ratio of the inertial to the Jeffery torque \citep{sheikh2020importance}. The
$O(\mathrm{Re}_{s})$ correction may be dropped when
$\mathrm{Re}_{s}\ll|\mathcal{K}\mathcal{F}_{p}|$, equivalently
$\mathrm{Re}_{p}\gg\sqrt{8\pi/|\mathcal{F}_{p}|}\;\mathrm{Re}_{s}$: the slip
due to settling must dominate the slip due to shear. This is the complement of
the regime analysed by \citet{subramanian2005inertial},
\citet{dabade2016effect} and \citet{marath2017effect}. Retaining only the
settling-induced torque, the resulting equation is
\begin{equation}\label{eq:pdoteq}
    \dot{p}_{i}=R_{ij}p_{j}
    +\mathcal{B}\left[S_{ij}p_{j}
    -p_{i}p_{j}p_{k}S_{jk}\right]
    -\mathcal{K}\mathcal{F}_{p}\,p_{k}\hat{g}_{k}
    \left(p_{i}p_{j}\hat{g}_{j}-\hat{g}_{i}\right).
\end{equation}

 A particle of semi-major axis $30\,\mu\mathrm{m}$ and
twice the density of water settles at $W\approx2\times10^{-3}\,\mathrm{m\,
s^{-1}}$, giving $\mathrm{Re}_{p}=0.06$; at $\gamma=0.1\,\mathrm{s}^{-1}$ this
yields $\mathcal{K}=1.5$ with $\mathrm{Re}_{s}=9\times10^{-5}$, so all three
conditions hold at once and the system sits above threshold. The inertial torque in (\ref{eq:pdoteq})
is derived as a leading-order correction in
$\mathrm{Re}_{p}$ and is therefore strictly valid for
$\mathcal{K}\geq O(10^{-2})$ with $\mathrm{Re}_{p}\ll 1$. In the
results that follow, we present calculations over a broader
range of $\mathcal{K}$ in order to expose the full bifurcation
structure and the qualitative changes in orientation statistics
across the different dynamical regimes. 


\subsection{Evolution equations in spherical coordinates}\label{sec:spherical}

The orientation vector is parameterised by the polar angle
$\theta$ (measured from the vorticity axis
$\hat{\boldsymbol{z}}$) and the azimuthal angle $\phi$
(measured from the flow direction $\hat{\boldsymbol{x}}$):
\begin{equation}\label{eq:uv}
    \boldsymbol{p}=\sin\theta\cos\phi\,\hat{\boldsymbol{x}}
    +\sin\theta\sin\phi\,\hat{\boldsymbol{y}}
    +\cos\theta\,\hat{\boldsymbol{z}}.
\end{equation}
The direction of gravity is similarly specified by angles
$\alpha$ and $\beta$:
\begin{equation}\label{eq:gdir}
    \boldsymbol{g}=-g\left(\sin\alpha\cos\beta\,
    \hat{\boldsymbol{x}}+\sin\alpha\sin\beta\,
    \hat{\boldsymbol{y}}+\cos\alpha\,
    \hat{\boldsymbol{z}}\right).
\end{equation}
Substituting (\ref{eq:uv}) and (\ref{eq:gdir}) into
(\ref{eq:pdoteq}) yields the coupled system
\begin{subequations}\label{eq:gtpeq}
\begin{multline}\label{eq:the}
    \dot{\theta}=\frac{\mathcal{B}}{4}\sin 2\theta\sin 2\phi
    -\frac{\mathcal{K}\mathcal{F}_{p}}{2}
    \left(\cos\alpha\cot\theta
    +\cos(\beta-\phi)\sin\alpha\right)\\
    \times\left(2\cos\alpha\sin^{2}\theta
    -\cos(\beta-\phi)\sin\alpha\sin 2\theta\right),
\end{multline}
\begin{equation}\label{eq:phe}
    \dot{\phi}=\frac{1}{2}\left[-1+\mathcal{B}\cos 2\phi
    +\mathcal{K}\mathcal{F}_{p}\left\{
    \cot\theta\sin 2\alpha\sin(\beta-\phi)
    +\sin^{2}\!\alpha\sin\bigl(2(\beta-\phi)\bigr)
    \right\}\right].
\end{equation}
\end{subequations}

The system (\ref{eq:gtpeq}) is more compact than it
appears. Introducing the orientation potential
\begin{equation}\label{eq:potential}
    \Chi(\boldsymbol{p})=\dfrac{1}{2}\left[
    \mathcal{B}\,\boldsymbol{p}\cdot\boldsymbol{S}\cdot\boldsymbol{p}
    +\mathcal{K}\mathcal{F}_{p}
    (\hat{\boldsymbol{g}}\cdot\boldsymbol{p})^{2}\right],
\end{equation}
equation (\ref{eq:pdoteq}) becomes
\begin{equation}\label{eq:gradflow}
    \dot{\boldsymbol{p}}
    =\boldsymbol{\omega}\times\boldsymbol{p}+\nabla_{s}\Chi,
    \qquad
    \boldsymbol{\omega}=\dfrac{1}{2}\nabla\times\boldsymbol{u}
    =-\dfrac{1}{2}\hat{\boldsymbol{z}},
\end{equation}
where $\nabla_{s}$ denotes the gradient on the unit
sphere. The orientation dynamics is therefore a rigid rotation at the
ambient angular velocity together with gradient ascent on a single
quadratic form. In the coordinates (\ref{eq:uv}) and (\ref{eq:gdir}),
(\ref{eq:gtpeq}) is equivalent to
\begin{subequations}\label{eq:gtpeqcompact}
    \begin{equation}
        \dot\theta=\frac{\partial\Chi}{\partial\theta},
        \qquad
        \dot\phi=-\frac{1}{2}
        +\frac{1}{\sin^{2}\theta}\frac{\partial\Chi}{\partial\phi},
    \end{equation}
    \begin{equation}
        \Chi=\frac{\mathcal{B}}{4}\sin^{2}\theta
        \sin 2\phi+\frac{\mathcal{K}\mathcal{F}_{p}}{2}
        \bigl(\cos\alpha\cos\theta
        +\sin\alpha\sin\theta\cos(\beta-\phi)\bigr)^{2}.
    \end{equation}
\end{subequations}

The gradient structure of the Jeffery term is due to
\citet{brenner1974transport}, who established that
$\nabla_{s}(\tfrac{1}{2}\boldsymbol{p}\cdot\boldsymbol{Q}\cdot\boldsymbol{p})
=\boldsymbol{Q}\cdot\boldsymbol{p}-(\boldsymbol{p}\cdot\boldsymbol{Q}\cdot
\boldsymbol{p})\boldsymbol{p}$ for any symmetric $\boldsymbol{Q}$, and used it
to build a rotary velocity potential for an axisymmetric Brownian particle in a
linear flow. What (\ref{eq:potential}) adds is that the settling-induced
inertial torque is quadratic in $\boldsymbol{p}$ as well, so a single potential
carries both: the two terms of (\ref{eq:pdoteq}) are of this form with
$\boldsymbol{Q}=\mathcal{B}\boldsymbol{S}$ and
$\boldsymbol{Q}=\mathcal{K}\mathcal{F}_{p}\hat{\boldsymbol{g}}
\hat{\boldsymbol{g}}$. The form (\ref{eq:gradflow}) is also not specific to
simple shear. The identity is pointwise in time, so
(\ref{eq:gradflow}) holds for an arbitrary linear flow
$\boldsymbol{\Gamma}(t)$ and a fluctuating slip direction, and hence for a
sub-Kolmogorov spheroid settling in turbulence, with $\boldsymbol{\omega}$ and
$\Chi$ evaluated at the instantaneous $\boldsymbol{S}(t)$. A steady
$\boldsymbol{\Gamma}$ is needed only for $\Chi$ to be a fixed potential.

We use (\ref{eq:gtpeq}) and (\ref{eq:gtpeqcompact})
interchangeably below. Since $\hat{\boldsymbol{g}}$ enters only through
$(\hat{\boldsymbol{g}}\cdot\boldsymbol{p})^{2}$, the sign convention in
(\ref{eq:gdir}) is immaterial. Two features are immediate. When gravity
lies in the shear plane ($\alpha=\pi/2$), every $\theta$-dependence of
$\Chi$ carries a factor $\sin^{2}\theta$, so $\dot\theta$ carries a
factor $\sin 2\theta$ and the equatorial circle $\theta=\pi/2$ is an
invariant circle of the deterministic dynamics. When gravity is
parallel to the vorticity axis ($\alpha=0$), the settling contribution
reduces to $(\mathcal{K}\mathcal{F}_{p}/2)\cos^{2}\theta$, which is
independent of $\phi$, so the azimuthal equation is that of Jeffery at
every $\mathcal{K}$.
Setting $\mathcal{K}=0$ recovers the classical Jeffery
dynamics. We analyse three canonical configurations:
gravity parallel to the vorticity axis ($\alpha=0$), to the
flow-gradient direction ($\alpha=\pi/2$, $\beta=\pi/2$), and
to the flow direction ($\alpha=\pi/2$, $\beta=0$).

We define
$\mathcal{K}_{p}=\mathcal{K}\mathcal{F}_{p}$. In the
deterministic analysis (\S\ref{sec:theory}), $\mathcal{K}$ is
preferred because it isolates the effect of settling from that
of the aspect ratio. In the stochastic analysis
(\S\ref{sec:sod}), we use $\mathcal{K}_{p}$, absorbing the
aspect-ratio dependence into a single parameter.

\section{Orientation dynamics due to hydrodynamic and inertial torques}\label{sec:theory}

 
 


At $\mathcal{K}=0$, the Jeffery equation admits a conserved
orbit constant $C(\theta,\phi)$, whose level sets are the
classical Jeffery orbits---a one-parameter family of closed
curves nested on $S^{2}$ \citep{jeffery1922motion}. The
existence of $C$ precludes isolated attractors or repellers:
all trajectories are periodic, and the long-time dynamics
depends on the initial orientation. General topological constraints on the dynamics, including the exclusion of chaos for the present autonomous system, are provided in appendix~\ref{appphase}.


A non-zero settling parameter $\mathcal{K}$ destroys the conserved orbit constant.
Without $C$, the system is no longer constrained to closed
orbits and can develop isolated attractors and repellers. The
settling-induced inertial torque thus plays a qualitatively
distinct role: it breaks the topological protection afforded by
the conserved quantity, permitting the bifurcations from
periodic orbits to stable equilibria identified in the following
subsections.
 
 
For the canonical configurations in which gravity lies in the
flow-gradient plane ($\alpha=\pi/2$ and $\beta=0$ or $\pi/2$), the $\dot{\theta}$ equation
factors as $\dot{\theta}=(\sin 2\theta/4)\,\Psi(\phi)$, where
$\Psi(\phi)=\mathcal{B}\sin 2\phi
+2\mathcal{K}\mathcal{F}_{p}\cos^{2}(\beta-\phi)$ is independent of $\theta$. The global
fixed points $\theta^{*}=0,\pi/2,\pi$ are therefore independent
of $\phi$, and the nontrivial dynamics resides in the azimuthal
equation. In both the flow-gradient ($\beta=\pi/2$) and
flow-direction ($\beta=0$) cases, this equation can be written in
the compact form
\begin{equation}\label{eq:phi_compact}
    \dot{\phi} = \frac{1}{2}\left[-1
    + \mathcal{R}\cos(2\phi-\psi)\right],
\end{equation}
where the effective amplitude
$\mathcal{R}=\sqrt{\mathcal{B}^{2}
+\mathcal{K}^{2}\mathcal{F}_{p}^{2}}$ and the phase shift
$\psi=\mathrm{atan2}(\pm\mathcal{K}\mathcal{F}_{p},\mathcal{B})$,
the upper sign for $\beta=\pi/2$ and the lower for $\beta=0$. The two-argument
form is needed because $\mathcal{B}<0$ for oblate spheroids, for which
$\arctan(\mathcal{K}\mathcal{F}_{p}/\mathcal{B})$ selects the wrong quadrant;
the threshold $\mathcal{R}=1$ is unaffected, but the branch locations and their
stability labels are not. Equation
(\ref{eq:phi_compact}) is the normal form for overdamped motion
in a tilted periodic potential
\citep{strogatz2015nonlinear}. Fixed points of (\ref{eq:phi_compact}) exist if and only if $\mathcal{R}\geq 1$. At $\mathcal{R}=1$ a single degenerate (semi-stable) fixed point appears on the invariant circle. This is a saddle-node bifurcation on an invariant circle (SNIC). For $\mathcal{R}>1$ it splits into a stable node and a saddle, and the rotation is arrested. Below threshold ($\mathcal{R}<1$) the azimuthal velocity never vanishes and the spheroid rotates continuously with period $T=2\pi/\sqrt{1-\mathcal{R}^{2}}$, which diverges as $(1-\mathcal{R})^{-1/2}$ as the bifurcation is approached from below.



\subsection{Gravity parallel to the vorticity axis ($\hat{\boldsymbol{z}}$ axis)}\label{sec:gz}

 In the first configuration, we consider the case where the gravity vector is parallel to the vorticity axis ($\alpha=0$ and $\boldsymbol{g}=-g\boldsymbol{\hat{z}}$). Such a flow configuration exists in a vertical cylindrical Couette flow and  was examined experimentally by \citet{di2024influence}. The governing equations (\ref{eq:gtpeq}) are accordingly modified, and analytical solutions are derived for a particular configuration (see appendix~\ref{appA}).

\begin{figure}
\hspace{-1cm}
\includegraphics[width=1.15\linewidth]{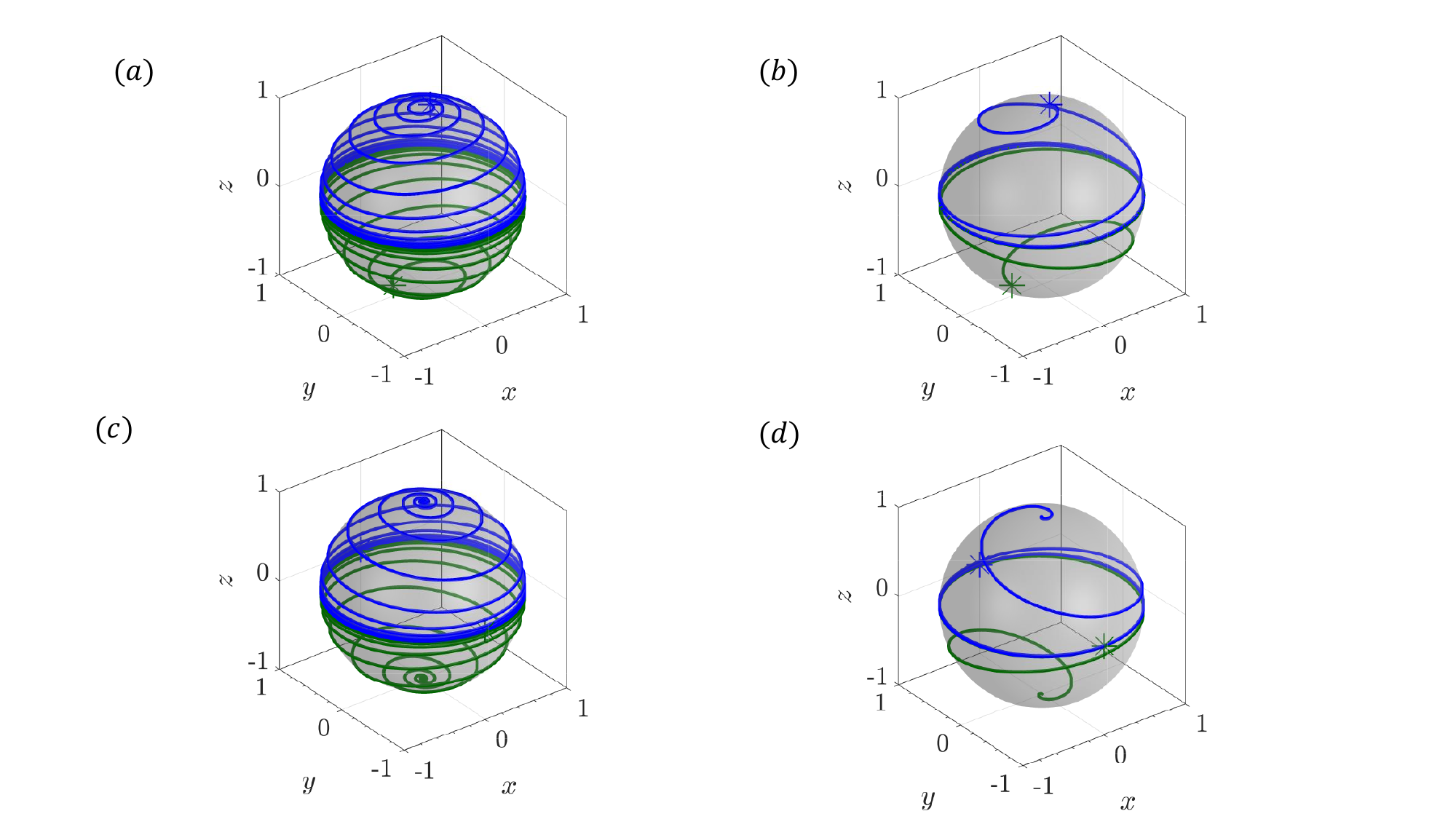}
\caption{Trajectories of orientation vector on the unit sphere when gravity vector is parallel to the vorticity vector. The Bretherton constant is fixed  $|\mathcal{B}|=0.1$ ($r=1.1$ for prolate spheroid and $r=0.9$ for an oblate spheroid). (a) $\mathcal{K}=0.1$ and prolate spheroid, (b) $\mathcal{K}=0.5$ and prolate spheroid, (c) $\mathcal{K}=0.14$ and oblate spheroid, and (d) $\mathcal{K}=0.7$ and oblate spheroid. } 
\label{fig:ns_gz_orbit}
\end{figure} 

For the case of vorticity-aligned ($\alpha=0$), the settling
parameter does not enter the azimuthal equation, and 
$\dot{\phi}$ reduces to the pure Jeffery form with no fixed
points (since $|\mathcal{B}|<1$ for all physical spheroids). The
settling instead acts solely on $\theta$, selecting a unique polar
angle from the degenerate family of Jeffery orbits. The resulting
attractor is a limit cycle (for a prolate spheroid), inherited from the Jeffery
rotation with a uniquely determined polar confinement.

 The rotational dynamics of a spheroid deviates from the classical Jeffery solution. In Jeffery's solution, the angle $\theta$ exhibits an oscillatory variation in time that depends on the initial orientation, causing the indeterminate rotational motion. In contrast, for a non-zero value of $\mathcal{K}$, $\theta$ converges to the steady state, leading to the rotational motion either in the flow-gradient plane or along the vorticity axis for all initial orientations. In figure \ref{fig:ns_gz_orbit}, we show the orbit trajectory of both prolate and oblate spheroids on the unit sphere. The Bretherton constant is kept fixed at $\mathcal{B}=0.1$ ($r=1.1$). One end of the orientation vector is fixed at the centre of the unit sphere, while allowing the other end to move, which results in rotational trajectories on the unit sphere. The trajectories of motion of the orientation vector are traced out on the unit sphere for two different initial orientations ($\theta_0,\phi_0$). In the first two panels, blue curve corresponds to $\theta_{0}$ and $\phi_{0}=0.1$, and the green curve corresponds to $\theta_{0}$ and $\phi_{0}=0.9\pi$. The blue and green crosses represent the position of the initial orientation of a spheroid on the
unit sphere. The results clearly demonstrate that the inclusion of inertial torque removes the degeneracy inherent in Jeffery orbits. In particular, trajectories of a prolate spheroid originating from different initial conditions converge to a common orbit in the flow--gradient plane, indicating a preferred rotational state induced by fluid inertia. When $\mathcal{K}=0.5$, the particle quickly converges to the preferred rotation with less oscillation of orientation in the transient.

In the next two panels, we show the trajectories of an oblate spheroid with $\mathcal{B}=-0.1$ ($r=0.9$). The initial orientations are $(\theta_{0},\phi_{0})=(\pi/2-\delta\theta,\pi/2)$ and $(\pi/2+\delta\theta,3\pi/2)$, corresponding to the blue and green crosses, respectively and $\delta\theta\approx10^{-4}$. In both cases, the trajectories converge towards alignment with the vorticity axis ($\theta^{*}=0$), and the particle exhibits log-rolling motion. As $\mathcal{K}$ increases, the amplitude of orientation oscillations decreases and the particle approaches the log-rolling state more rapidly. It should be noted that we observe similar tumbling (log-rolling) behaviour for prolate (oblate) spheroid even for large values of $\mathcal{K}$ (not shown in the figure). Increasing $|\mathcal{B}|$ has the same qualitative effect as
increasing $\mathcal{K}$: the amplitude of the transient orientation
oscillations decreases for both particle shapes.

A related study by \citet{subramanian2006inertial} has examined the orientation dynamics of nearly spherical particles by accounting for the combined effects of particle and fluid inertia, the latter including contributions from background shear and inertial torque. In their analysis, anisotropy in the shape of a particle was incorporated via an asymptotic expansion in a small parameter measuring the deviation of the aspect ratio from unity. For a similar flow configuration, in which the gravity and vorticity vectors are aligned, they have shown that even a slight departure from sphericity leads to a tumbling motion in the flow-gradient plane. In the present work, by contrast, we isolate the contributions of inertial torque and hydrodynamic drag. We find that qualitatively similar dynamics persist when inertial torque dominates, demonstrating that the tumbling of prolate spheroids and the log-rolling of oblate spheroids are governed primarily by inertial torque. Moreover, we extend the analysis to spheroids of arbitrary aspect ratio and provide analytical solutions that confirm this behaviour and clarify the role of aspect ratio in governing the dynamics.

Recent studies by \citet{cui2024effect} and \citet{cui2025stability} have also examined orientation dynamics in the presence of particle and fluid inertia, the latter arising from shear- and slip-induced effects. Their slip is generated by particle inertia in the shear flow itself and is therefore tied to the local velocity gradient, so its direction is fixed by the flow. Here the slip is set by an external body force, and its direction relative to the vorticity vector is an independent parameter. That single difference is what the present work exploits: the angle $\alpha$ between gravity and the vorticity axis is a control parameter rather than an outcome, and the classification of \S\ref{sec:theory} is organised by it. Where the two problems overlap, at fixed slip direction, we recover the same competition between the Jeffery torque and an inertial torque quadratic in the slip; what is new here is that the character of the attractor, a limit cycle on an invariant great circle for $\alpha=0$ against fixed points off it for $\alpha=\pi/2$, changes with that angle, and that the change is opposite for prolate and oblate particles. As a result, the dynamics is directly influenced by the direction of settling relative to the vorticity vector, leading to distinct alignment behaviour.

\subsection{Gravity parallel to the flow gradient direction ($\hat{\boldsymbol{y}}$ axis)}\label{sec:gy}

In this section, we consider the configuration in which the gravity vector is aligned in the flow-gradient direction. Particle dynamics in such settings has been extensively studied. Classic examples include two-dimensional Couette and Poiseuille flows, among the most widely investigated cases. In geophysical contexts, several theoretical studies have explored particle motion in comparable configurations, for instance, the motion of a particle in surface gravity waves \citep{eames2008settling, mishra2025transport}. These two-dimensional flows continue to provide a fundamental framework for analysing the translational dynamics of spheroids. For the present configuration, with $\alpha=\beta=\pi/2$, the equations of motion mentioned in (\ref{eq:gtpeq}) get modified and analytical solutions are given in appendix~\ref{appA}. 
  For given values of $\alpha$ and $\beta$, the equations of motion yield the fixed points in terms of $(\theta^{*},\phi^{*})$ which are given by

\begin{equation}\label{eq:fpphigy}
                \phi^{*}=\tan^{-1}\left[\dfrac{\mathcal{K}\mathcal{F}_{p}\pm \mathcal{G}_{1}}{1+\mathcal{B}}\right], \quad\text{and}\quad \theta^{*}=0,   \dfrac{\pi}{2}.
\end{equation}

The fixed points $\phi^*$ do not exist when $(\mathcal{B}^{2}+\mathcal{K}^{2}\mathcal{F}^{2}_{p})<1$ (or $\mathcal{R}^2 <1$). When $\mathcal{G}_{1}$ becomes complex (imaginary) identity, a particle undergoes a continuous rotational motion, leading to an oscillatory variation of $\phi$ over time. The condition $\mathcal{G}_{1}^{2}=
\mathcal{B}^{2}+\mathcal{K}^{2}\mathcal{F}_{p}^{2}-1=0$
marks a saddle-node bifurcation on an invariant circle (SNIC).
Below this threshold, $\dot{\phi}$ does not vanish and the spheroid undergoes continuous rotation. At the
threshold, a saddle-node pair appears on the rotating orbit,
arresting the rotation. Approaching the bifurcation from below, the
tumbling period diverges as
$T=2\pi/\sqrt{1-\mathcal{R}^{2}}\sim
(1-\mathcal{B}^{2}-\mathcal{K}^{2}\mathcal{F}_{p}^{2})^{-1/2}$, that is
as $|\mathcal{G}_{1}|^{-1}$, which is the universal scaling of a SNIC
bifurcation. The same square root controls the approach from above: in
the analytical solution (\ref{eq:phi_yeq}) the $\tanh$ argument involves
$\mathcal{G}_{1}t$, so as $\mathcal{G}_{1}\to 0$ the time taken to
reach the equilibrium orientation diverges as $\mathcal{G}_{1}^{-1}$.
The period below the threshold and the relaxation time above it are
distinct quantities on either side of the same bifurcation, and both
scale as $|\mathcal{R}^{2}-1|^{-1/2}$. Below the threshold, no fixed point of $\phi$ exists and the particle rotates
continuously: a prolate spheroid log-rolls for every initial orientation, while
an oblate spheroid tumbles in the flow-gradient plane.

The nature of the fixed points can be further examined from the linear stability analysis. We investigate the stability of a fixed point by evaluating the eigenvalues of the Jacobian matrix $\mathbb{J}$. The Jacobian matrix is a $2\times 2$ matrix and is given by

\begin{equation}\label{eq:jacobian}
    \mathbb{J}=\begin{bmatrix}
        \dfrac{\partial \dot{\theta}}{\partial \theta} & \dfrac{1}{\sin\theta^{*}}\dfrac{\partial \dot{\theta}}{\partial \phi}\\[0.3cm]
         \sin\theta^{*}\dfrac{\partial \dot{\phi}}{\partial \theta} & \dfrac{\partial \dot{\phi}}{\partial \phi}
    \end{bmatrix}_{(\theta^{*},\phi^{*})}
\end{equation}
Here the displacement from the fixed point is measured in the orthonormal
tangent basis, $(\delta\theta,\ \sin\theta^{*}\delta\phi)$, so that
(\ref{eq:jacobian}) includes the metric of the unit sphere; the remaining metric
term, $\cos\theta\,\dot\phi$, vanishes at a fixed point. The matrix
(\ref{eq:jacobian}) equals $\mathsf{M}\mathbb{J}_{c}\mathsf{M}^{-1}$, where
$\mathbb{J}_{c}$ is the Jacobian of $(\dot\theta,\dot\phi)$ with respect to
$(\theta,\phi)$ and $\mathsf{M}=\mathrm{diag}(1,\sin\theta^{*})$, so its trace,
determinant and eigenvalues are those of $\mathbb{J}_{c}$, and the two coincide at
the equatorial fixed points, where $\sin\theta^{*}=1$. At the poles the chart is
singular and (\ref{eq:jacobian}) is not defined; there we linearise in local
Cartesian coordinates $(p_{x},p_{y})$, as in (\ref{eq:lin_pole_gx}).

\begin{figure}
    \centering
    \includegraphics[width=1.0\linewidth]{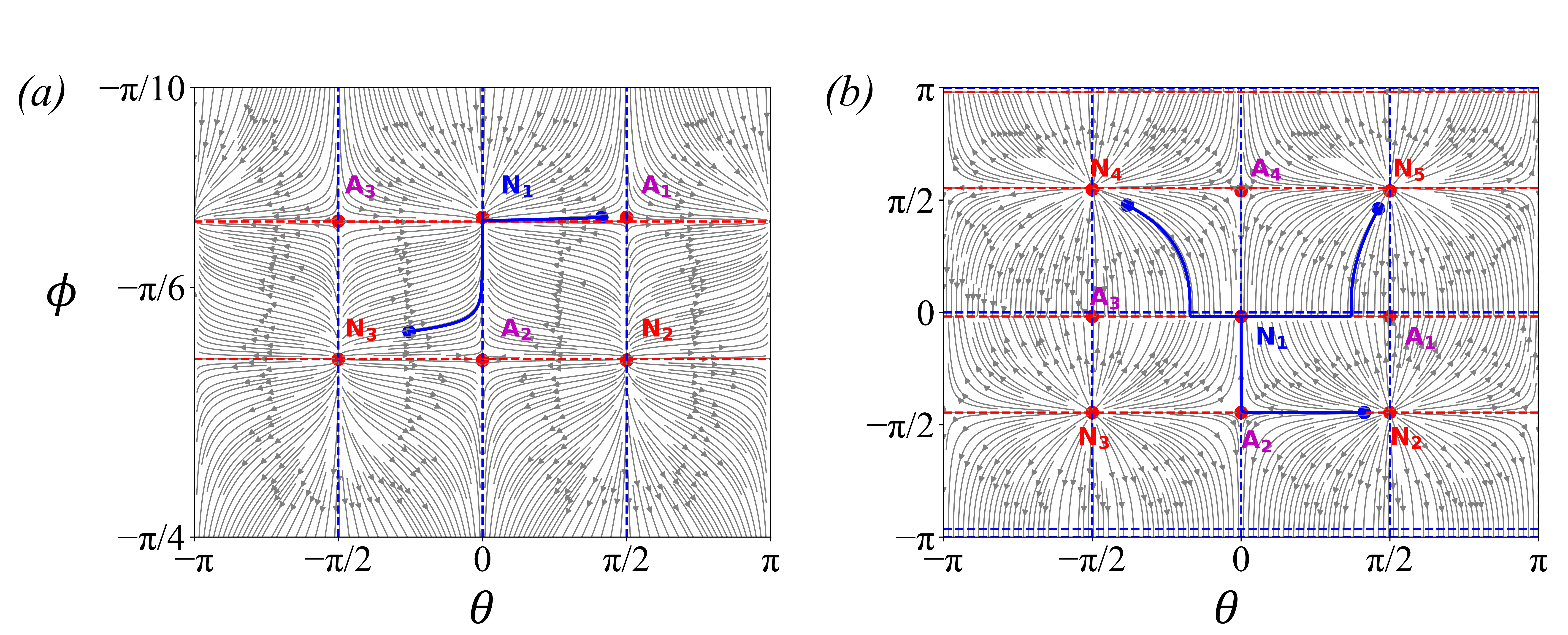}
    \caption{Phase portrait diagrams in the $\theta-\phi$ for a prolate spheroid with gravity aligned along the flow--gradient direction $\alpha=\beta=\pi/2$. The solid blue lines denote the particle trajectories, with the blue dots indicating the initial orientations. The dashed blue and red lines denote the $\theta$- and $\phi$-nullclines, respectively. Further, $N$ and $A$ represent the node and saddle fixed points, respectively. A stable node is indicated by blue colour, an unstable node is represented by red colour, and a saddle point is indicated by magenta colour. The Bretherton constant is fixed, $\mathcal{B}=0.5$ ($r=1.73$). (a)  $\mathcal{K}=0.2$, and (b) $\mathcal{K}=1$.}
    \label{fig:pp_proob_pi_2_pro}
\end{figure}

\begin{figure}
    \centering
    \includegraphics[width=1.05\linewidth]{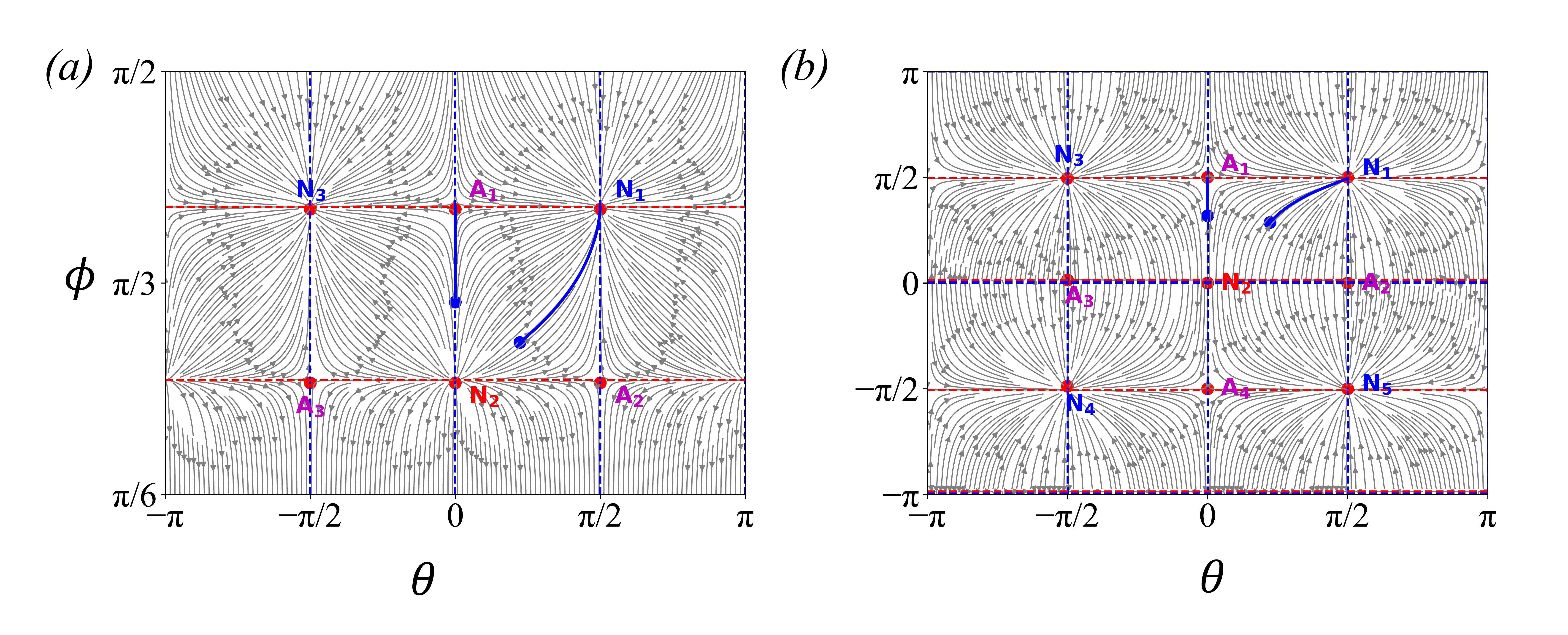}
    \caption{Phase portrait diagrams in the $\theta$--$\phi$ plane
for an oblate spheroid with gravity aligned along the
flow--gradient direction $\alpha=\beta=\pi/2$. The blue lines are the trajectories, and the blue dot indicates the initial orientation.  The dashed blue and red lines denote the $\theta$- and $\phi$-nullclines, respectively. Also, $N$ and $A$ represent the node and saddle fixed points, respectively. A stable node is indicated by blue colour, an unstable node is represented by red colour, and a saddle point is indicated by magenta colour. The Bretherton constant is fixed, $\mathcal{B}=-0.5$ ($r=0.58$). (a)  $\mathcal{K}=0.3$, and (b) $\mathcal{K}=5$.}
    \label{fig:pp_proob_pi_2_ob}
\end{figure}


We next examine the phase portraits for two representative values of $\mathcal{K}$, selected according to their location relative to the bifurcation point. For a prolate spheroid, we consider $\mathcal{K}=0.2$ and $1$. The smaller value, $\mathcal{K}=0.2$, lies close to the bifurcation point, where the phase portrait undergoes qualitative changes associated with the emergence of new fixed points. In contrast, $\mathcal{K}=1$ lies well beyond the bifurcation, where two fixed points exist (see (\ref{eq:fpphigy})). For an oblate spheroid, we consider $\mathcal{K}=0.3$ and $5$.

In figure~\ref{fig:pp_proob_pi_2_pro}, we present phase portraits in the $\theta$--$\phi$ plane for a prolate spheroid with $\mathcal{B}=0.5$. In figures, the nodes and saddles are denoted by $N$ and $A$, respectively. The solid blue lines denote the trajectory, with the blue circle representing the initial orientation. The blue and red dashed lines correspond to $\theta$- and $\phi$-nullclines, respectively. The intersection of the $\theta$- and $\phi$-nullclines results in a fixed point. Note that the orientations $(\theta,\phi)=(-\pi/2,\phi)$ and $(\pi/2,\phi+\pi)$ are equivalent.

 In the first panel, we fix the parameter $\mathcal{K}=0.2$. Here, over the plotted ranges of $\theta$ and $\phi$, the
$\theta$- and $\phi$-nullclines intersect at six points. The plot is a
coordinate-chart continuation rather than a chart of $S^{2}$: the rows
$\theta=\pi/2$ and $\theta=-\pi/2$ are the same great circle, so those six intersections are five distinct points of $S^2$ ($N_1$ and $A_2$ being
the same pole), and under the nematic identification
$\boldsymbol{p}\leftrightarrow-\boldsymbol{p}$ three distinct physical orientations. Appendix~\ref{appphase} sets out the index bookkeeping. We observe two fixed points of $\phi^{*}$ for each $\theta^{*}=0$, $\pi/2$ and $-\pi/2$. The determinant of the Jacobian $\Delta$ at $N_{1}$ is positive, and the trace $\tau$ is negative, corresponding to a stable node fixed point. Similarly, $\Delta$ at the point $A_{2}$ is negative, and hence $A_{2}$ is a saddle point of the $(\theta,\phi)$ chart. From a similar analysis, $A_{1}$ and $N_{2}$ are the saddle and unstable node, respectively. The points $N_{1}$ and $A_{2}$, however,
both lie on $\theta=0$, where the chart is singular, and are therefore images of the single orientation $\boldsymbol{p}=\hat{\boldsymbol{z}}$ rather than distinct equilibria. Linearising in local Cartesian coordinates $(p_x,p_y)$ gives $\operatorname{tr}\mathbb{J}=\mathcal{K}_p$ and $\det\mathbb{J}=(1-\mathcal{B}^2)/4$, so the vorticity pole is a single stable node on $S^2$, with eigenvalues $(\mathcal{K}_p\pm\mathcal{G}_1)/2$; $N_{1}$ and $A_{2}$ mark its slow and fast eigendirections, respectively, and $A_{2}$ is a saddle of the
chart, not of the dynamics on $S^2$. As the magnitude of $\mathcal{K}$ increases, the $\phi$-nullcline containing the fixed points $N_{1}$, $A_1$ and $A_{3}$ approaches the $\theta$-nullcline at $\phi^{*}=0$, thereby reducing the region between the two nullclines. This narrowing region is illustrated in the second panel for $\mathcal{K}=1$, where the fixed points $N_{1}$,$A_1$ and $A_{3}$ lie close to $\phi^{*}\approx 0$. The trajectories within this narrow region exhibit a distinct direction of motion.
 
 As $|\mathcal{K}_{p}|$ grows, the region between the
nullclines narrows (at $\phi^{*}\approx 0$) and the dynamics separates into
two time scales. This is clearest in the potential form
(\ref{eq:gtpeqcompact}). With $\alpha=\beta=\pi/2$ the projection
$\hat{\boldsymbol{g}}\cdot\boldsymbol{p}=\sin\theta\sin\phi$, so
\begin{equation}\label{eq:chi_gy}
    \Chi=\sin^{2}\theta\left[\frac{\mathcal{B}}{4}\sin
    2\phi+\frac{\mathcal{K}_{p}}{2}\sin^{2}\phi\right],
\end{equation}
and, because every $\theta$-dependence of $\Chi$ carries the
same factor $\sin^{2}\theta$, the factor $1/\sin^{2}\theta$ in
(\ref{eq:gtpeqcompact}) cancels and the azimuthal equation closes on itself,
\begin{equation}\label{eq:skew_gy}
    \dot\phi=-\frac{1}{2}+\frac{\mathcal{B}}{2}\cos 2\phi
    +\frac{\mathcal{K}_{p}}{2}\sin 2\phi,
    \qquad
    \dot\theta=\sin 2\theta\left[\frac{\mathcal{B}}{4}\sin 2\phi
    +\frac{\mathcal{K}_{p}}{2}\sin^{2}\phi\right].
\end{equation}
The system is therefore a skew product: $\phi$ evolves
autonomously and drives $\theta$. At large $|\mathcal{K}_{p}|$ the azimuthal
equation is dominated by its settling term, so $\phi$ relaxes on the fast
time scale $|\mathcal{K}_{p}|^{-1}$ to the root of $\dot\phi=0$ nearest the
origin,
\begin{equation}\label{eq:slowman_gy}
    \phi_{s}=\frac{1-\mathcal{B}}{2\mathcal{K}_{p}}
    +O(\mathcal{K}_{p}^{-3}),
    \qquad
    \left.\frac{\partial\dot\phi}{\partial\phi}\right|_{\phi_{s}}
    =\mathcal{K}_{p}+O(\mathcal{K}_{p}^{-1}),
\end{equation}
which is attracting for a prolate spheroid
($\mathcal{K}_{p}<0$) and repelling for an oblate one. The curve
$\phi=\phi_{s}$ is a slow manifold: substituting
$\phi_{s}$ into $\dot\theta$ and expanding gives a residual drift along it,
\begin{equation}\label{eq:slowdrift_gy}
    \dot\theta=\frac{1-\mathcal{B}^{2}}{8\mathcal{K}_{p}}
    \sin 2\theta+O(\mathcal{K}_{p}^{-3}),
\end{equation}
so that near $\theta=0$ the polar angle relaxes at the slow
rate $(1-\mathcal{B}^{2})/(4|\mathcal{K}_{p}|)$. The two rates differ by the
factor $4\mathcal{K}_{p}^{2}/(1-\mathcal{B}^{2})$, and rises quadratically thereafter.
The equilibrium is thus unique, $\theta^{*}=0$ for a prolate spheroid, but the
time taken to reach it grows in proportion to $|\mathcal{K}_{p}|$, so over any
finite observation window a rapidly settling spheroid appears to hold a broad
range of polar angles on the broadside great circle.

In figure~\ref{fig:pp_proob_pi_2_ob}, we present the phase portraits for an oblate spheroid with the same values of $\alpha$ and $\beta$. Applying the same stability analysis, we find that the stability of the fixed points is reversed compared to the prolate case at the larger value of $\mathcal{K}$, while the saddles and nodes interchange their positions at the smaller value of $\mathcal{K}$ and for fixed $\theta^{*}$. In particular, the second panel reveals four stable node-type fixed points and four saddle points within the specified ranges of $\theta$ and $\phi$. Overall, the phase portraits are qualitatively similar to those for the prolate spheroid, except for the reversed stability of the fixed points.




\subsection{Gravity parallel to a flow direction ($\hat{\boldsymbol{x}}$ axis)}\label{sec:gx}

In this section, we consider the case where the gravity vector is aligned with the flow direction. A common example of such a configuration is the vertical channel flow, where the mean flow lies in the direction of gravity. The primary interest lies in evaluating the effect of gravity on the dynamics of non-spherical particles \citep{gavze1996sedimentation,broday1998motion}. In such flows, a non-neutrally buoyant, non-spherical particle exhibits a lateral drift even when the effects of fluid and particle inertia are negligible. The mentioned lateral drift depends on the aspect ratio and orientation of a particle. To analyse the orientation dynamics of a spheroid in the present case, we fix $\alpha=\pi/2$ and $\beta=0$ in (\ref{eq:gtpeq}) and analytical solutions are given in appendix~\ref{appA}.  Similar to previous configurations, the global fixed points are $\theta^{*}=0$ and $\pi/2$. The other fixed points depend on the parameters $\mathcal{B}$ and $\mathcal{K}$ and are given as

\begin{subequations}\label{eq:gxfixphi}
    \begin{equation}
        \phi^{*}_1=-\tan^{-1}\left[\dfrac{\mathcal{K}\mathcal{F}_{p}+\mathcal{G}_{1}}{\mathcal{B}+1}\right],
    \end{equation}
    \begin{equation}\label{eq:fp_xd2}
        \phi^{*}_2=\tan^{-1}\left[\dfrac{-\mathcal{K}\mathcal{F}_{p}+\mathcal{G}_{1}}{\mathcal{B}+1}\right].
    \end{equation}
\end{subequations}

The critical value of $\mathcal{K}$ at which bifurcation occurs remains the same as that shown in (\ref{eq:fpphigy}) for the corresponding values of $\mathcal{B}$. However, the locations of the stable fixed points interchange for prolate and oblate spheroids when compared to the previous case ($\alpha=\beta=\pi/2$). 
 


\begin{figure}
    \centering
    \includegraphics[width=1.05\linewidth]{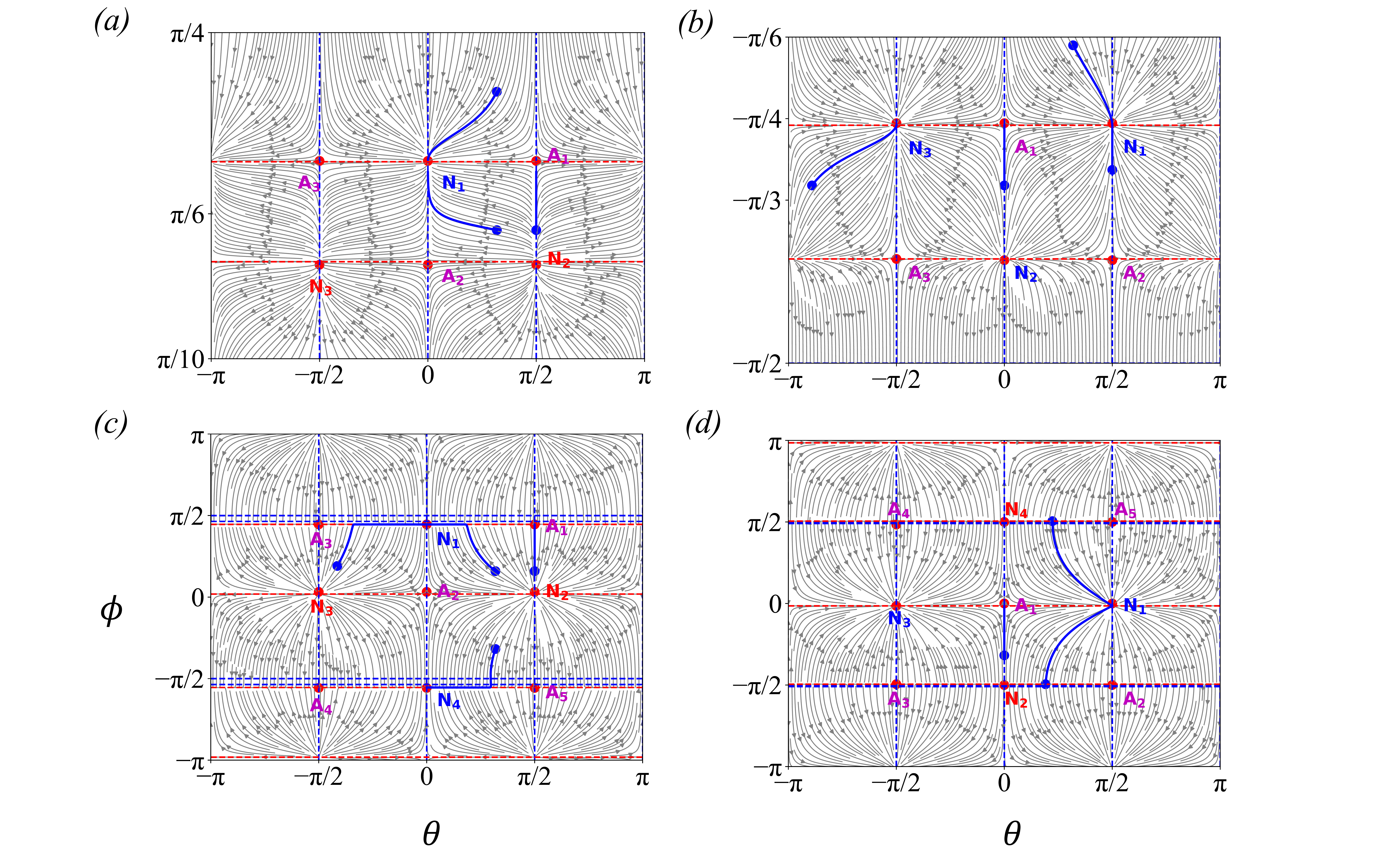}
    \caption{Phase portrait diagrams in $\theta-\phi$ plane when gravity is aligned in a flow  direction $\alpha=\pi/2$ and $\beta=0$. A Bretherton constant is fixed at $|\mathcal{B}|=0.5$ ($r=1.73$ for prolate spheroid and $r=0.58$ for an oblate spheroid). Left-hand side figures are shown for a prolate spheroid, and right-hand side figures correspond to an oblate spheroid. Trajectories are shown in a blue line, with a blue dot representing an initial condition. The dashed blue and red lines denote the $\theta$- and $\phi$-nullclines, respectively. Also, $N$ and $A$ represent the node and saddle fixed points, respectively. A stable node is indicated by blue colour, an unstable node is represented by red colour, and a saddle point is indicated by magenta colour. (a) $\mathcal{K}=0.2$ (Prolate spheroid), (b) $\mathcal{K}=0.3$ (Oblate spheroid), (c) $\mathcal{K}=1$ (Prolate spheroid) and (d) $\mathcal{K}=5$ (Oblate spheroid). }
    \label{fig:ppgxx}
\end{figure}

Figure~\ref{fig:ppgxx} presents the phase portraits for
this configuration at $|\mathcal{B}|=0.5$, with the values of $\mathcal{K}$
selected above the bifurcation point; the
left-hand panels are prolate and the right-hand panels oblate. Their structure
is that of figures~\ref{fig:pp_proob_pi_2_pro} and
\ref{fig:pp_proob_pi_2_ob}, with the fixed points displaced but the
classification unchanged, so we do not repeat the panel-by-panel discussion.
The same transverse eigenvalue (\ref{eq:lamperp}) decides which branch is an
attractor on $S^{2}$. As the magnitude of $\mathcal{K}$ increases, the
$\phi$-nullcline moves towards $\phi^{*}=\pi/2$ and approaches the
$\theta$-nullcline. The two do not merge. Here
$\hat{\boldsymbol{g}}\cdot\boldsymbol{p}=\sin\theta\cos\phi$, so
$\Chi=\sin^{2}\theta[(\mathcal{B}/4)\sin 2\phi
+(\mathcal{K}_{p}/2)\cos^{2}\phi]$ and the azimuthal equation again closes on
itself, giving the same two-time-scale structure as in \S\ref{sec:gy} with the
slow manifold displaced to
$\phi_{s}=\pi/2+(1+\mathcal{B})/(2\mathcal{K}_{p})+O(\mathcal{K}_{p}^{-3})$.
The fast relaxation rate is again $\mathcal{K}_{p}$ and the drift along the
manifold is again (\ref{eq:slowdrift_gy}), so the two in-plane gravity
directions differ only in where the broadside great circle lies, not in the
structure of the dynamics on it.

The preceding analysis characterised the deterministic orientation dynamics of a settling spheroid for the three canonical gravity--vorticity configurations. It is well known that, in the presence of inertial torque, a spheroid tends to adopt broadside-on alignment. However, our results show that a complete description of the rotational dynamics requires two angles. For each configuration, we identify the preferred rotational states---tumbling, log-rolling, or steady equilibrium---and determine the bifurcations governing the transitions between them as functions of the settling parameter $\mathcal{K}$ and the Bretherton constant $\mathcal{B}$. In the rapid-settling regime a slow manifold emerges in the two configurations
with gravity in the shear plane. Settling draws the orientation onto the broadside great circle normal to
gravity at rate $|\mathcal{K}_{p}|$, and shear then selects the orientation on
that circle at the slower rate $(1-\mathcal{B}^{2})/(4|\mathcal{K}_{p}|)$, a
separation of $O(\mathcal{K}_{p}^{2})$. The normal eigenvalue
along that branch is $\mathcal{K}_{p}+O(\mathcal{K}_{p}^{-1})$, so the
broadside manifold attracts for prolate spheroids and repels for oblate ones;
for an oblate spheroid the rapid stage instead aligns the symmetry axis with
gravity. The equilibrium is unique, but the time taken to
reach it grows in proportion to $|\mathcal{K}_{p}|$, so a wide range of polar
angles persists over any finite observation window, depending on the initial
orientation. Identifying this requires the full dynamics in both $\theta$ and
$\phi$. To our knowledge this slow broadside manifold, and the divergence of
its selection time with settling strength, have not been identified in previous
studies. Because the plateau is set by a rate rather than by a degeneracy, it survives
perturbations that would destroy a line of equilibria. Its observed width is
not an intrinsic property: in the deterministic problem it depends on the
initial ensemble and the observation window, and the slow restoring rate. Quantifying it therefore requires a
stated ensemble and a defined measure, which we do not attempt here.

A natural question is how these deterministic structures manifest when the dynamics is perturbed by stochastic noise, as occurs in suspensions due to Brownian diffusion, particle-particle interactions, or turbulent fluctuations. In particular, one expects the bifurcation structure to produce sharp transitions in the orientation moments as functions of $\mathrm{Pe}$ and $\mathcal{K}_{p}$. In \S\ref{sec:sod}, we investigate this connection by solving the steady Fokker-Planck equation for the two flow configurations.

\section{Orientation dynamics of a spheroid in the presence of noise}\label{sec:sod}

In this section, we analyse the orientation dynamics of a spheroid in the presence of stochastic noise. Thermal fluctuations give rise to Brownian rotational diffusion, while athermal sources such as particle-particle collisions and turbulent velocity fluctuations introduce additional randomness into the orientation dynamics. In either case, the alignment of a spheroid is characterised by the orientation distribution function $f(\theta,\phi)$. We note that a closely related approach has been developed in the context of polymer dynamics, where the tumbling and orientation statistics of a dumbbell in a mean shear flow with superimposed random velocity fluctuations have been studied analytically and numerically \citep{chertkov2005polymer,turitsyn2007polymer,celani2005dynamical,puliafito2005numerical}. In those studies, the turbulent velocity gradient is modelled as a short-correlated Gaussian noise, leading to a Fokker-Planck description of the orientation dynamics that is formally analogous to the one employed here. Their results demonstrated that the orientation PDF develops algebraic tails around the shear-preferred direction, with tumbling time statistics that depend on the ratio of the noise intensity to the mean shear rate. The present analysis extends this framework by incorporating the settling-induced inertial torque, which introduces an additional deterministic drift that modifies both the preferred alignment and the structure of the orientation distribution. The probability of finding a spheroid with an angle between $\theta_{1}$ and $(\theta_{1}+d\theta)$, and $\phi_{1}$ and $(\phi_{1}+d\phi)$ is then given by

\begin{equation}
    \mathcal{P}(\theta_{1} \leq \theta \leq \theta_{1}+d\theta,\phi_{1}\leq\phi\leq\phi_{1}+d\phi)=f(\theta_{1},\phi_{1})\sin\theta_{1}d\theta d\phi.
\end{equation}

The orientation distribution function $f(\theta,\phi)$ should satisfy the normalisation condition given by

\begin{equation}
    \int^{2\pi}_{0}\int^{\pi}_{0}f(\theta,\phi)\sin\theta d\theta d\phi=1.
\end{equation}

The evolution of the orientation distribution function is described by the Fokker-Planck equation as
\begin{equation}\label{eq:fpsto}
    \dfrac{\partial f}{\partial t}+\nabla_{p}\cdot (\dot{\boldsymbol{p}}f)=\mathcal{D}_{r}\nabla^{2}_{p}f.
\end{equation}

Here, $\nabla_{p}$ represents the gradient on the unit sphere with
respect to $\boldsymbol{p}$, the term
$(\dot{\boldsymbol{p}}f)$ is the deterministic flux in orientation
space, and the right-hand side represents rotary-diffusion with
coefficient $\mathcal{D}_{r}$.

The diffusion coefficient $\mathcal{D}_{r}$ can have either a thermal
or an athermal origin. For Brownian particles, it arises from thermal fluctuations and is given by the Stokes-Einstein
relation $\mathcal{D}_{r}=k_{B}T/\zeta_{r}$, where $k_{B}$ is
the Boltzmann constant, $T$ the absolute temperature and
$\zeta_{r}=8\pi\mu \ell^{3}Y_{C}$ the rotational friction coefficient about an
axis transverse to the symmetry axis. In turbulent or
fluctuating flows, however, the dominant source of rotational
diffusion is athermal: the rapidly varying velocity gradient
experienced by the particle acts as an effective noise that randomises
its orientation. In this case, the rotary diffusivity is set by the
integral time scale and variance of the velocity gradient
fluctuations. Specifically, if the fluctuating component of the
velocity gradient tensor is $\boldsymbol{G}'(t)$ with zero mean and
autocorrelation time $\tau_{\eta}$, the effective rotary diffusivity
scales as \citep{krushkal1988orientation,shin2005rotational}
\begin{equation}\label{eq:Dr_turb}
    \mathcal{D}_{r}^{\mathrm{turb}} \sim
    \langle |\boldsymbol{G}'|^{2}\rangle\,\tau_{\eta},
\end{equation}
\noindent where $\langle |\boldsymbol{G}'|^{2}\rangle$ is the
variance of the velocity gradient fluctuations. In the Kraichnan
(white-noise) limit, where $\tau_{\eta}\to 0$ with
$\langle |\boldsymbol{G}'|^{2}\rangle\,\tau_{\eta}$ held fixed, the
Fokker-Planck description (\ref{eq:fpsto}) becomes exact. This is the
same modelling framework employed in studies of polymer tumbling in
turbulent shear, where the chaotic velocity gradient is represented as
a short-correlated random process superimposed on a mean shear. The reduction of a fluctuating velocity gradient to a scalar
multiple of the Laplace-Beltrami operator is not automatic. A short-correlated
gradient produces an orientation-dependent diffusion tensor whose form depends
on the rank-four covariance of the gradient fluctuations and on particle shape.
It collapses to isotropic rotary diffusion only when that covariance is itself
isotropic, as it is in homogeneous isotropic turbulence, and even then the
coefficient carries a shape factor,
$\mathcal{D}_{r}^{\mathrm{turb}}=\langle|\boldsymbol{G}'|^{2}\rangle\tau_{\eta}
(5+3\mathcal{B}^{2})/60$, which varies by a factor of $8/5$ between a sphere
and a slender particle. In the present work we therefore treat
$\mathcal{D}_{r}$ as a given isotropic rotary diffusivity, and the results
apply to thermal noise and to isotropic effective rotary noise; extending them
to anisotropic gradient statistics requires the tensorial generator rather than
the scalar one used here.

The evolution of $f(\theta,\phi)$ in (\ref{eq:fpsto}) is governed by
the combined effects of hydrodynamic torque, settling-induced inertial
torque, and rotary diffusion. The relative importance of these effects
is characterised by two dimensionless ratios of the three relevant
time scales:

\begin{equation}\label{eq:fp}
    \mathrm{Pe}=\dfrac{\tau_{d}}{\tau_{c}} ~  \text{and}~ \mathcal{K}_{p}=\dfrac{\tau_{c}}{\tau_{k}}.
\end{equation}

Here, $\tau_{d}$ is a diffusive time scale, $\tau_{d}=1/\mathcal{D}_{r}$. Moreover, $\tau_{c}$ is a convective time scale, $\tau_{c}=1/\gamma$ and $\tau_{k}$ is a settling time scale defined as
$\tau_{k}=8\pi\mu/(\rho_{f}\tau_{p}^{2}g^{2}|\mathcal{F}_{p}|)$,
which is positive for both particle shapes. Since $\mathcal{F}_{p}$ changes
sign between prolate and oblate spheroids, $\mathcal{K}_{p}$ is a signed rate
ratio rather than a ratio of time scales,
$\mathcal{K}_{p}=\mathrm{sgn}(\mathcal{F}_{p})\,\tau_{c}/\tau_{k}$, negative
for prolate and positive for oblate particles. Its sign carries the physics:
it fixes which of the two invariant sets on $S^{2}$ attracts. When $\mathrm{Pe}\ll 1$, diffusion dominates and the orientation distribution is nearly isotropic. When $\mathrm{Pe}\gg 1$, the deterministic dynamics (shear and settling) governs the orientation, and diffusion acts as a weak perturbation. The parameter $\mathcal{K}_{p}$ quantifies the relative strength of settling and shear: when $|\mathcal{K}_{p}|\gg 1$, the inertial torque dominates and the particle approaches its settling-induced preferred alignment even in the presence of noise. The product $\mathrm{Pe}\,\mathcal{K}_{p}=\tau_{d}/\tau_{k}$ measures the importance of settling relative to diffusion directly; when $\mathrm{Pe}\,|\mathcal{K}_{p}|\gg 1$, the settling-induced alignment is robust against diffusive randomisation.

Following the steady assumption, the evolution of the orientation distribution function is governed by the steady dimensionless Fokker--Planck equation, which can be expressed using (\ref{eq:fp}) as

\begin{equation}\label{eq:stfp}
    \nabla^{2}_{p}f-\mathrm{Pe}\nabla_{p}\cdot( \dot{\boldsymbol{p}}f)=0.
\end{equation}

In the above equation, one can substitute the expression of the evolution of the orientation vector $\boldsymbol{\dot{p}}$ in the drift term to evaluate $f(\theta,\phi)$. More conveniently, the drift term in (\ref{eq:stfp}) is written in the spherical coordinate system as

\begin{equation}\label{eq:dsc}
    \nabla_{p}\cdot (\dot{\boldsymbol{p}}f)=\dfrac{1}{\sin\theta}\dfrac{\partial\{ f\dot{\theta}\sin\theta \}}{\partial \theta}+\dfrac{\partial \{ f\dot{\phi} \}}{\partial \phi}.
\end{equation}

Incorporating the expressions of the rate of change of orientation from (\ref{eq:gtpeq}) into the above equation, we write the drift term for non-zero $\mathcal{K}_{p}$ as follows

\begin{multline}
    \nabla_{p}\cdot (\dot{\boldsymbol{p}}f)= \dfrac{1}{\sin\theta}\left[\mathcal{K}_{p} \cos^{2}(\beta-\phi)\sin^{2}\alpha-\mathcal{K}_{p}\cos^{2}\alpha+\mathcal{B}\sin\phi\cos\phi \right]\dfrac{\partial f \sin^{2}\theta \cos\theta}{\partial \theta}+\\
    \dfrac{\mathcal{K}_{p}}{2}\dfrac{\sin2\alpha \cos(\beta-\phi)}{\sin\theta}\dfrac{\partial(f \sin\theta\cos^{2} \theta)}{\partial \theta}-\dfrac{\mathcal{K}_{p}}{2}\dfrac{\sin2\alpha \cos(\beta-\phi)}{\sin\theta}\dfrac{\partial(f \sin^{3}\theta)}{\partial \theta}+\\
    \dfrac{(\mathcal{B}-1)}{2}\dfrac{\partial f}{\partial \phi}-\mathcal{B} \dfrac{\partial (f\sin^{2}\phi)}{\partial \phi}+\dfrac{\mathcal{K}_{p}}{2}\left[\cot\theta \sin2\alpha \dfrac{\partial (f \sin(\beta-\phi))}{\partial \phi}  +\sin^{2}\alpha \dfrac{\partial (f\sin(2\beta-2\phi))}{\partial \phi}\right].
\end{multline}
In substituting $\mathcal{K}_{p}=0$ and $\mathcal{B}=1$ in the above, the drift of a dumbbell is recovered. Furthermore, the first term in (\ref{eq:stfp}) can also be written in spherical coordinates as

\begin{equation}\label{eq:lapsm}
    \nabla^{2}_{p}f=\Lambda f=\dfrac{1}{\sin\theta}\dfrac{\partial }{\partial \theta}\left(\sin\theta\dfrac{\partial f}{\partial \theta}\right)+\dfrac{1}{\sin^{2}\theta}\dfrac{\partial^{2}f}{\partial\phi^{2}}.
\end{equation}

Here, $\Lambda$ denotes the linear operator. Incorporating stochastic noise into the orientation dynamics has a direct impact on the bulk rheology of the suspension. Specifically, properties such as effective viscosity and bulk stresses are of particular interest. To evaluate these rheological properties, we first need to determine the orientation distribution function. After that, we compute various orientation moments, which are subsequently used to assess the fluid stress tensor. Hence, our primary objective is to evaluate the distribution function $f(\theta, \phi)$  and use it to compute the orientation moments for different angles $\alpha$ and $\beta$. The first-order moment (mean) of any quantity can be evaluated by taking its integral over all possible orientations. If an arbitrary quantity is denoted by $A$, its average is expressed as

\begin{equation}\label{eq:avg}
    \langle A\rangle=\int^{2\pi}_{0}\int^{\pi}_{0} A f(\theta,\phi) \sin\theta d\theta d\phi.
\end{equation}

The orientation moments $\langle \boldsymbol{p}\boldsymbol{p}\rangle$ and $\langle \boldsymbol{p}\boldsymbol{p}\boldsymbol{p}\boldsymbol{p}\rangle$, are fundamental quantities for evaluating the particle stresslet, which in turn determines the suspension shear stress and effective viscosity \citep{brenner1974rheology,kim2013microhydrodynamics}. The relation between the stress tensor with the average orientation moments are given by

\begin{equation}
    \boldsymbol{\tau_s}= 2\mu \Phi \left[2\mathbb{A}\boldsymbol{S}:\langle\boldsymbol{p}\boldsymbol{p}\boldsymbol{p}\boldsymbol{p}\rangle+2\mathbb{B}\left(\boldsymbol{S}\cdot\langle\boldsymbol{p}\boldsymbol{p}\rangle+\langle\boldsymbol{p}\boldsymbol{p}\rangle\cdot\boldsymbol{S}-\dfrac{2}{3}\boldsymbol{\delta}\boldsymbol{S}:\langle\boldsymbol{p}\boldsymbol{p}\rangle\right)+\mathbb{E}\boldsymbol{S}+\mathbb{F}\mathcal{D}_{r}\left(\langle \boldsymbol{p}\boldsymbol{p}\rangle-\dfrac{1}{3}\boldsymbol{\delta}\right)\right]. \label{eq:part_stress}
\end{equation}

Here, $\mathbb{A}, \mathbb{B}, \mathbb{E}$ and $\mathbb{F}$ depend on aspect ratio of the particle. Also, $\boldsymbol{\delta}$ is the Kronecker delta and $\Phi$ is a volume concentration of particles. Several components of these orientation moments have been studied extensively in the literature, including $\langle p_z^2\rangle$, $\langle p_x^2p_y^2\rangle$, and $\langle p_xp_y\rangle$ \citep{chen1996rheology,chen1999orientation,asokan2002novel}. In the following, we focus on the moments $\langle p_z^2\rangle$ and $\langle p_x^2p_y^2\rangle$, which could be relevant to the rheological properties of the suspension.

\subsection{Asymptotic analysis in the small $\mathrm{Pe}$ limit} \label{sec:lowPe}

In the limit $\mathrm{Pe}\rightarrow 0$ diffusion dominates and the orientation
distribution is nearly isotropic. A weak mean shear then enters as regular perturbations of (\ref{eq:stfp}),

\begin{equation}\label{eq:serexp}
    f(\theta,\phi)\approx\dfrac{1}{4\pi}\left[1+\mathrm{Pe}\,f_{1}(\theta,\phi)
    +\mathrm{Pe}^{2}f_{2}(\theta,\phi)+\cdots\right],
\end{equation}

with $f_{0}=1/4\pi$ fixed by normalisation \citep{stewart1972hydrodynamic} and
$\int f_{n}\sin\theta\,d\theta\,d\phi=0$ for $n\geqslant 1$. Collecting powers
of $\mathrm{Pe}$ gives the hierarchy

\begin{subequations}\label{eq:hierarchy}
    \begin{equation}\label{eq:hierarchy1}
        \nabla^{2}_{p}f_{1}=\nabla_{p}\cdot\dot{\boldsymbol{p}},
    \end{equation}
    \begin{equation}\label{eq:hierarchy2}
        \nabla^{2}_{p}f_{2}
        =\nabla_{p}\cdot\left(f_{1}\dot{\boldsymbol{p}}\right).
    \end{equation}
\end{subequations}

These are inverted term by term in the spherical harmonics $P^{m}_{n}c_{m}$ and
$P^{m}_{n}s_{m}$, where $c_{m}$ and $s_{m}$ denote $\cos m\phi$ and
$\sin m\phi$, and $P^{m}_{n}$ are the associated Legendre polynomials of
argument $\cos\theta$, with $P^{m}_{n}=0$ for $m>n$. The Laplacian
(\ref{eq:lapsm}) is diagonal in this basis,

\begin{subequations}\label{eq:eglap}
    \begin{equation}
        \Lambda P^{m}_{n}c_{m}=-n(n+1)P^{m}_{n}c_{m},
    \end{equation}
    \begin{equation}
        \Lambda P^{m}_{n}s_{m}=-n(n+1)P^{m}_{n}s_{m}.
    \end{equation}
\end{subequations}

\subsubsection{Tensor form of the drift}

Equation (\ref{eq:pdoteq}) has the generic form
$\dot{p}_{i}=A_{ij}p_{j}-p_{i}p_{j}A_{jk}p_{k}$, which preserves
$|\boldsymbol{p}|=1$ for any second-order tensor $\boldsymbol{A}$. Term-by-term
comparison identifies

\begin{equation}\label{eq:Atensor}
    \boldsymbol{A}=\boldsymbol{R}+\mathcal{B}\boldsymbol{S}
    +\mathcal{K}_{p}\,\hat{\boldsymbol{g}}\hat{\boldsymbol{g}},
\end{equation}

so the settling torque is a rank-one addition to the effective velocity
gradient. For any drift of this form,

\begin{equation}\label{eq:divgeneric}
    \nabla_{p}\cdot\dot{\boldsymbol{p}}
    =\mathrm{tr}\,\boldsymbol{A}
    -3\,\boldsymbol{p}\cdot\boldsymbol{A}\cdot\boldsymbol{p}.
\end{equation}

Since $\boldsymbol{p}\cdot\boldsymbol{R}\cdot\boldsymbol{p}=0$, while the
isotropic part of $\mathcal{K}_{p}\hat{\boldsymbol{g}}\hat{\boldsymbol{g}}$
cancels against $\mathrm{tr}\,\boldsymbol{A}=\mathcal{K}_{p}$, equation
(\ref{eq:divgeneric}) reduces to

\begin{equation}\label{eq:divp}
    \nabla_{p}\cdot\dot{\boldsymbol{p}}
    =-3\,\boldsymbol{p}\cdot\boldsymbol{M}\cdot\boldsymbol{p},
\end{equation}

where $\boldsymbol{M}$ is the deviatoric symmetric part of $\boldsymbol{A}$,

\begin{equation}\label{eq:Mtensor}
    M_{ij}=\mathcal{B}S_{ij}
    +\mathcal{K}_{p}\left(\hat{g}_{i}\hat{g}_{j}
    -\dfrac{\delta_{ij}}{3}\right).
\end{equation}

Every low-$\mathrm{Pe}$ result below therefore depends on the shear and the
settling only through the single symmetric traceless tensor $\boldsymbol{M}$.
Written out in the spherical coordinates of (\ref{eq:uv}) and (\ref{eq:gdir}),
the $O(\mathrm{Pe})$ balance produces eight separate settling terms, which
collapse into a single Legendre polynomial,

\begin{equation}\label{eq:peorcompact}
    \nabla_{p}^{2}f_{1}+2\mathcal{K}_{p}P_{2}(\zeta)
    +\dfrac{3\mathcal{B}\sin^{2}\theta\sin 2\phi}{2}=0,
\end{equation}

where $\zeta=\cos\alpha\cos\theta+\sin\alpha\sin\theta\cos(\beta-\phi)$
is the cosine of the angle between the symmetry axis and the direction
$(\alpha,\beta)$, which is $-\hat{\boldsymbol{g}}$ by the convention of
(\ref{eq:gdir}) and $P_2$ is a Legendre polynomial. Since $P_{2}$ is even,
$P_{2}(\zeta)=P_{2}(\boldsymbol{p}\cdot\hat{\boldsymbol{g}})$ and the sign is
immaterial, reflecting the quadratic dependence of the inertial torque on
$\hat{\boldsymbol{g}}$. Note that $P_{2}(\zeta)$ is zonal about
$\hat{\boldsymbol{g}}$ rather than $\hat{\boldsymbol{z}}$, so by the addition
theorem it carries all five components $m=-2,\ldots,2$ in the $(\theta,\phi)$
frame.

\subsubsection{First-order distribution and moments}

Both forcing terms in (\ref{eq:peorcompact}) are surface harmonics of degree
two, for which (\ref{eq:eglap}) gives $\Lambda=-6$. The equation therefore
inverts by inspection,

\begin{equation}\label{eq:f1tensor}
    f_{1}=\dfrac{M_{ij}p_{i}p_{j}}{2}
    =\dfrac{\mathcal{K}_{p}}{3}P_{2}(\zeta)
    +\dfrac{\mathcal{B}}{4}\sin^{2}\theta\sin 2\phi,
\end{equation}

so that

\begin{equation}\label{eq:pertmomcompact}
    f(\theta,\phi)\approx\dfrac{1}{4\pi}\left[1+\mathrm{Pe}\left(
    \dfrac{\mathcal{K}_{p}}{3}P_{2}(\zeta)
    +\dfrac{\mathcal{B}P^{2}_{2}s_{2}}{12}\right)\right].
\end{equation}

Equation (\ref{eq:f1tensor}) is the orientation potential of
\S\ref{sec:spherical}. Since
$\boldsymbol{p}\cdot\boldsymbol{R}\cdot\boldsymbol{p}=0$, the potential
(\ref{eq:potential}) is $\Chi=\tfrac{1}{2}\boldsymbol{p}\cdot\boldsymbol{A}
\cdot\boldsymbol{p}$ with $\boldsymbol{A}$ of (\ref{eq:Atensor}), and removing
its isotropic part gives
$\Chi-\langle\Chi\rangle=\tfrac{1}{2}M_{ij}p_{i}p_{j}=f_{1}$, where
$\langle\Chi\rangle=\mathcal{K}_{p}/6$ is the average over the sphere. The
first-order distribution is therefore the potential itself,
\begin{equation}\label{eq:f1chi}
    f=\dfrac{1}{4\pi}\left[1+\mathrm{Pe}
    \left(\Chi-\langle\Chi\rangle\right)\right]+O(\mathrm{Pe}^{2}),
\end{equation}
which is the linearisation of the exact solution  $f\propto\exp(\mathrm{Pe}\,\Chi)$ that holds when the imposed flow is irrotational. The small-$\mathrm{Pe}$ expansion and the no-flow
limit are thus the same statement at successive orders, and the shear enters
the difference only through the single non-gradient term
$-\tfrac{1}{2}\partial f/\partial\phi$. This term cannot be absorbed into any
potential, since the rotary flux vanishes if and only if the ambient flow is
irrotational \citep{brenner1974transport}, which the present one is not.
\citet{kramers1946behavior} noted the same structure for a chain molecule in
physical space: an irrotational flow enters as a potential energy and yields a
Boltzmann distribution, whereas in simple shear the potential alone would
predict alignment at $45^{\circ}$ to the flow, the orientation turning towards
the flow direction as the shear strengthens. The inversion is immediate because
$\Chi$ is quadratic in $\boldsymbol{p}$ and hence purely of degree two, so
$\nabla_{p}^{2}(\Chi-\langle\Chi\rangle)=-6(\Chi-\langle\Chi\rangle)$.

At this order the shear and settling contributions are independent: the shear adds a term that depends on $\phi$, while the settling adds a term that depends only on the angle between $\boldsymbol{p}$ and $\hat{\boldsymbol{g}}$. Setting $\mathcal{K}_{p}=0$ leaves a correction
depending solely on $\mathcal{B}$, obtained previously for $\mathcal{B}=1$
\citep{kim1984,bird_book} and for arbitrary $\mathcal{B}$ by
\citet{talbot2024exploring}.

Because $f_{1}$ is quadratic in $\boldsymbol{p}$, the settling-modified moments
follow by contraction with the isotropic fourth- and sixth-order tensors and
require no further integration. Using
$\mathrm{tr}\,\boldsymbol{M}=0$, the second moment is

\begin{equation}\label{eq:genmom1}
    \langle p_{i}p_{j}\rangle=\dfrac{\delta_{ij}}{3}
    +\dfrac{\mathrm{Pe}}{15}M_{ij}+O(\mathrm{Pe}^{2}),
\end{equation}

and the fourth moment is the analogous contraction, its
$O(\mathrm{Pe})$ part being $\Sigma_{ijkl}/105$, where $\Sigma_{ijkl}$ is the
sum of the six products $M_{ij}\delta_{kl}$ over distinct index pairings.

These hold for arbitrary $\mathcal{B}$. The second-order terms follow from the moment hierarchy rather than from
$f_{2}$ directly. Multiplying (\ref{eq:stfp}) by $p_{i}p_{j}$, integrating over
the unit sphere and using
$\Lambda(p_{i}p_{j})=-6\left(p_{i}p_{j}-\delta_{ij}/3\right)$ gives the exact
relation

\begin{equation}\label{eq:momhier}
    \langle p_{i}p_{j}\rangle=\dfrac{\delta_{ij}}{3}
    +\dfrac{\mathrm{Pe}}{6}\left\langle \dot{p}_{i}p_{j}
    +p_{i}\dot{p}_{j}\right\rangle,
\end{equation}

which closes at $O(\mathrm{Pe}^{2})$ because the right-hand side needs the
fourth moment only to $O(\mathrm{Pe})$. Substituting (\ref{eq:genmom1}),

\begin{equation}\label{eq:genmom2}
    \langle p_{i}p_{j}\rangle=\dfrac{\delta_{ij}}{3}
    +\dfrac{\mathrm{Pe}}{15}M_{ij}+\mathrm{Pe}^{2}\left[\dfrac{\left[\boldsymbol{R},
    \boldsymbol{M}\right]_{ij}}{90}
    +\dfrac{1}{105}\left(M_{ik}M_{kj}
    -\dfrac{\delta_{ij}}{3}M_{kl}M_{kl}\right)\right]
    +O(\mathrm{Pe}^{3}),
\end{equation}

where the commutator $[\boldsymbol{R},\boldsymbol{M}]
=\boldsymbol{R}\boldsymbol{M}-\boldsymbol{M}\boldsymbol{R}$ is symmetric
because $\boldsymbol{R}$ is antisymmetric and $\boldsymbol{M}$ symmetric.
Equation (\ref{eq:genmom2}) holds for any orientation of gravity and supplies
the second-order orientation tensor entering the particle stress
(\ref{eq:part_stress}); the fourth moment at $O(\mathrm{Pe}^{2})$ is obtained
in the same way from the projection of $f_{2}$ onto harmonics of degree $0$,
$2$ and $4$.

Evaluating the components of $\boldsymbol{M}$ with $\hat{\boldsymbol{g}}$
specified by $(\alpha,\beta)$, and noting that
$[\boldsymbol{R},\boldsymbol{M}]_{zz}=0$ identically because $\boldsymbol{R}$
has support only in the flow-gradient plane, the moments for arbitrary $\alpha$
and $\beta$ are

\begin{subequations}\label{eq:genmomexp}
\begin{equation}\label{eq:genmompz}
    \langle p^{2}_{z}\rangle\approx\dfrac{1}{3}
    +\dfrac{2\mathcal{K}_{p}}{45}P_{2}(\cos\alpha)\,\mathrm{Pe}+\left[\dfrac{2\mathcal{K}^{2}_{p}}{945}P_{2}(\cos\alpha)
    -\dfrac{\mathcal{B}^{2}}{630}
    -\dfrac{\mathcal{B}\mathcal{K}_{p}}{315}\sin^{2}\alpha\sin2\beta
    \right]\mathrm{Pe}^{2},
\end{equation}
\begin{multline}\label{eq:genmompxpy}
    \langle p^{2}_{x}p^{2}_{y}\rangle\approx\dfrac{1}{15}
    -\dfrac{2\mathcal{K}_{p}}{315}P_{2}(\cos\alpha)\,\mathrm{Pe}\\
    +\dfrac{1}{4725}\left[3\mathcal{B}^{2}-\mathcal{K}^{2}_{p}
    +6\mathcal{B}\mathcal{K}_{p}\sin^{2}\alpha\sin2\beta
    +\dfrac{15\mathcal{K}^{2}_{p}}{4}\sin^{4}\alpha\sin^{2}2\beta
    \right]\mathrm{Pe}^{2},
\end{multline}
\begin{equation}\label{eq:genmomshear}
    \langle p_{x}p_{y}\rangle\approx
    \dfrac{\mathrm{Pe}}{30}\left(\mathcal{B}
    +\mathcal{K}_{p}\sin^{2}\alpha\sin2\beta\right)+O(\mathrm{Pe}^{2}).
\end{equation}
\end{subequations}

Three features of (\ref{eq:genmomexp}) are not visible in the individual
configurations. First, $\langle p^{2}_{z}\rangle$ and
$\langle p^{2}_{x}p^{2}_{y}\rangle$ are independent of $\mathcal{B}$ and of
$\beta$ at $O(\mathrm{Pe})$, a consequence of $S_{zz}=0$ and
$M_{xx}+M_{yy}=-M_{zz}$, and are locked to each other,

\begin{equation}\label{eq:locking}
    \langle p^{2}_{x}p^{2}_{y}\rangle-\dfrac{1}{15}
    =-\dfrac{1}{7}\left(\langle p^{2}_{z}\rangle-\dfrac{1}{3}\right)
    +O(\mathrm{Pe}^{2}),
\end{equation}

independently of $\mathcal{B}$, $\mathcal{K}_{p}$, $\alpha$ and $\beta$, the
ratio being set by the isotropic sixth-order tensor alone. Second, the settling
contribution to both moments vanishes at first order when $P_{2}(\cos\alpha)=0$,
that is at $\alpha_{m}=\arccos(1/\sqrt{3})\approx 54.74^{\circ}$, for every
$\mathcal{K}_{p}$. Third, the azimuthal placement of gravity
enters these two moments only at second order, through
$\sin^{2}\alpha\sin2\beta=2\,\boldsymbol{S}\!:\!\hat{\boldsymbol{g}}
\hat{\boldsymbol{g}}$, which measures the alignment of gravity with the
extensional axis of the shear and is maximal at $\beta=\pi/4$. It enters
$\langle p_{x}p_{y}\rangle$, and hence the suspension shear stress, already at
first order. All three canonical configurations of \S\ref{sec:theory} have
$\sin^{2}\alpha\sin2\beta=0$.

\subsubsection{The canonical configurations}

When $\alpha=0$ we have $P_{2}(\cos\alpha)=1$ and $\zeta=\cos\theta$, so
(\ref{eq:pertmomcompact}) becomes

\begin{equation}\label{eq:rpsgzmom}
    f(\theta,\phi)=\dfrac{1}{4\pi}\left[1+\dfrac{\mathrm{Pe}}{3}\left(
    \dfrac{\mathcal{B}P^{2}_{2}s_{2}}{4}
    +\mathcal{K}_{p}P^{0}_{2}c_{0}\right)\right]+O(\mathrm{Pe}^{2}),
\end{equation}

and (\ref{eq:genmomexp}) reduces to

\begin{subequations}\label{eq:momgzexp}
    \begin{equation}
        \langle p^{2}_{z}\rangle\approx\dfrac{1}{3}
        +\dfrac{2\mathcal{K}_{p}}{45}\mathrm{Pe}
        +\dfrac{(4\mathcal{K}^{2}_{p}-3\mathcal{B}^{2})}{1890}\mathrm{Pe}^{2},
        \quad\text{and}
    \end{equation}
    \begin{equation}
        \langle p^{2}_{x}p^{2}_{y}\rangle\approx\dfrac{1}{15}
        -\dfrac{2\mathcal{K}_{p}}{315}\mathrm{Pe}
        -\dfrac{(\mathcal{K}^{2}_{p}-3\mathcal{B}^{2})}{4725}\mathrm{Pe}^{2}.
    \end{equation}
\end{subequations}

When gravity lies in the flow-gradient plane, $P_{2}(\cos\alpha)=-1/2$ and
$\sin^{2}\alpha\sin2\beta=0$ for both $\beta=0$ and $\beta=\pi/2$. For the
flow-aligned case ($\alpha=\pi/2$, $\beta=0$), $\zeta=\sin\theta\cos\phi$ and

\begin{equation}\label{eq:rpsgxdist}
    f(\theta,\phi)\approx\dfrac{1}{4\pi}\left[1+\dfrac{\mathrm{Pe}}{12}\left\{
    \mathcal{B}P^{2}_{2}s_{2}
    +\mathcal{K}_{p}\left(P^{2}_{2}c_{2}
    -2P^{0}_{2}c_{0}\right)\right\}\right],
\end{equation}

with the corresponding moments

\begin{subequations}\label{eq:momgxexp}
    \begin{equation}
        \langle p^{2}_{z}\rangle\approx\dfrac{1}{3}
        -\dfrac{\mathcal{K}_{p}}{45}\mathrm{Pe}
        -\dfrac{(2\mathcal{K}^{2}_{p}+3\mathcal{B}^{2})}{1890}\mathrm{Pe}^{2},
        \quad\text{and}
    \end{equation}
    \begin{equation}
        \langle p^{2}_{x}p^{2}_{y}\rangle\approx\dfrac{1}{15}
        +\dfrac{\mathcal{K}_{p}}{315}\mathrm{Pe}
        -\dfrac{(\mathcal{K}^{2}_{p}-3\mathcal{B}^{2})}{4725}\mathrm{Pe}^{2}.
    \end{equation}
\end{subequations}

The flow-gradient case ($\alpha=\beta=\pi/2$) gives the same two moments at
both orders, the two configurations being distinguished only through
$\langle p_{x}p_{y}\rangle$. Comparing (\ref{eq:momgzexp}) and
(\ref{eq:momgxexp}), the $P_{2}(\cos\alpha)$ dependence cancels identically at
$O(\mathrm{Pe}^{2})$ in the fourth moment but not in the second. Setting
$\mathcal{K}_{p}=0$ recovers the pure Jeffery limit,

\begin{subequations}\label{eq:jefferylimit}
    \begin{equation}
        \langle p^{2}_{z}\rangle\approx\dfrac{1}{3}
        -\dfrac{\mathcal{B}^{2}}{630}\mathrm{Pe}^{2},
        \quad\text{and}
    \end{equation}
    \begin{equation}
        \langle p^{2}_{x}p^{2}_{y}\rangle\approx\dfrac{1}{15}
        +\dfrac{\mathcal{B}^{2}}{1575}\mathrm{Pe}^{2},
    \end{equation}
\end{subequations}

independent of $\alpha$ and $\beta$. Higher orders are
progressively less tractable, since the moment hierarchy no longer closes. The
predictions (\ref{eq:momgzexp}) and (\ref{eq:momgxexp}) are compared against
the spherical-harmonic solutions of \S\ref{sec:numerics} in the insets of
figures~\ref{fig:gz_pzsq} and \ref{fig:gx_pzsq}.

\subsection{Asymptotic analysis in the $\mathrm{Pe}\rightarrow\infty$ limit}\label{sec:largePe}
 
In the preceding subsection we have performed the asymptotic calculations in low $\mathrm{Pe}$ limit. In this subsection, we complement the low-$\mathrm{Pe}$ expansion of \S\ref{sec:lowPe} with an asymptotic analysis valid in the opposite limit, $\mathrm{Pe}\gg 1$, where rotary diffusion acts as a weak perturbation to the deterministic orientation dynamics. The nature of the asymptotic analysis depends on the structure of the deterministic attractor: a stable fixed point leads to a Gaussian concentration of the distribution, whereas a periodic orbit leads to a boundary-layer structure in which the distribution is sharply peaked around the orbit. We treat each case in turn.
 
The large-$\mathrm{Pe}$ regime has been extensively studied for the
classical Jeffery problem (without settling). For the dumbbell model
($\mathcal{B}=1$, $\mathcal{K}_{p}=0$), \citet{leal1971effect} have
applied singular perturbation methods to analyse the effect of weak
Brownian diffusion on tumbling dynamics, showing that diffusion acts
on a slow time scale $O(\mathrm{Pe})$ to randomise the Jeffery orbit
constant. \citet{hinch1972effect} have extended this analysis to obtain the
stationary distribution over orbit constants and the resulting
rheological properties. For spheroids of arbitrary aspect ratio,
\citet{chen1996rheology} have obtained numerical solutions at large $\mathrm{Pe}$ and compared them with singular perturbation
predictions. In the polymer dynamics literature,
\citet{chertkov2005polymer} and \citet{turitsyn2007polymer} have analysed the analogous problem of a dumbbell in mean shear with random velocity fluctuations, obtaining the large-Weissenberg-number asymptotics of the orientation and extension statistics. The present analysis extends these approaches to include the settling-induced inertial torque.
 
\subsubsection{Concentration near periodic orbits}

When gravity is parallel to
the vorticity axis and the spheroid is prolate, the attractor is the
tumbling orbit at $\theta=\pi/2$ for every $\mathcal{K}$. The orientation distribution then
concentrates in a boundary layer around this orbit, with a width
that scales as $O(\mathrm{Pe}^{-1/2})$ in the transverse
direction.

 Setting $\theta=\pi/2+\eta$ with
$|\eta|\ll 1$, the evolution equation (\ref{eq:gtpeq}a) yields,
at leading order,
\begin{equation}\label{eq:eta_drift}
   \dot{\eta} \approx \mathcal{K}\mathcal{F}_{p}\,\eta
- \frac{\mathcal{B}}{2}\eta\sin 2\phi.
\end{equation}
The first term provides a restoring force toward $\eta=0$ (for
$\mathcal{K}\mathcal{F}_{p}<0$, i.e.\ prolate spheroids), while
the second introduces a periodic modulation from the Jeffery
strain-rate coupling. The effective transverse restoring rate,
\begin{equation}\label{eq:lambda_phi}
    \lambda(\phi) = |\mathcal{K}\mathcal{F}_{p}|
    + \frac{\mathcal{B}}{2}\sin 2\phi,
\end{equation}
strengthens the confinement near the extensional quadrant ($\sin 2\phi>0$) and weakens it near the compressional quadrant ($\sin 2\phi<0$). At large $\mathrm{Pe}$, the distribution factorises at leading order as
\begin{equation}\label{eq:f_factor}
    f(\theta,\phi) \approx h(\eta;\phi)\,\rho(\phi),
\end{equation}
where $\rho(\phi)$ is the marginal distribution along the orbit
and $h(\eta;\phi)$ describes the transverse profile. For the
transverse direction, the local Fokker-Planck equation in $\eta$
at fixed $\phi$ yields a Gaussian profile:
\begin{equation}\label{eq:h_gaussian}
    h(\eta;\phi) \propto
    \exp\!\left(-\frac{\mathrm{Pe}}{2}\,
    \frac{\eta^{2}}{\sigma^{2}(\phi)}\right),
\end{equation}
with $\phi$-dependent variance $\sigma^{2}(\phi)$ determined by
the local restoring rate (\ref{eq:lambda_phi}).

Due to the nematic symmetry of the spheroid
($\boldsymbol{p}$ and $-\boldsymbol{p}$ represent the same
physical orientation), the azimuthal angle is restricted to
$\phi\in[0,\pi)$. Over this domain, the large-$\mathrm{Pe}$
stationary density is inversely proportional to the angular
velocity
\citep{leal1971effect,hinch1972effect,brenner1974rheology}:
\begin{equation}\label{eq:rho_jeffery}
    \rho(\phi) = \frac{1}{T_{J}\,|\dot{\phi}|} =
    \frac{1}{T_{J}}\,\frac{2}{|1-\mathcal{B}\cos 2\phi|},
    \quad \phi\in[0,\pi),
\end{equation}
where $T_{J}=2\pi/\sqrt{1-\mathcal{B}^{2}}$ is the Jeffery
half-period corresponding to a rotation of $\pi$ in $\phi$. This
classical result remains valid at leading order when $\alpha=0$
and $\mathcal{K}\neq 0$, since the azimuthal dynamics is
unchanged from the Jeffery equation in this configuration. The settling parameter enters
only through the transverse confinement.

Using (\ref{eq:f_factor})--(\ref{eq:rho_jeffery}), the moment
$\langle p^{2}_{z}\rangle=\langle\cos^{2}\theta\rangle
\approx\langle\eta^{2}\rangle$ (since $\theta^{*}=\pi/2$) is
obtained by integrating the local variance over the orbit:
\begin{equation}\label{eq:pz2_orbit}
    \langle p^{2}_{z}\rangle \approx
    \frac{1}{\mathrm{Pe}}\int_{0}^{\pi}
    \sigma^{2}(\phi)\,\rho(\phi)\,d\phi.
\end{equation}
When the settling-induced restoring rate dominates the periodic
modulation, $|\mathcal{K}\mathcal{F}_{p}|\gg\mathcal{B}/2$,
the variance is approximately uniform,
$\sigma^{2}\approx 1/|\mathcal{K}\mathcal{F}_{p}|$, and the
integral reduces to
\begin{equation}\label{eq:pz2_orbit_simple}
    \langle p^{2}_{z}\rangle \approx
    \frac{1}{\mathrm{Pe}\,|\mathcal{K}\mathcal{F}_{p}|},
    \quad \mathrm{Pe}\gg 1,\quad
    |\mathcal{K}\mathcal{F}_{p}|\gg\mathcal{B}/2.
\end{equation}
This result predicts that $\langle p_{z}^{2}\rangle$ decreases
as $\mathrm{Pe}^{-1}$ at large $\mathrm{Pe}$, with a prefactor
inversely proportional to $|\mathcal{K}\mathcal{F}_{p}|$: as
the inertial torque increases, the spheroid is confined more
tightly to the flow-gradient plane, reducing the variance in
$\theta$. 

However, the simplified result (\ref{eq:pz2_orbit_simple}) has
a restricted regime of validity. For $\mathcal{B}=0.9$, the
condition $|\mathcal{K}\mathcal{F}_{p}|\gg\mathcal{B}/2$
requires $|\mathcal{K}_{p}|\gg 0.45$. When this condition is
not met, the effective restoring rate (\ref{eq:lambda_phi})
changes sign during the compressional phase of the Jeffery
orbit ($\sin 2\phi<0$), and the transverse boundary layer
inflates periodically. The resulting variance involves a
$\phi$-resolved Floquet problem: the local variance satisfies
\begin{equation}\label{eq:variance_ode}
\dot{\phi}\,\frac{d\sigma^{2}}{d\phi}
= -2\lambda(\phi)\,\sigma^{2} + 2,
\end{equation}
subject to periodicity over $[0,\pi)$, and the orbit-averaged
moment (\ref{eq:pz2_orbit}) retains the $\mathrm{Pe}^{-1}$
scaling but with a prefactor
$\mathcal{C}(\mathcal{B},\mathcal{K}_{p})$ that can be much
larger than $1/|\mathcal{K}\mathcal{F}_{p}|$. In the limit $|\mathcal{K}_{p}|\to 0$ this prefactor
diverges. The divergence does not recover the finite Hinch-Leal plateau; it
marks the failure of the limit $\mathrm{Pe}\to\infty$ taken at fixed nonzero
$\mathcal{K}_{p}$. The two limits do not commute, and the Jeffery plateau is reached only in a
distinguished crossover in which $\mathrm{Pe}$ and $\mathcal{K}_{p}$ are taken
together with $\mathrm{Pe}|\mathcal{K}_{p}|$ held fixed. Numerical solutions of
(\ref{eq:stfp}) confirm the collapse: at $\mathcal{B}=0.9$, $\alpha=0$ and
$\mathrm{Pe}|\mathcal{K}_{p}|=2$, $\langle p_{z}^{2}\rangle$ takes the values
$0.207$, $0.203$, $0.202$ and $0.202$ as $\mathrm{Pe}$ increases from $40$ to
$320$ by successive doublings, whereas at fixed $\mathrm{Pe}\mathcal{K}_{p}^{2}$
it drifts monotonically over the same range.
The crossover behaviour is visible in
figure~\ref{fig:gz_pzsq}(a): curves at
$\mathcal{K}_{p}=0$ remain near the Hinch-Leal
plateau throughout the plotted range of $\mathrm{Pe}$, while
those with $|\mathcal{K}_{p}|\gg\mathcal{B}/2$ display the predicted
$\mathrm{Pe}^{-1}$ decay.

The behaviour of the fourth-order moment
$\langle p^{2}_{x}p^{2}_{y}\rangle$ is controlled by two
competing effects. On the periodic orbit ($\theta=\pi/2$),
$p^{2}_{x}p^{2}_{y}=(1/4)\sin^{2}2\phi$. Averaging over the
Jeffery density (\ref{eq:rho_jeffery}) gives the limiting value
\begin{equation}\label{eq:pxpy_orbit}
    \langle p^{2}_{x}p^{2}_{y}\rangle_{\infty} =
    \frac{1}{4T_{J}}\int^{\pi}_{0}
    \frac{\sin^{2}2\phi}{|\dot{\phi}|}\,d\phi,
\end{equation}
which is a finite $O(1)$ constant depending on $\mathcal{B}$.
However, the approach to this limit is non-monotonic
(figure~\ref{fig:gz_pzsq}b). At intermediate $\mathrm{Pe}$,
the settling torque confines $\theta$ toward $\pi/2$, pushing
$\sin^{4}\theta$ toward unity and thereby increasing
the moment relative to its isotropic value $1/15$. At larger
$\mathrm{Pe}$, the Jeffery density (\ref{eq:rho_jeffery})
concentrates $\phi$ near $0$ and $\pi$ (where
$\sin^{2}2\phi\approx 0$), and the moment decreases. The peak
at intermediate $\mathrm{Pe}$ marks the crossover between
these two effects. For $\mathcal{K}_{p}=0$, only the
$\phi$-concentration operates and
$\langle p^{2}_{x}p^{2}_{y}\rangle$ decreases monotonically
at large $\mathrm{Pe}$. For non-zero $\mathcal{K}_{p}$, the
peak is more pronounced because the settling torque enhances
the $\theta$-confinement before the $\phi$-concentration takes
over.

More precisely, the leading correction to
(\ref{eq:pxpy_orbit}) from the finite boundary-layer width
arises from expanding $\sin^{4}\theta=\cos^{4}\eta
\approx 1-2\eta^{2}$:
\begin{equation}\label{eq:pxpy_correction}
    \langle p^{2}_{x}p^{2}_{y}\rangle \approx
    \langle p^{2}_{x}p^{2}_{y}\rangle_{\infty}
    - \frac{2}{\mathrm{Pe}}
    \frac{1}{4T_{J}}\int^{\pi}_{0}
    \frac{\sigma^{2}(\phi)\sin^{2}2\phi}{|\dot{\phi}|}\,d\phi
    + O(\mathrm{Pe}^{-2}).
\end{equation}
The negative $O(\mathrm{Pe}^{-1})$ correction reflects the fact
that fluctuations away from the equatorial plane reduce the
in-plane anisotropy.

The scaling $\langle p_{z}^{2}\rangle\sim\mathrm{Pe}^{-1}$
(decaying to zero) for periodic orbits contrasts with the
approach to a finite deterministic constant for stable fixed
points. This distinction provides a
direct link between the deterministic bifurcation structure
identified in \S\ref{sec:theory} and the stochastic orientation
statistics: the transition from a periodic orbit to a stable
equilibrium as $\mathcal{K}$ crosses the SNIC bifurcation
threshold manifests as a qualitative change in the
large-$\mathrm{Pe}$ scaling of $\langle p_{z}^{2}\rangle$.

\subsubsection{Concentration near stable fixed points}\label{sec:asym_gx_largeprsec}
 
When the settling parameter $\mathcal{K}$ exceeds the critical
bifurcation value, the
deterministic dynamics possesses a stable node at
$(\theta^{*},\phi^{*})$, and the orientation distribution concentrates
around this fixed point at large $\mathrm{Pe}$. To analyse this
regime, we introduce local coordinates $\delta\theta =
\theta-\theta^{*}$ and $\delta\phi = \phi-\phi^{*}$, and linearise
the deterministic drift about the fixed point. Let
$\boldsymbol{\delta}=(\delta\theta,\sin\theta^{*}\,\delta\phi)^{T}$
denote the displacement vector in the tangent plane (the factor
$\sin\theta^{*}$ accounts for the spherical metric), and let
$\mathbb{J}$ denote the Jacobian matrix of the linearised system
evaluated at $(\theta^{*},\phi^{*})$. The steady Fokker-Planck
equation~(\ref{eq:stfp}), when written in terms of the local
coordinates, reduces at leading order to the stationary equation for a
two-dimensional Ornstein-Uhlenbeck process:
 
\begin{equation}\label{eq:OUsteady}
    \nabla^{2}_{\delta}f + \mathrm{Pe}\,\nabla_{\delta}\cdot
    \left(\mathbb{A}\,\boldsymbol{\delta}\,f\right) = 0,
\end{equation}
 
\noindent where $\mathbb{A}=-\mathbb{J}$ is the restoring-force matrix
(positive stable when the fixed point is stable, that is with
eigenvalues of positive real part, equivalently $\mathbb{J}$ Hurwitz; for a
nonsymmetric $\mathbb{A}$ this is weaker than positive definiteness) and
$\nabla_{\delta}$ denotes differentiation with respect to the local
coordinates. The stationary solution of (\ref{eq:OUsteady}) is a
bivariate Gaussian \citep{risken1996fokker}:
 
\begin{equation}\label{eq:fGauss}
 f(\boldsymbol{\delta}) =
\frac{\mathrm{Pe}}{2\pi\sqrt{\det\mathbb{C}}}
\exp\left(-\frac{\mathrm{Pe}}{2}\,
\boldsymbol{\delta}^{T}\mathbb{C}^{-1}\boldsymbol{\delta}\right),
\end{equation}
 
\noindent where the covariance matrix $\mathbb{C}$ satisfies the
Lyapunov equation:
 
\begin{equation}\label{eq:lyapunov}
    \mathbb{A}\,\mathbb{C} + \mathbb{C}\,\mathbb{A}^{T} = 2\,\mathbb{I}.
\end{equation}

 The Lyapunov solutions are given as
 \begin{equation}
     \mathbb{C}=\frac{(\det\mathbb{A})\,\mathbb{I}+\operatorname{adj}(\mathbb{A})\operatorname{adj}(\mathbb{A})^{T}}{\operatorname{tr}\mathbb{A}\,\det\mathbb{A}},
\qquad
\operatorname{tr}\mathbb{C}=\frac{2\det\mathbb{A}+\lVert\mathbb{A}\rVert_{F}^{2}}{\operatorname{tr}\mathbb{A}\,\det\mathbb{A}}.
 \end{equation}
 
Using the Gaussian approximation, the orientation moments can be
evaluated analytically. For a moment whose gradient vanishes at the fixed point of the form $\langle
g(\theta,\phi)\rangle$, we expand $g$ about the fixed point and
integrate against the Gaussian distribution. At leading order, the
moment reduces to its deterministic value $g(\theta^{*},\phi^{*})$,
with $O(\mathrm{Pe}^{-1})$ corrections arising from the variance of
the distribution. Specifically,
 
\begin{equation}\label{eq:pz2largePe_fp}
 \langle p^{2}_{z}\rangle = \cos^{2}\theta^{*} -
\frac{1}{\mathrm{Pe}}\left(\cos 2\theta^{*}C_{11}
\right) + O(\mathrm{Pe}^{-2}).
\end{equation}

 We apply this result to the flow-aligned configuration
($\alpha=\pi/2$, $\beta=0$) analysed in \S\ref{sec:gxfp}. At
large $\mathcal{K}$, a prolate spheroid aligns along the
vorticity axis ($\theta^{*}=0$), corresponding to the broadside-on
configuration with the symmetry axis perpendicular to the settling
direction. Since the $(\theta,\phi)$ coordinate system is singular
at the pole $\theta^{*}=0$, we reformulate the analysis in local
Cartesian coordinates
$(p_{x},p_{y})\approx(\theta\cos\phi,\theta\sin\phi)$, where
$p_{z}\approx 1-\theta^{2}/2$ and hence
$\langle p^{2}_{z}\rangle\approx
1-(\langle p^{2}_{x}\rangle+\langle p^{2}_{y}\rangle)$.
Linearising the evolution equation (\ref{eq:pdoteq}) about
$\boldsymbol{p}=\hat{\boldsymbol{z}}$ with
$\hat{\boldsymbol{g}}=-\hat{\boldsymbol{x}}$ yields the
two-dimensional system

\begin{equation}\label{eq:lin_pole_gx}
    \dot{p}_{x} \approx \mathcal{K}\mathcal{F}_{p}
    p_{x}
    + \dfrac{\mathcal{B}+1}{2}\,p_{y},
    \qquad
    \dot{p}_{y} \approx  \dfrac{\mathcal{B}-1}{2}\,p_{x}.
\end{equation}

 The corresponding Ornstein-Uhlenbeck analysis gives
\begin{equation}\label{eq:pz2gx_largePe}
    \langle p^{2}_{z}\rangle \approx 1
    - \dfrac{\mathrm{tr}~\mathbb{C}}{\mathrm{Pe}},
    \quad \mathrm{Pe}\gg \mathrm{tr}~\mathbb{C},\quad
    \mathcal{F}_{p}<0.
\end{equation}

Here, $\mathrm{tr}~\mathbb{C}=4(\mathcal{K}_{p}^{2}+1)/(|\mathcal{K}_{p}|(1-\mathcal{B}^{2}))$ . There is no threshold in $\mathcal{K}$ here: for
$\mathcal{F}_{p}<0$ the vorticity pole attracts at every nonzero settling
strength, and crossing $\mathcal{R}=1$ changes it from a stable focus to a
stable node without creating or destroying it. The conditions for
(\ref{eq:pz2gx_largePe}) are $\mathcal{F}_{p}<0$ and
$\mathrm{Pe}\gg\mathrm{tr}~\mathbb{C}$, with the second failing as
$\mathcal{K}_{p}\to 0$ because $\mathrm{tr}~\mathbb{C}$ then diverges. Note that
$\mathrm{tr}~\mathbb{C}$ is not the sum of the reciprocal eigenvalues,
which would be $4|\mathcal{K}_{p}|/(1-\mathcal{B}^{2})$; the additional
$+1$ in the numerator is the square of the antisymmetric part of
(\ref{eq:lin_pole_gx}), that is the vorticity of the imposed shear,
which rotates fluctuations between the two transverse directions and so
inflates the variance beyond what the decay rates alone would give. At
$\mathcal{B}=0.9$ and $\mathcal{K}_{p}=-0.5$ the two differ by a factor
of five, $52.6$ against $10.5$.  For the fourth-order moment, the fixed-point prediction differs
qualitatively from the periodic-orbit case. At
$\theta^{*}=0$, $p_{x}=p_{y}=0$, and the Gaussian fluctuations
give $p_{x}\sim O(\mathrm{Pe}^{-1/2})$,
$p_{y}\sim O(\mathrm{Pe}^{-1/2})$. Consequently,
\begin{equation}\label{eq:pxpy_gx_largePe}
    \langle p^{2}_{x}p^{2}_{y}\rangle
    \sim O(\mathrm{Pe}^{-2}),
    \quad \mathrm{Pe}\gg 1,\quad \mathcal{K}>\mathcal{K}_{c},
\end{equation}
decaying to zero rather than approaching a finite plateau.
This contrasts sharply with the periodic-orbit result
(\ref{eq:pxpy_orbit}), where
$\langle p^{2}_{x}p^{2}_{y}\rangle$ remains $O(1)$. The
transition from $O(1)$ to $O(\mathrm{Pe}^{-2})$ as
$\mathcal{K}$ crosses the SNIC bifurcation threshold provides
a second stochastic signature of the bifurcation, complementing
the $\langle p^{2}_{z}\rangle$ diagnostic.



\subsection{Numerical evaluation of diffusion equation} \label{sec:numerics}

Next, we investigate the orientation dynamics of a spheroid at arbitrary $\mathrm{Pe}$. Several studies have examined the orientation distribution function $f(\theta,\phi)$ over a wide range of $\mathrm{Pe}$. \cite{stewart1972hydrodynamic} have evaluated $f(\theta,\phi)$ for a dumbbell model ($\mathcal{B}=1$) in a steady shear flow. Later, \cite{strand1987computation} extended this analysis to find  unsteady distribution function in a similar flow configuration. Both investigations have evaluated the Fokker-Planck equation using the spherical harmonics method. Subsequently, \cite{asokan2002novel} and \cite{talbot2024exploring} have carried out the investigation to study the orientation distribution function of a spheroid at arbitrary $\mathcal{B}$. 

To study the dynamics at arbitrary $\mathrm{Pe}$, we expand the orientation distribution function in spherical harmonics. We solve (\ref{eq:stfp}) by expanding $f(\theta,\phi)$ on the basis of spherical harmonics as

\begin{equation}\label{eq:smt}
    f(\theta,\phi)=\dfrac{1}{4\pi}\left[\sum^{\mathcal{N}}_{n=0}\sum^{n}_{m=0}\left\{A^{m}_{n}P^{m}_{n}(\cos\theta)\cos m\phi+B^{m}_{n}P^{m}_{n}(\cos\theta)\sin m\phi\right\}\right].
\end{equation}

Here, $A^{m}_{n}$ and $B^{m}_{n}$ are the coefficients to be found. Incorporating the above equation in (\ref{eq:stfp}) leads to a diffusion equation which depends on $\mathcal{B}$, $\mathrm{Pe}$, and $\mathcal{K}$. The numerical solution is obtained following the procedure outlined by \citet{stewart1972hydrodynamic}, \citet{bird_book}, \cite{strand1987computation} and \citet{strand1992dynamics}.


\subsubsection{When the gravity vector aligns with the vorticity vector}

First, we analyse the case where the vorticity and gravity vectors are aligned parallel to each other. In this case, the steady Fokker-Planck equation is modified as

\begin{equation}
    \nabla^{2}_{p} f-\mathrm{Pe}\left[\dfrac{1}{\sin\theta}\left\{\mathcal{B} \sin\phi\cos\phi-\mathcal{K}_{p}\right\}\dfrac{\partial (f \sin^{2}\theta \cos\theta)}{\partial \theta} +\dfrac{(\mathcal{B}-1)}{2}\dfrac{\partial f}{\partial \phi}-\mathcal{B}\dfrac{\partial(f \sin^{2}\phi)}{\partial \phi}\right]=0.
\end{equation}

The above equation is written in terms of the operators $\Lambda(f)$, $\Omega(f)$, and $\mathcal{G}(f)$. The expressions of these operators are given as

\begin{subequations}\label{eq:ggz}
    \begin{equation}
        \Lambda(f)=\dfrac{1}{\sin\theta}\dfrac{\partial }{\partial \theta}\left(\sin\theta\dfrac{\partial f}{\partial \theta}\right)+\dfrac{1}{\sin^{2}\theta}\dfrac{\partial^{2}f}{\partial\phi^{2}},
    \end{equation}
    \begin{equation}
        \Omega(f)=\dfrac{\sin\phi\cos\phi}{\sin\theta}\dfrac{\partial}{\partial \theta}\left( f \sin^{2}\theta\cos\theta\right)-\dfrac{\partial }{\partial \phi}\left(f \sin^{2}\phi\right),
    \end{equation}
    \begin{equation}
    \mathcal{G}(f)=-\sin\theta\cos\theta\dfrac{\partial f}{\partial \theta}-\dfrac{\{1+3\cos2\theta\}f}{2}.
\end{equation}
\end{subequations}

Subsequently, the steady Fokker-Planck equation in terms of these operators can be written as

\begin{equation}\label{eq:fpz}
    \Lambda (f)-\mathrm{Pe}\left[\hat{\Omega}(f)+\mathcal{K}_{p}\mathcal{G}(f)\right]=0.
\end{equation}

Here the operator $\hat{\Omega}$ is defined as follows

\begin{equation}
    \hat{\Omega}(f)=\mathcal{B}\Omega(f)+\dfrac{(\mathcal{B}-1)}{2}\dfrac{\partial f}{\partial \phi}.
\end{equation}

In (\ref{eq:fpz}), when $\mathcal{K}_{p}=0$ and $\mathcal{B}=1$, then the equation is simplified and reduced to the expression given for the dumbbell model \citep{stewart1972hydrodynamic}. The operators $\hat{\Omega}(f)$ and $\mathcal{G}(f)$ need to be written in terms of spherical harmonics. Following the method shown in \cite{strand1989dynamic}, on substituting the cosine spherical harmonics $P^{m}_{n}(\cos\theta)\cos(m\phi)$ in the operator $\mathcal{G}(f)$, we get

\begin{equation}
    \mathcal{G}(P^{m}_{n}c_{m})=-\sin\theta \cos\theta c_{m}\dfrac{d P^{m}_{n}}{d\theta}-\left(\dfrac{1+3\cos2\theta}{2}\right)P^{m}_{n}c_{m}.
\end{equation}

Moreover, the $\theta$ terms are evaluated using the recurrence relations for the associated Legendre polynomials. After evaluating the operator $\mathcal{G}$ algebraically, we get

\begin{multline}\label{eq:gzsh}
    \mathcal{G}(P^{m}_{n}c_{m})=\left[\left\{ \dfrac{(n-2)(n+m)(n+m-1)}{(2n+1)(2n-1)}  \right\}P^{m}_{n-2}+\left\{\dfrac{3m^2 -n(n+1)}{(2n+3)(2n-1)} \right\}P^{m}_{n}+\right.\\
    \left.\left\{\dfrac{-(n+3)(n-m+1)(n-m+2)}{(2n+1)(2n+3)} \right\}P^{m}_{n+2}\right]c_{m}.
\end{multline}

When sine spherical harmonics ($P^{m}_{n}(\cos\theta)\sin m\phi$) are substituted in the expression of $\mathcal{G}$, we obtain the same coefficients of $P^{m}_{n-2}$, $P^{m}_{n}$ and $P^{m}_{n+2}$ as shown in (\ref{eq:gzsh}).  Therefore, the operator $\mathcal{G}$ acts on the spherical harmonics and is written in the form of 

\begin{subequations}\label{eq:gzc}
    \begin{equation}
        \mathcal{G}(P^{m}_{n}c_{m})=\sum^{n+2}_{q=n-2}\sum^{m+2}_{p=m-2}\mathcal{Z}^{mp}_{nq}P^{p}_{q}c_{p},
    \end{equation}
    \begin{equation}
        \mathcal{G}(P^{m}_{n}s_{m})=\sum^{n+2}_{q=n-2}\sum^{m+2}_{p=m-2}\mathcal{Z}^{mp}_{nq}P^{p}_{q}s_{p}.
    \end{equation}
\end{subequations}

Here, the coefficients $\mathcal{Z}^{mp}_{nq}$ are numerically evaluated for various combinations of $m,n,p$ and $q$ using (\ref{eq:gzsh}). To verify the correctness of the algebraic expressions for $\mathcal{Z}^{m,p}_{n,q}$, various forms of spherical harmonics are substituted into (\ref{eq:ggz}) to express the results in terms of Legendre polynomials. The resulting expressions can then be validated by applying (\ref{eq:gzc}) with the same forms of spherical harmonics, following the approach outlined by \cite{strand1989dynamic}.

\begin{table}
  \begin{center}
\def~{\hphantom{0}}
  \begin{tabular}{c}
     $\mathcal{Z}^{m,m}_{n,n}=\dfrac{3m^2 -n(n+1)}{(2n+3)(2n-1)}$ \\[0.5cm]
     $\mathcal{Z}^{m,m}_{n,n+2}=\dfrac{-(n+3)(n-m+1)(n-m+2)}{(2n+1)(2n+3)}$ \\[0.5cm]
     $\mathcal{Z}^{m,m}_{n,n-2}=\dfrac{(n-2)(n+m)(n+m-1)}{(2n+1)(2n-1)} $\\[0.5cm]
  \end{tabular}
  \caption{Tabulation of coefficients $Z^{m,p}_{n,q}$ from (\ref{eq:gzc}). All other coefficients of $\mathcal{Z}$ are zero.}
  \label{tab:gzcot}
  \end{center}
\end{table}

For the present case, the diffusion equations are expanded in spherical harmonics and solved for $q=1,\ldots,\mathcal{N}$ and $p=0,\ldots,q$, yielding the following set of equations
\begin{subequations}
    \begin{equation}
        -q(q+1) A^{p}_{q}-\mathrm{Pe}\left[\mathcal{B}\sum^{\mathcal{N}}_{n=0}\sum_{m=0}^{n}\left(B^{m}_{n}(\mathrm{b w})^{mp}_{nq}\right)+\dfrac{(\mathcal{B}-1)}{2}p B^{p}_{q}+\mathcal{K}_{p}\sum^{\mathcal{N}}_{n=0}\sum_{m=0}^{n}\left(A^{m}_{n}\mathcal{Z}_{nq}^{mp}\right)\right]=0,
    \end{equation}
     \begin{equation}
        -q(q+1) B^{p}_{q}+\mathrm{Pe}\left[\mathcal{B}\sum^{\mathcal{N}}_{n=0}\sum_{m=0}^{n}\left(A^{m}_{n}(\mathrm{b w})^{mp}_{nq}\right)+\dfrac{(\mathcal{B}-1)}{2}p A^{p}_{q}-\mathcal{K}_{p}\sum^{\mathcal{N}}_{n=0}\sum_{m=0}^{n}\left(B^{m}_{n}\mathcal{Z}_{nq}^{mp}\right)\right]=0.
    \end{equation}    
\end{subequations}

In the above expression, we set $A^{0}_{0}=1$. Also, $\mathcal{N}$ is an order of spherical harmonics. Further, $(\mathrm{bw})$ are the Bird-Warner coefficients and their expressions are given in \cite{bird_book}. The numerical evaluation requires solving the resulting set of coupled equations to determine the coefficients  $A^{m}_{n}$ and $B^{m}_{n}$ for various combinations of $m$ and $n$. The total number of equations (or unknowns) to be solved is denoted by $\mathcal{N}_{e}$, where $\mathcal{N}_{e}=\mathcal{N}(\mathcal{N}+2)$. For the dumbbell model ($\mathcal{K}_{p}=0$), value of $\mathcal{N}$  required to obtain a converged solution depends on the value of $\mathrm{Pe}$. At lower values of $\mathrm{Pe}$, solving a small number of equations is sufficient to obtain the converged solution. In contrast, at higher $\mathrm{Pe}$, a larger system (with greater $\mathcal{N}_{e}$) must be solved, and convergence is achieved at higher values of $\mathcal{N}$. In the present problem ($\mathcal{K}_{p} \neq 0$), the convergence and accuracy of the numerical solution also depend on $\mathcal{K}_{p}$ and $\mathcal{B}$. As the magnitude of $\mathcal{K}_{p}$ increases, the numerical solution tends to lose accuracy at relatively lower values of $\mathrm{Pe}$ than in the case $\mathcal{K}_{p}=0$. A similar loss of accuracy is observed for larger magnitudes of $\mathcal{B}$, indicating that both parameters play an  important role in obtaining a converged numerical solution.

\begin{figure}
     \hspace{-2.5cm}
    \includegraphics[width=1.2\linewidth]{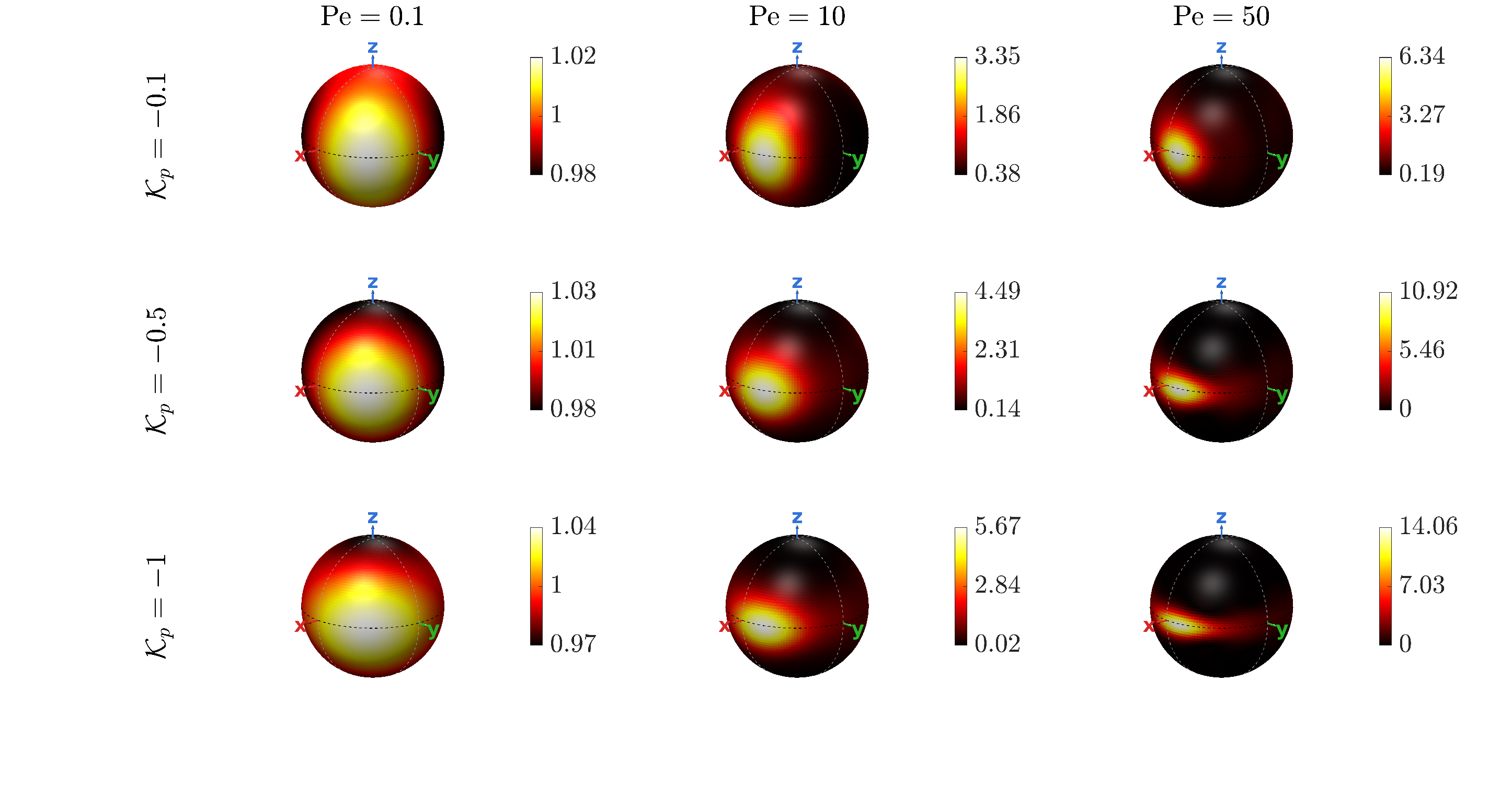}
    \caption{Contour plot of the  orientation distribution function $4\pi f$ on the unit sphere for a prolate spheroid with $\mathcal{B}=0.9$ ($r=4.36$). The results correspond to the case where the gravity vector is aligned with the vorticity direction ($\alpha=0$).}
    \label{fig:ffgz}
\end{figure}

In figure~\ref{fig:ffgz}, we present the  orientation distribution function $4\pi f$ on the unit sphere for $\mathcal{B}=0.9$, corresponding to the case where the gravity vector is aligned with the vorticity direction. As both $|\mathcal{K}_{p}|$ and $\mathrm{Pe}$ increase, the distribution becomes increasingly concentrated towards flow direction. In particular, the orientations cluster predominantly within the flow--gradient plane, with a strong preference near the $\hat{\boldsymbol{x}}$-axis. At higher values of $\mathcal{K}_{p}$, this preferential alignment is evident even at relatively low $\mathrm{Pe}$.

\begin{figure}
\centering  
\subfigure[]{\includegraphics[width=0.49\linewidth]{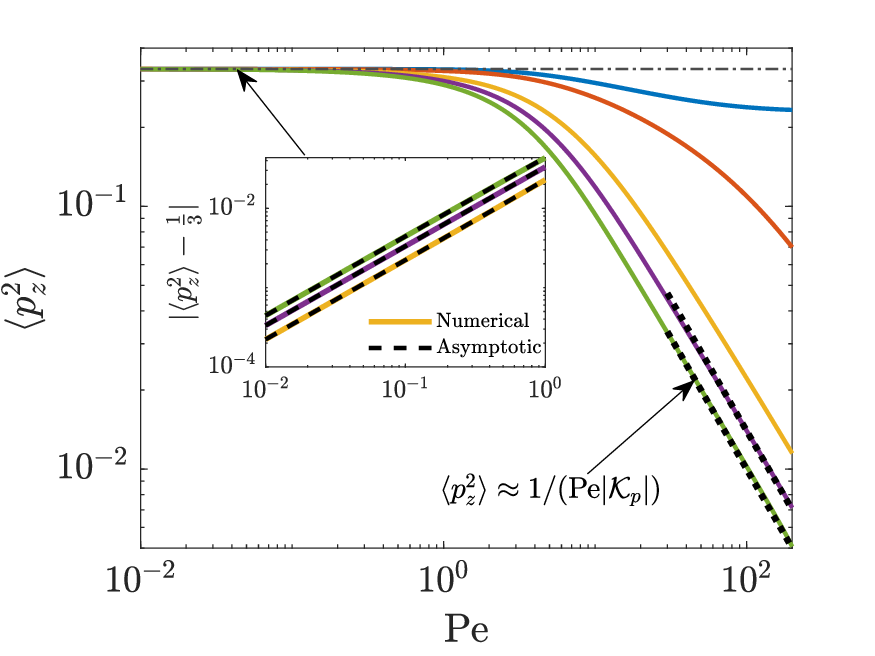}}
\subfigure[]{\includegraphics[width=0.49\linewidth]{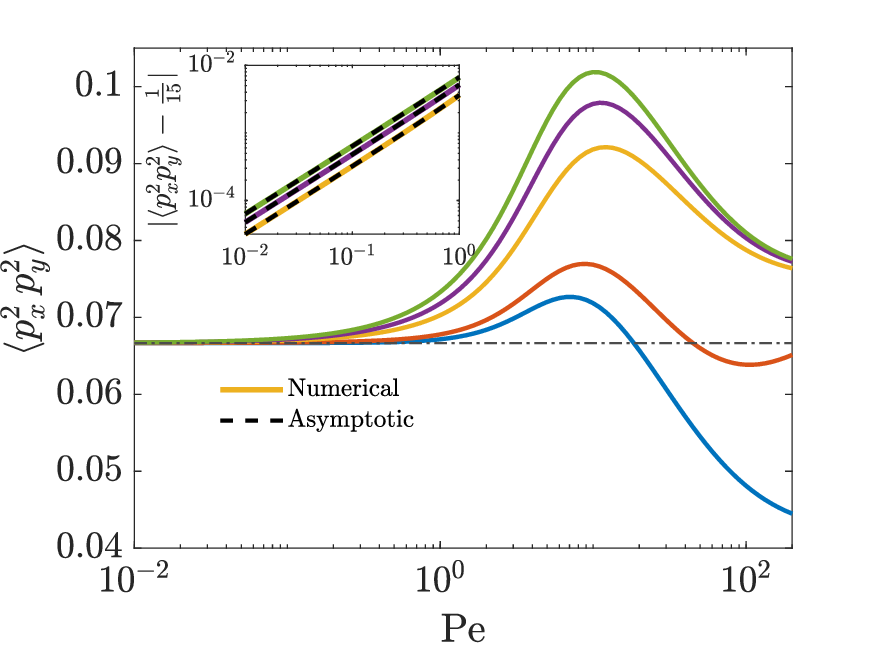}}
\caption{Plot of orientation moments $\langle p^2_z \rangle$ and $\langle p^2_x p^2_y\rangle$  versus $\mathrm{Pe}$ for the case when gravity is aligned in the vorticity direction ($\alpha=0$) and for various values of $\mathcal{K}_{p}$:
(\textcolor[rgb]{0.000,0.447,0.741}{\rule[0.5ex]{1em}{1.5pt}}) $\mathcal{K}_{p}=0$,
(\textcolor[rgb]{0.850,0.325,0.098}{\rule[0.5ex]{1em}{1.5pt}}) $\mathcal{K}_{p}=-0.1$,
(\textcolor[rgb]{0.929,0.694,0.125}{\rule[0.5ex]{1em}{1.5pt}}) $\mathcal{K}_{p}=-0.5$,
(\textcolor[rgb]{0.494,0.184,0.556}{\rule[0.5ex]{1em}{1.5pt}}) $\mathcal{K}_{p}=-0.75$, and
(\textcolor[rgb]{0.466,0.674,0.188}{\rule[0.5ex]{1em}{1.5pt}}) $\mathcal{K}_{p}=-1$. (a) Variation of the orientation moment $\langle p_{z}^{2}\rangle$ with $\mathrm{Pe}$, and
(b) variation of $\langle p_{x}^{2}p_{y}^{2}\rangle$ with $\mathrm{Pe}$. Both panels present a comparison between the asymptotic calculations and the numerical solution. The inset shows a comparison between the asymptotic
solution (dashed lines) in a low-$\mathrm{Pe}$ limit and numerical results (solid lines).
In the main figure of the left panel, dotted lines correspond to the high-$\mathrm{Pe}$ limit with
different values of $\mathcal{K}_{p}$. Dash-dotted line corresponds to the moments for the isotropic orientation distribution. All results are presented for
$\mathcal{B}=0.9$ ($r=4.36$) and $\mathcal{N}=34$.
} 
\label{fig:gz_pzsq}
\end{figure}

In figure \ref{fig:gz_pzsq} (first panel), we plot $\langle p^{2}_{z}\rangle$ as a function of $\mathrm{Pe}$.  A similar result has previously been reported for the dumbbell model ($\mathcal{B}=1$ and $\mathcal{K}_{p}=0$)(\cite{chen1996rheology,asokan2002novel}). Plots in both panels are obtained at $\mathcal{N}=34$.  For comparison, results are also shown for $\mathcal{K}_{p}=0$. When $\mathcal{B}=0.9$ and at zero $\mathcal{K}_{p}$, we observe a decrease in moment with an increasing value of $\mathrm{Pe}$. This qualitative behaviour of the moment is consistent with the case of the dumbbell model ($\mathcal{B} = 1$), although at higher $\mathrm{Pe}$  the present result for $\mathcal{B}=0.9$ yields slightly higher values of the moment comparatively. This difference highlights the influence of aspect ratio of particles on the orientation moments. Further, for $\mathcal{K}_{p}=0$ and $\mathcal{B}=1$, we compare the numerically obtained moment with the corresponding result from the dumbbell model in Appendix~\ref{appbmom}, providing an independent validation of our numerical implementation. For $\mathcal{K}_{p}\neq 0$, a further reduction in $\langle p_{z}^{2}\rangle$ is observed, with the lowest values obtained at larger magnitudes of $\mathcal{K}_{p}$. As discussed in the previous section, we demonstrate that for a simple shear flow without noise, a spheroid undergoes a tumbling motion in the flow plane ($\hat{\boldsymbol{x}}$-$\hat{\boldsymbol{y}}$) when $\mathcal{K}_{p} \neq 0$ ($\alpha=0$). As a result, $\langle p^{2}_{z}\rangle$ is expected to decrease more rapidly at higher magnitudes of $\mathcal{K}_{p}$ as a particle exhibits more fluctuating tumbling motion away from the $\hat{\boldsymbol{z}}$ axis. A convergence study of the orientation moments with respect to $\mathcal{N}$ is presented in Appendix~\ref{appbmom} (see figure~\ref{fig:conv_gz_gx}). Consequently, for $\alpha=0$ case, we restrict our analysis to $\mathrm{Pe}=200$.

In the second panel, we show the plot of the moment $\langle p^{2}_{x}p^{2}_{y}\rangle$ versus $\mathrm{Pe}$. A pronounced peak in the moment is observed for non-zero $\mathcal{K}_{p}$  when compared with the case of $\mathcal{K}_{p}=0$.  The increase in moment can be attributed to the pronounced fluctuations in the tumbling motion of a prolate spheroid in the flow plane. It is worth noting that, in contrast to the behaviour observed for $\langle p^{2}_{z}\rangle$, a non-zero $\mathcal{K}_{p}$ leads to higher orientational moment compared with the  $\mathcal{K}_{p}=0$ case.

In both panels, we compare the numerical results with the corresponding asymptotic solutions in the low- and high-$\mathrm{Pe}$ limits. The inset compares the numerical results with the low-$\mathrm{Pe}$ asymptotic expression, shown in dashed lines and given by (\ref{eq:momgzexp}), showing good agreement. In the main figure (left panel), the dotted lines denote the high-$\mathrm{Pe}$ asymptotic results given by (\ref{eq:pz2_orbit_simple}). For the moment $\langle p_x^2p_y^2\rangle$, the high-$\mathrm{Pe}$ asymptotic expression does not capture the non-monotonic variation at moderate values of $\mathrm{Pe}$ for fixed $\mathcal{K}_{p}$ and $\mathcal{B}$. However, the expression in (\ref{eq:pxpy_correction}) accurately captures the high-$\mathrm{Pe}$ behaviour and agrees well with the numerical results at $\mathrm{Pe}=200$.

\begin{figure}
    \centering
    \includegraphics[width=0.95\linewidth]{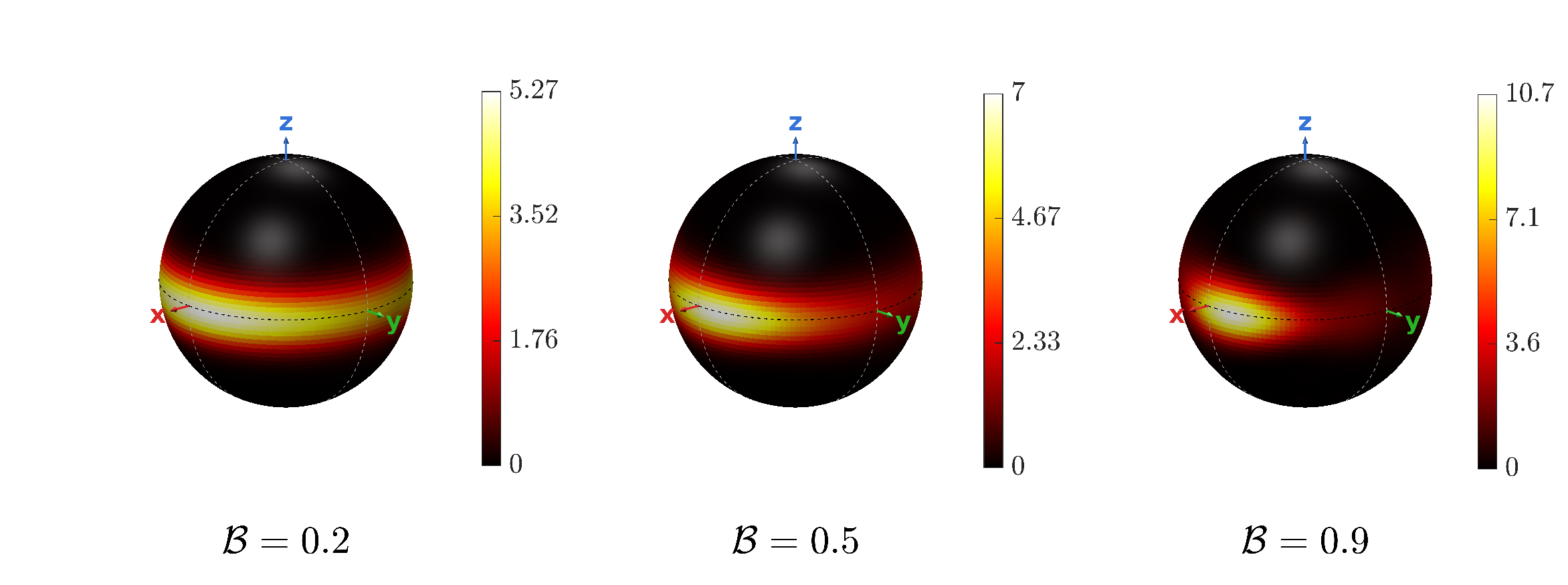}
    \caption{Orientation distribution function $4\pi f$ on the unit sphere for the gravity--vorticity aligned configuration ($\alpha=0$) with $\mathcal{K}_{p}=-1$ and $\mathrm{Pe}=30$.}
    \label{fig:us_varyb_gz}
\end{figure}

Next, we examine the influence of the Bretherton constant $\mathcal{B}$ on the orientation distribution function for a fixed settling parameter, $\mathcal{K}_{p}=-1$. Figure~\ref{fig:us_varyb_gz} shows the orientation distribution function $4\pi f$ on the unit sphere for different values of $\mathcal{B}$ at $\mathrm{Pe}=30$ and $\alpha=0$. For a nearly spherical prolate spheroid (small $\mathcal{B}$), the distribution is broadly spread over the flow-gradient plane. As $\mathcal{B}$ increases, the distribution becomes progressively more concentrated toward the $\hat{\boldsymbol{x}}$-axis, indicating stronger preferential alignment with the flow direction. No significant effect of $\mathcal{B}$ on the distribution function has been observed at $\mathrm{Pe}<10$.

\subsubsection{When the gravity vector aligns with the flow direction}\label{sec:gxfp}

When the flow direction and the gravity vector coincide ($\alpha=\pi/2$ and $\beta=0$), the drift term in the Fokker-Planck equation gets modified to

\begin{multline}
    \nabla_{p}\cdot (\dot{\boldsymbol{p}}f)=\dfrac{1}{\sin\theta}\left\{\mathcal{B}\sin\phi\cos\phi+\mathcal{K}_{p}\cos^{2}\phi\right\}\dfrac{\partial (f \sin^{2}\theta\cos\theta)}{\partial \theta}+\\
    \dfrac{\mathcal{B}}{2}\dfrac{\partial}{\partial \phi}\left\{f\cos2\phi \right\}-\dfrac{1}{2}\dfrac{\partial f}{\partial \phi}   -\dfrac{\mathcal{K}_{p}}{2}\dfrac{\partial}{\partial \phi}\{f\sin2\phi\}.
\end{multline}

This equation is more involved than the previous case because the $\phi$ derivative terms are now depend explicitly on $\mathcal{K}_{p}$. The steady Fokker-Planck equation can then be written in terms of a differential operator as

\begin{equation}\label{eq:gxfp}
    \Lambda(f)-\mathrm{Pe}\left[\hat{\Omega}(f)+\mathcal{K}_{p}\mathcal{X}(f)\right]=0.
\end{equation}

In the above equation, the differential operator $\mathcal{X}(f)$ is given as
\begin{equation}\label{eq:gxk}
    \mathcal{X}(f)=\left\{\left(2\cos^{2}\theta-\sin^{2}\theta\right)\cos^{2}\phi-\cos2\phi\right\}f+\sin\theta\cos\theta\cos^{2}\phi\dfrac{\partial f}{\partial \theta}-\dfrac{\sin2\phi}{2}\dfrac{\partial f}{\partial \phi}.
\end{equation}

On substituting $f=P^{m}_{n}c_{m}$ into (\ref{eq:gxfp}), we write the algebraic equation in terms of spherical harmonics. The conversion of the $\mathcal{X}(f)$ operator in terms of spherical harmonics is

\begin{equation}
    \mathcal{X}(P^{m}_{n}c_{m})=\left\{\left(2\cos^{2}\theta-\sin^{2}\theta\right)\dfrac{(1+c_{2})}{2}-c_{2}\right\}P^{m}_{n}c_{m}+\sin\theta\cos\theta\left(\dfrac{1+c_{2}}{2}\right)\dfrac{d P^{m}_{n}}{d\theta}c_{m}+\dfrac{ms_{m}s_{2}P^{m}_{n}}{2}.
\end{equation}

Simplifying the above equation yields terms proportional to $c_m$, $c_{m+2}$, and $c_{m-2}$. The coefficients associated with each of these terms are evaluated separately using the recurrence relations of the associated Legendre polynomials, resulting in a system of algebraic equations. The $\phi$ terms are evaluated using the identities listed below,
\begin{equation}
    c_{2}c_{m}=\dfrac{c_{m+2}+c_{m-2}}{2}, \quad\quad s_{2}s_{m}=\dfrac{-c_{m+2}+c_{m-2}}{2}.
\end{equation}

Consequently, the operator $\mathcal{X}$, expressed in the spherical harmonic basis $P_n^m c_m$, can be written as

\begin{multline}\label{eq:gxae}
    \mathcal{X}(P^{m}_{n})=\left[\left\{\dfrac{n(n+1)-3m^{2}}{2(2n-1)(2n+3)}\right\}P^{m}_{n}+\left\{\dfrac{(n+3)(n-m+1)(n-m+2)}{2(2n+1)(2n+3)}\right\}P^{m}_{n+2}+\right.\\
    \left.\left\{\dfrac{-(n-2)(n+m)(n+m-1)}{2(2n+1)(2n-1)}\right\}P^{m}_{n-2}\right]c_{m}+    \left[\left\{\dfrac{3(1+\delta_{m0})}{4(2n-1)(2n+3)}\right\}P^{m+2}_{n}+\right.\\
    \left.\left\{\dfrac{-(n+3)(1+\delta_{m0})}{4(2n+1)(2n+3)}\right\}P^{m+2}_{n+2}+\left\{\dfrac{(n-2)(1+\delta_{m0})}{4(2n+1)(2n-1)}\right\}P^{m+2}_{n-2}\right]c_{m+2}+\\
    \left[\left\{\dfrac{3(n+m)(n+m-1)(m-n-2)(m-n-1)(1-\delta_{m0})}{4(2n+3)(2n-1)}\right\}P^{m-2}_{n}+\right.\\
    \left.\left\{\dfrac{(n+3)(n-m+2)(n-m+3)(n-m+4)(m-n-1)(1-\delta_{m0})}{4(2n+1)(2n+3)}\right\}P^{m-2}_{n+2}+\right.\\
    \left.\left\{\dfrac{(n-2)(n+m)(n+m-1)(n+m-2)(n+m-3)(1-\delta_{m0})}{4(2n+1)(2n-1)} \right\}P^{m-2}_{n-2}\right]c_{m-2}.
\end{multline}

Similar calculations can be repeated on substituting the sine spherical harmonics like before. On substituting the sine spherical harmonics $(P^{m}_{n}s_{m})$ in (\ref{eq:gxk}), we get the same coefficients as shown in (\ref{eq:gxae}). As in the preceding case, the operator acts on the cosine and
sine harmonics through the same coefficients, and

\begin{subequations}\label{eq:gxc}
    \begin{equation}
        \mathcal{X}(P^{m}_{n}c_{m})=\sum^{n+2}_{q=n-2}\sum^{m+2}_{p=m-2}\mathcal{D}^{mp}_{nq}P^{p}_{q}c_{p},
    \end{equation}
    \begin{equation}
        \mathcal{X}(P^{m}_{n}s_{m})=\sum^{n+2}_{q=n-2}\sum^{m+2}_{p=m-2}\mathcal{D}^{mp}_{nq}P^{p}_{q}s_{p}.
    \end{equation}
\end{subequations}

The expression of coefficients is tabulated for a specific range of $m$ and
$n$. All other values of the coefficients $\mathcal{D}$ will be zero (see table
\ref{tab:gxcot}). One boundary deserves comment. At $m=1$ the
$p=m-2$ block lands on $p=-1$, which is not itself a basis index; with the
standard conventions $P^{-1}_{q}=-[(q-1)!/(q+1)!]P^{1}_{q}$,
$c_{-1}=\cos(-\phi)=c_{1}$ and $s_{-1}=\sin(-\phi)=-s_{1}$ it folds back onto
$p=1$ with opposite signs in the two sectors, so that the assembled $p=1$
coefficients differ between them although the tabulated
$\mathcal{D}^{mp}_{nq}$ do not. We have verified the expansion by evaluating
the operator on a Gauss-Legendre grid for $0\leqslant m\leqslant 8$ and
$m\leqslant n\leqslant 11$ in both sectors, obtaining a worst relative residual
of $6\times10^{-15}$. In practice the question does not arise: the nematic
symmetry $\boldsymbol{p}\leftrightarrow-\boldsymbol{p}$ restricts the solution
to even $n$ and even $m$, all four operators map $n\to n,n\pm2$ and
$m\to m,m\pm2$, and the remaining three parity blocks are identically zero. We
impose that restriction explicitly in the solver. The two coefficient tables are
also not independent: for every $n$ and every admissible $q$ the $m$-diagonal
entries satisfy $\mathcal{D}^{m,m}_{n,q}=-\tfrac{1}{2}\mathcal{Z}^{m,m}_{n,q}$,
as may be read directly off tables \ref{tab:gzcot} and \ref{tab:gxcot}. We have
confirmed that this holds identically for $2\leqslant n\leqslant 13$, which
provides a check on both sets of entries at once.

\begin{table}
  \begin{center}
\def~{\hphantom{0}}
  \begin{tabular}{c}
     $\mathcal{D}^{m,m}_{n,n}=\dfrac{n(n+1)-3m^{2}}{2(2n-1)(2n+3)}$ \\[0.5cm]
     $\mathcal{D}^{m,m}_{n,n+2}=\dfrac{(n+3)(n-m+1)(n-m+2)}{2(2n+1)(2n+3)}$ \\[0.5cm]
     $\mathcal{D}^{m,m}_{n,n-2}=\dfrac{-(n-2)(n+m)(n+m-1)}{2(2n+1)(2n-1)} $\\[0.5cm]
    $\mathcal{D}^{m,m+2}_{n,n}=\dfrac{3(1+\delta_{m0})}{4(2n-1)(2n+3)} $\\[0.5cm]
    $\mathcal{D}^{m,m+2}_{n,n+2}=\dfrac{-(n+3)(1+\delta_{m0})}{4(2n+1)(2n+3)} $\\[0.5cm]
    $\mathcal{D}^{m,m+2}_{n,n-2}=\dfrac{(n-2)(1+\delta_{m0})}{4(2n+1)(2n-1)}$\\[0.5cm]
     $\mathcal{D}^{m,m-2}_{n,n}=\dfrac{3(n+m)(n+m-1)(m-n-2)(m-n-1)(1-\delta_{m0})}{4(2n+3)(2n-1)}$\\[0.5cm]
     $\mathcal{D}^{m,m-2}_{n,n+2}=\dfrac{(n+3)(n-m+2)(n-m+3)(n-m+4)(m-n-1)(1-\delta_{m0})}{4(2n+1)(2n+3)}$\\[0.5cm]
     $\mathcal{D}^{m,m-2}_{n,n-2}=\dfrac{(n-2)(n+m)(n+m-1)(n+m-2)(n+m-3)(1-\delta_{m0})}{4(2n+1)(2n-1)}$\\[0.5cm]
  \end{tabular}
  \caption{Tabulation of coefficients $D^{m,p}_{n,q}$ from (\ref{eq:gxc}). All other coefficients of $\mathcal{D}^{m,p}_{n,q}$ will be zero.}
  \label{tab:gxcot}
  \end{center}
\end{table}

For the present case ($\alpha=\pi/2$ and $\beta=0$), the steady diffusion equations are reduced to

\begin{subequations}
    \begin{equation}
        -q(q+1) A^{p}_{q}-\mathrm{Pe}\left[\mathcal{B}\sum^{\mathcal{N}}_{n=0}\sum_{m=0}^{n}\left(B^{m}_{n}(\mathrm{bw})^{mp}_{nq}\right)+\dfrac{(\mathcal{B}-1)}{2}p B^{p}_{q}+\mathcal{K}_{p}\sum^{\mathcal{N}}_{n=0}\sum_{m=0}^{n}\left(A^{m}_{n}\mathcal{D}_{nq}^{mp}\right)\right]=0,
    \end{equation}
     \begin{equation}
        -q(q+1) B^{p}_{q}+\mathrm{Pe}\left[\mathcal{B}\sum^{\mathcal{N}}_{n=0}\sum_{m=0}^{n}\left(A^{m}_{n}(\mathrm{bw})^{mp}_{nq}\right)+\dfrac{(\mathcal{B}-1)}{2}p A^{p}_{q}-\mathcal{K}_{p}\sum^{\mathcal{N}}_{n=0}\sum_{m=0}^{n}\left(B^{m}_{n}\mathcal{D}_{nq}^{mp}\right)\right]=0.
    \end{equation}    
\end{subequations}

The above equation must be solved numerically, following the same procedure as before for the case when $\alpha=0$.

\begin{figure}
     \hspace{-2.5cm}
    \includegraphics[width=1.2\linewidth]{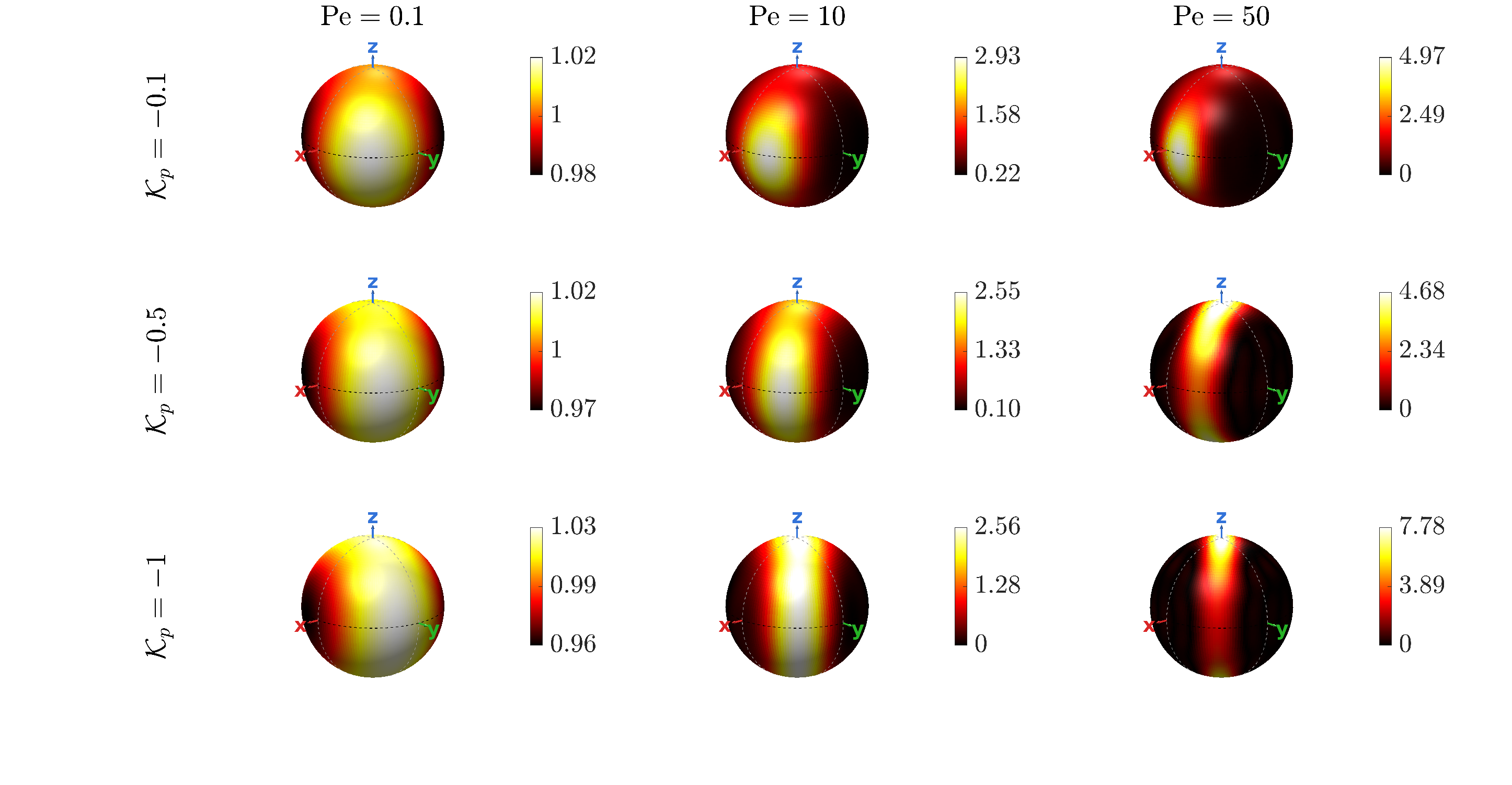}
    \caption{Contour plot of the orientation distribution function $4\pi f$ on the unit sphere for a prolate spheroid with $\mathcal{B}=0.9$ ($r=4.36$). The results correspond to the case where the gravity vector is aligned with the flow direction ($\alpha=\pi/2$ and $\beta=0$).}
    \label{fig:ffgx}
\end{figure}

In figure~\ref{fig:ffgx}, we present the orientation distribution $4\pi f$ on the unit sphere for different values of $\mathcal{K}_{p}$ and $\mathrm{Pe}$, and when $\alpha=\pi/2$ and $\beta=0$. As in the previous case, both parameters induce deviations from an initially isotropic distribution. With increasing magnitude of $\mathcal{K}_{p}$, the distribution becomes progressively concentrated in the vorticity direction, as illustrated for $\mathcal{K}_{p}=-1$. This trend is consistently observed across all range of $\mathrm{Pe}$.

\begin{figure}
\centering  
\subfigure[]{\includegraphics[width=0.49\linewidth]{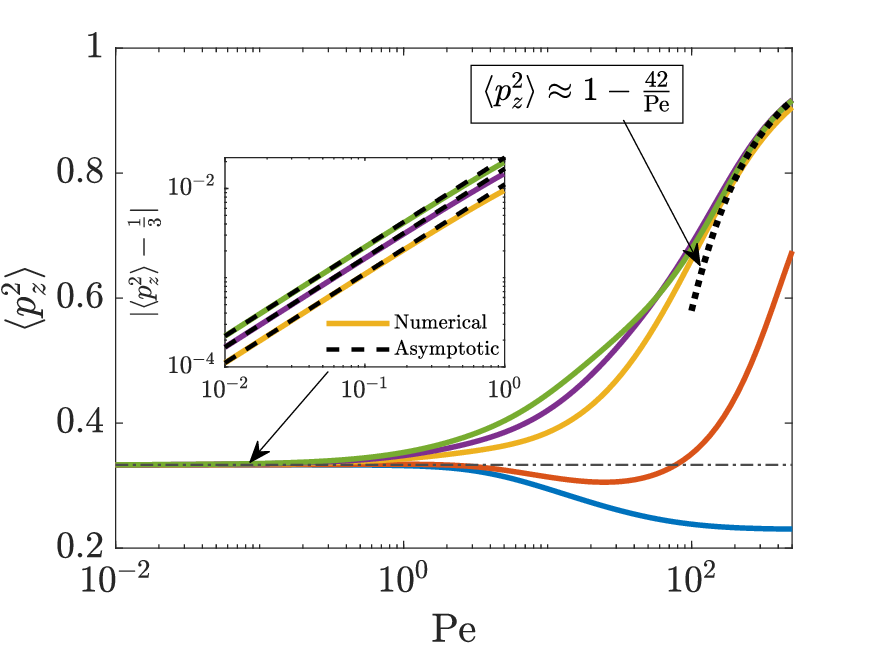}}
\subfigure[]{\includegraphics[width=0.49\linewidth]{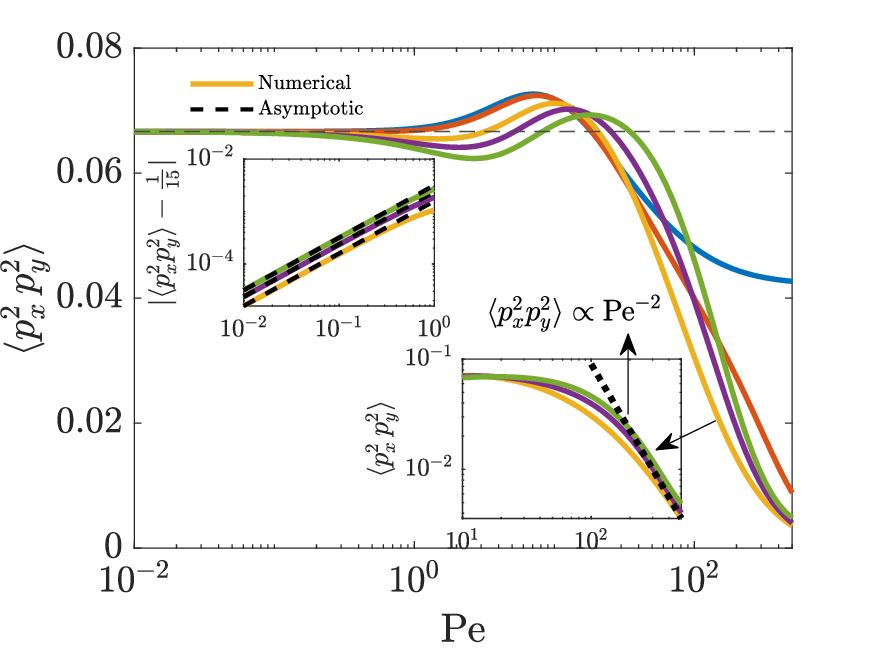}}
\caption{Plot of orientation moments $\langle p^2_z \rangle$ and $\langle p^2_x p^2_y\rangle$ versus $\mathrm{Pe}$  for the case when gravity is aligned in the flow direction ($\alpha=\pi/2$ and $\beta=0$) and for various values of $\mathcal{K}_{p}$:
(\textcolor[rgb]{0.000,0.447,0.741}{\rule[0.5ex]{1em}{1.5pt}}) $\mathcal{K}_{p}=0$,
(\textcolor[rgb]{0.850,0.325,0.098}{\rule[0.5ex]{1em}{1.5pt}}) $\mathcal{K}_{p}=-0.1$,
(\textcolor[rgb]{0.929,0.694,0.125}{\rule[0.5ex]{1em}{1.5pt}}) $\mathcal{K}_{p}=-0.5$,
(\textcolor[rgb]{0.494,0.184,0.556}{\rule[0.5ex]{1em}{1.5pt}}) $\mathcal{K}_{p}=-0.75$, and
(\textcolor[rgb]{0.466,0.674,0.188}{\rule[0.5ex]{1em}{1.5pt}}) $\mathcal{K}_{p}=-1$. (a) Variation of the orientation moment $\langle p_{z}^{2}\rangle$ with $\mathrm{Pe}$, and (b) variation of $\langle p_{x}^{2}p_{y}^{2}\rangle$ with $\mathrm{Pe}$. We compare numerically obtained $\langle p_{z}^{2}\rangle$ and $\langle p_{x}^{2}p_{y}^{2}\rangle$, respectively, with their asymptotic low- and high-$\mathrm{Pe}$ limits. Insets show a comparison between the asymptotic solution in the low-$\mathrm{Pe}$ limit (dashed lines) and numerical results (solid lines). The dotted lines in both panels represent the asymptotic expressions in the high-$\mathrm{Pe}$ limit given in (\ref{eq:pz2gx_largePe}) and (\ref{eq:pxpy_gx_largePe}). In first panel of the main figure, at $\mathcal{K}_{p}=-1$ and $\mathcal{B}=0.9$ we get $\mathrm{tr}~\mathbb{C}\approx 42$, as indicated by the dotted line. Dash-dotted line corresponds to the moments for the isotropic orientation distribution. All results are presented for $\mathcal{B}=0.9$ ($r=4.36$) and $\mathcal{N}=30$.} 
\label{fig:gx_pzsq}
\end{figure}

In figure~\ref{fig:gx_pzsq} (first panel), we show the orientation moment $\langle p_z^2\rangle$ as a function of $\mathrm{Pe}$ for different values of $\mathcal{K}_{p}$ ($\mathcal{N}=30$). As discussed in \S\ref{sec:theory}, when gravity is aligned with the flow direction, a prolate spheroid preferentially aligns within the $\hat{\boldsymbol{y}}$-$\hat{\boldsymbol{z}}$ plane. The orientation moment increases with increasing $|\mathcal{K}_{p}|$, reflecting the enhanced tendency of the spheroid to align with the $\hat{\boldsymbol{z}}$ axis as settling becomes increasingly dominant. For $|\mathcal{K}_{p}|\geq0.5$, beyond the bifurcation point, $\langle p_z^2\rangle$ approaches unity as $\mathrm{Pe}$ increases. 

In the second panel, we show $\langle p_x^2p_y^2\rangle$ as a function of $\mathrm{Pe}$. The moment initially exhibits a non-monotonic variation before decreasing with increasing $\mathrm{Pe}$. This decrease is significantly more pronounced for $\mathcal{K}_{p}\neq0$. Also, we compare the numerically obtained orientation moments with the corresponding asymptotic results in the low- and high-$\mathrm{Pe}$ limits. The insets with dashed lines compare the numerical results with the low-$\mathrm{Pe}$ asymptotic predictions, while the high-$\mathrm{Pe}$ asymptotic expressions, denoted as dotted black lines are shown in the main figure in the left panel and in the inset figure in the right panel. The numerical results show good agreement with the asymptotic predictions in both limits. For the high-$\mathrm{Pe}$ regime, we use equations~(\ref{eq:pz2gx_largePe}) and~(\ref{eq:pxpy_gx_largePe}) for comparison. The evaluation of $\langle p_z^2\rangle$ requires $\mathrm{tr}~\mathbb{C}$, which is $\mathrm{tr}~\mathbb{C}\approx42$ for $\mathcal{K}_{p}=-1$ and $\mathcal{B}=0.9$. The asymptote then matches the computed moment to $1\%$ at $\mathrm{Pe}=200$ but is $15\%$ low at $\mathrm{Pe}=100$, which is the quantitative content of the condition $\mathrm{Pe}\gg\mathrm{tr}~\mathbb{C}$ stated with (\ref{eq:pz2gx_largePe}).

\begin{figure}
    \centering
    \includegraphics[width=0.95\linewidth]{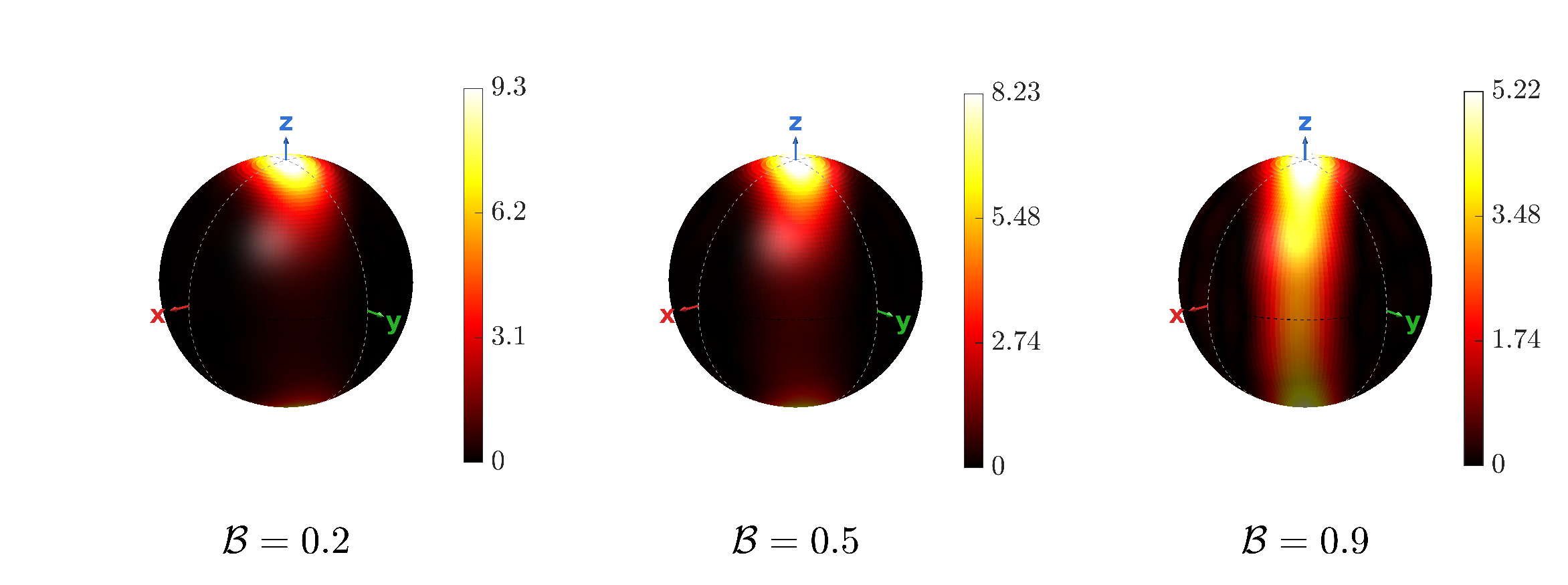}
    \caption{Orientation distribution function $4\pi f$ on the unit sphere for $\mathcal{K}_{p}=-1$ and $\mathrm{Pe}=30$, with the gravity vector aligned with the flow direction.}
    \label{fig:con_gx_varyb}
\end{figure}

Next, in figure~\ref{fig:con_gx_varyb}, we present the orientation distribution function $4\pi f$ on the unit sphere for $\alpha=\pi/2$ and $\beta=0$. For $\mathcal{B}=0.9$, the distribution is concentrated between the $\hat{\boldsymbol{x}}$- and $\hat{\boldsymbol{y}}$-axes, with a strong preference for alignment along the $\hat{\boldsymbol{z}}$-axis. As $\mathcal{B}$ decreases, the distribution progressively shifts towards the $\hat{\boldsymbol{z}}$-axis, with the spread in the $\hat{\boldsymbol{x}}$-$\hat{\boldsymbol{y}}$ plane becoming increasingly suppressed. For sufficiently small $\mathcal{B}$, corresponding to a nearly spherical particle, the distribution becomes strongly concentrated along the $\hat{\boldsymbol{z}}$-axis. The trend is therefore the opposite of the one that might be expected: it is
the more nearly spherical particles, for which the Jeffery torque is weakest,
whose orientation is most strongly concentrated by settling. The
mechanism can be made quantitative. The attractor here is the vorticity axis
$\theta=0$, whose linearisation is (\ref{eq:lin_pole_gx}); solving the associated
Lyapunov equation for the covariance $\mathbb{C}$ in the tangent plane gives
$\det\mathbb{C}=4(\mathcal{K}_{p}^{2}+1)/
[\mathcal{K}_{p}^{2}(1-\mathcal{B}^{2})]$, the counterpart of the trace quoted
in \S\ref{sec:largePe}. The Gaussian estimate of the peak is then
\begin{equation}\label{eq:peakgauss}
    4\pi f_{\max}\simeq\dfrac{\mathrm{Pe}}{\sqrt{\det\mathbb{C}}}
    =\dfrac{\mathrm{Pe}\,|\mathcal{K}_{p}|}{2}
          \sqrt{\dfrac{1-\mathcal{B}^{2}}{\mathcal{K}_{p}^{2}+1}},
\end{equation}
which falls with $\mathcal{B}$ as $\sqrt{1-\mathcal{B}^{2}}$ and so reproduces
the trend.

\subsection{Role of noise: diffusion over orbits versus
Kramers escape between equilibria}\label{sec:kramers}
 
The large-$\mathrm{Pe}$ analysis of the preceding subsection
reveals the typical (Gaussian) fluctuations around deterministic
attractors. However, the qualitative role of noise differs
fundamentally between the pure Jeffery problem
($\mathcal{K}_{p}=0$) and the settling problem
($\mathcal{K}_{p}\neq 0$). In this subsection, we contrast these
two roles and show that settling introduces a new class of
noise-induced phenomena, Kramers escape and phase
slips, which are absent in the classical Hinch-Leal theory.
The form these events take is set by the particle
shape: for an oblate spheroid the escape is a one-dimensional
barrier crossing in the azimuthal angle, whereas for a prolate
spheroid it is an excursion out of the vorticity-aligned state
along its weakly attracting direction.
 
\subsubsection{Pure shear ($\mathcal{K}_{p}=0$): diffusion over
the Jeffery orbit constant}
 
In the absence of settling, the deterministic dynamics supports a one-parameter family of Jeffery orbits
labelled by the orbit constant $C$. There are no isolated
attractors, and noise cannot drive transitions between
``states'' because no distinct states exist. Instead, as shown by
\citet{leal1971effect} and \citet{hinch1972effect}, weak
Brownian diffusion acts on the slow time scale
$O(\mathrm{Pe})$ to randomise the orbit constant $C$. The
resulting steady-state distribution over $C$ is independent of
initial conditions and can be obtained from a one-dimensional
Fokker-Planck equation for the orbit-averaged dynamics. The
orientation moments receive $O(1)$ corrections (not
exponentially small) from this diffusion, because the orbit
constant parametrizes qualitatively different tumbling modes
(spinning, tumbling, kayaking) that are all equally accessible
under noise. The noise plays a regularizing role:
it selects a unique steady state from the degenerate continuum
of Jeffery orbits.
 
\subsubsection{Settling ($\mathcal{K}_{p}\neq 0$,
$\mathcal{R}>1$): Kramers escape and phase slips}
 
When the settling parameter exceeds the SNIC bifurcation
threshold ($\mathcal{R}>1$), isolated equilibria appear
on the equatorial circle $\theta=\pi/2$,
separated by saddle points. For the
configurations in which the azimuthal dynamics decouples from
$\theta$ ($\alpha=\pi/2$), the reduced $\phi$ equation takes the
form $\dot{\phi}=-U'(\phi)$, where the quasi-potential
\begin{equation}\label{eq:quasipot}
    U(\phi) = \frac{\phi}{2}
    - \frac{\mathcal{R}}{4}\sin(2\phi-\psi),
\end{equation}
with
$\mathcal{R}=\sqrt{\mathcal{B}^{2}
+\mathcal{K}^{2}\mathcal{F}^{2}_{p}}$ and
$\tan\psi=\pm\mathcal{K}\mathcal{F}_{p}/\mathcal{B}$ (sign
depending on the configuration), is a tilted periodic potential
with local minima (stable equilibria) and maxima (saddle points).
The barrier separating adjacent wells is
\begin{equation}\label{eq:barrier}
    \Delta U = \frac{1}{2}\left[
    \sqrt{\mathcal{R}^{2}-1}
    - \cos^{-1}\!\left(\frac{1}{\mathcal{R}}\right)\right],
\end{equation}
which vanishes at $\mathcal{R}=1$ and grows linearly as
$\Delta U\approx\mathcal{R}/2-\pi/4$ for $\mathcal{R}\gg 1$.
Because the potential is tilted, $U$ satisfies
$U(\phi+\pi)=U(\phi)+\pi/2$, so the two barriers on either side of
a minimum are unequal; (\ref{eq:barrier}) is the smaller of the
two, crossed in the direction of the deterministic drift, and is
therefore the one that controls the escape.
 
The reduction to a single degree of freedom is
legitimate only if the equatorial circle is itself attracting.
Linearising (\ref{eq:pdoteq}) about one of these equilibria in the
local tangent plane gives two eigenvalues. The azimuthal one,
$\lambda_{\phi}=-\mathcal{G}_1$ with
$\mathcal{G}_1=\sqrt{\mathcal{R}^{2}-1}$, is the one already
contained in $U''$ above. The second measures the growth of a
perturbation out of the shear plane,
\begin{equation}\label{eq:lamperp}
    \lambda_{\perp}=-\frac{\mathcal{K}_{p}+\mathcal{G}_1}{2}.
\end{equation}
Since $\mathcal{G}^{2}_1-\mathcal{K}^{2}_{p}=\mathcal{B}^{2}-1<0$
for every physical spheroid, $\mathcal{G}_1<|\mathcal{K}_{p}|$
whenever $\mathcal{R}>1$, and the sign of $\lambda_{\perp}$ is
fixed by the sign of $\mathcal{K}_{p}$ alone. It is negative for
an oblate spheroid ($\mathcal{K}_{p}>0$) and positive for a
prolate one ($\mathcal{K}_{p}<0$). The washboard description
therefore applies to oblate spheroids, for which the orientation
collapses onto the equatorial circle on the time scale
$|\lambda_{\perp}|^{-1}$ and thereafter executes the
one-dimensional dynamics above. For prolate spheroids the same
equilibria exist but are saddles of the full dynamics on $S^{2}$,
and the attractor is instead the vorticity-aligned orientation
$\boldsymbol{p}=\pm\hat{\boldsymbol{z}}$ identified in
\S\ref{sec:asym_gx_largeprsec}. We treat the two cases in turn.
 
We consider first an oblate spheroid, for which
the azimuthal reduction is exact at leading order. Isotropic
rotary diffusion perturbs this potential dynamics.
On the equatorial circle $\sin\theta=1$, so the
azimuthal component of the isotropic rotary diffusion reduces to
additive noise of intensity $1/\mathrm{Pe}$, and the corresponding
Langevin equation is
\begin{equation}\label{eq:langevin_phi}
    d\phi = -U'(\phi)\,dt + \sqrt{2/\mathrm{Pe}}\,d\mathcal{W},
\end{equation}
which describes overdamped motion in a tilted
periodic potential with noise intensity $1/\mathrm{Pe}$, where
$\mathcal{W}$ is the Wiener process. The Kramers escape rate from a
stable potential well is \citep{kramers1940brownian,risken1996fokker}
\begin{equation}\label{eq:kramers_rate}
    r_{e} = \frac{\sqrt{\mathcal{R}^{2}-1}}{2\pi}
    \exp\!\left(-\mathrm{Pe}\,\Delta U\right).
\end{equation}
On the equatorial circle $\sin\theta=1$, so the azimuthal noise in
(\ref{eq:langevin_phi}) is additive with intensity $1/\mathrm{Pe}$ and
independent of $\theta$. For that one-dimensional model the escape rate is
available in closed form and not only in the Kramers limit. We stress that the
reduction is exact for a process constrained to the circle, and is the leading
large-$\mathrm{Pe}$ reduction of the full problem on $S^{2}$: isotropic rotary
diffusion has a transverse component, so the circle is invariant under the
deterministic drift but not under the noise. The resulting error is
quantifiable and small when the circle is strongly attracting. Comparing the
one-dimensional rate below with equator crossings measured from the full
It\^o dynamics (\ref{eq:lethtphi}) for an oblate spheroid at
$\mathcal{B}=-0.5$, the ratio of the two is $1.03$ and $1.02$ at
$\mathrm{Pe}=30$ and $50$ for $\mathcal{K}_{p}=1$, and $1.14$ at
$\mathrm{Pe}=30$ for $\mathcal{K}_{p}=1.5$. For
overdamped motion in a periodic potential satisfying
$U(\phi+\pi)=U(\phi)+\pi/2$ with additive noise of intensity
$D=1/\mathrm{Pe}$, the stationary solution carries a constant
probability current, and the mean rate of $\pi$-slips is
\begin{equation}\label{eq:exactrate}
    r_{e}^{\mathrm{ex}}
    =\frac{D\left(e^{\pi/2D}-1\right)}
    {\displaystyle\int_{0}^{\pi}\!\mathrm{d}\phi
    \int_{\phi}^{\phi+\pi}\!\mathrm{d}\phi'\,
    \exp\!\left[\frac{U(\phi')-U(\phi)}{D}\right]},
\end{equation}
a single quadrature over one period
\citep{risken1996fokker}. Evaluating the double integral by its saddle
point for $\mathrm{Pe}\,\Delta U\gg 1$ recovers
(\ref{eq:kramers_rate}), so (\ref{eq:exactrate}) is the parent result
and (\ref{eq:kramers_rate}) its large-barrier limit. Direct quadrature of (\ref{eq:exactrate}) with the potential
(\ref{eq:quasipot}) gives relative differences
$(r_{e}-r_{e}^{\mathrm{ex}})/r_{e}^{\mathrm{ex}}$ of $+0.74\%$ at
$\mathcal{R}=5$, $+3.4\%$ at $\mathcal{R}=2$ and $+12.5\%$ at
$\mathcal{R}=1.12$ for $\mathrm{Pe}=30$, and $+0.44\%$, $+1.9\%$ and $+21\%$
respectively for $\mathrm{Pe}=50$. The controlling parameter is
$\mathrm{Pe}\,\Delta U$, not $\mathrm{Pe}$ alone: at $\mathcal{R}=5$ and $2$ it
is $53$ and $11$ at $\mathrm{Pe}=30$, and the saddle-point evaluation is
accurate, whereas at $\mathcal{R}=1.12$ it is only $0.56$ and the harmonic approximation is far
from its asymptotic regime. The approach to that regime is not monotonic: at
$\mathcal{R}=1.12$ the relative difference is $+12.5\%$, $+20.6\%$, $+20.7\%$,
$+11.5\%$, $+6.8\%$ and $+1.3\%$ at $\mathrm{Pe}=30$, $50$, $100$, $200$, $300$
and $1200$, rising through a maximum near $\mathrm{Pe}\,\Delta U\approx 2$
before decaying as it must once the barrier is well resolved. Equation
(\ref{eq:exactrate}) should therefore be used whenever $\mathrm{Pe}\,\Delta U$
is not large, and (\ref{eq:kramers_rate}) only well beyond that maximum. Equation (\ref{eq:exactrate}) remains valid
throughout, and shows that the slip rate approaches a finite,
$\mathrm{Pe}$-dependent limit at the bifurcation rather than vanishing
as the prefactor of (\ref{eq:kramers_rate}) would suggest.
 
Each escape event constitutes a phase slip: a sudden
$\pi$-rotation of the azimuthal orientation (tumbling events), after which the
particle settles into the adjacent potential well. The mean
residence time between successive slips,
$\langle\tau\rangle\sim r_{e}^{-1}
\sim\exp(\mathrm{Pe}\,\Delta U)$, is exponentially sensitive
to both the noise intensity and the barrier height.
 
This Kramers mechanism is entirely absent in the pure-shear
Jeffery problem ($\mathcal{K}_{p}=0$), where no potential
barriers exist. It represents a qualitative change in the role
of noise: rather than diffusing smoothly over a continuum of
orbits as in the Hinch-Leal theory, the
particle dwells near a single equilibrium for exponentially long
times before executing a sudden reorientation. These phase slips
are extreme events in the sense of large deviation theory
\citep{freidlin1998random}: rare fluctuations that overcome the
deterministic potential barrier, with residence times drawn from
an exponential distribution
$P(\tau)\sim\exp(-r_{e}\tau)$.
 
\begin{figure}
    \centering
    \includegraphics[width=\linewidth]{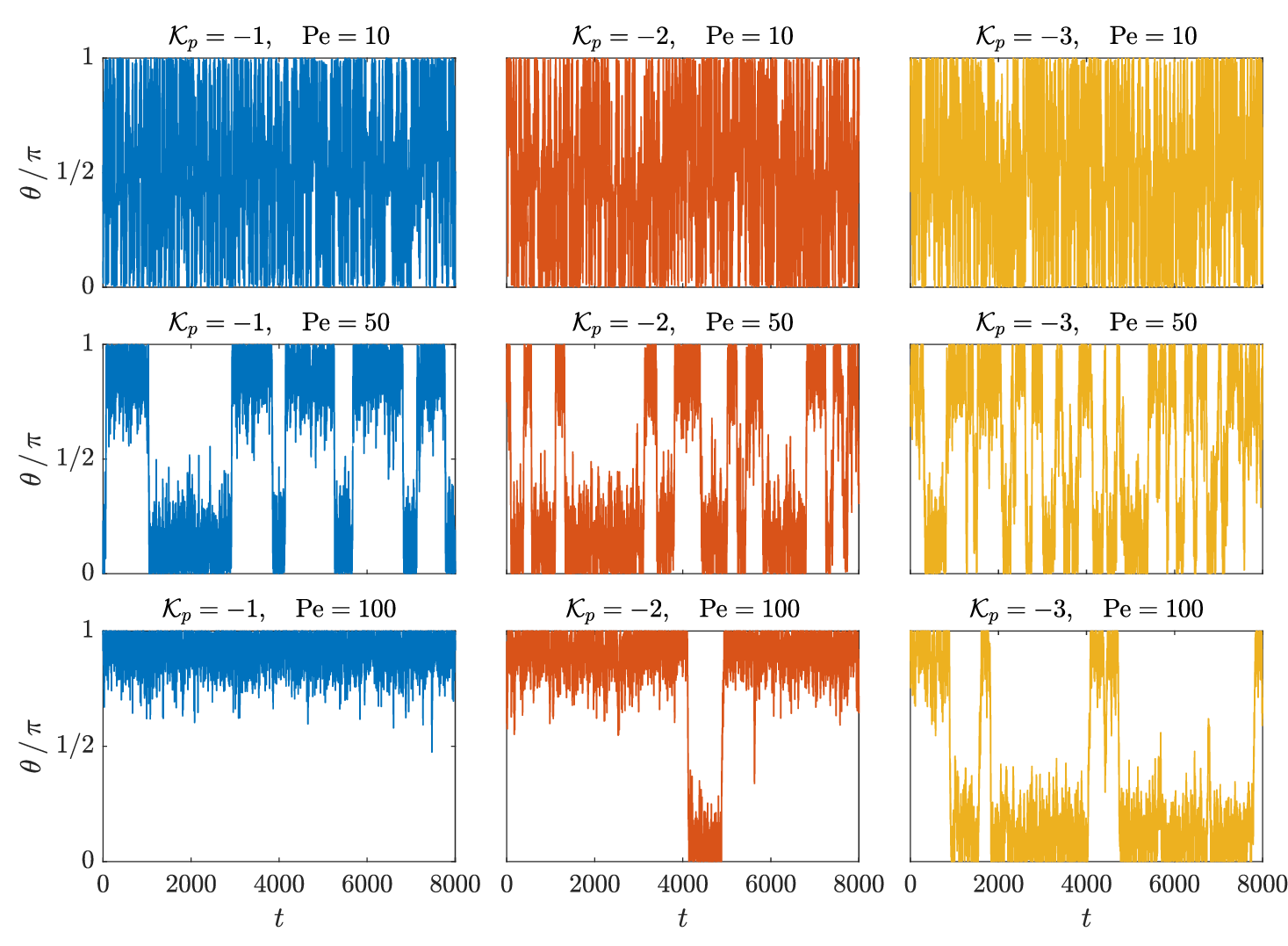}
    \caption{Polar trajectories $\theta(t)$ from Langevin
    simulations of the flow-aligned configuration
    ($\alpha=\pi/2$, $\beta=0$) and $\mathcal{B}=0.5$. Columns
    correspond to increasing $|\mathcal{K}_{p}|$
    ($|\mathcal{K}_{p}|=1$, $2$, $3$); rows to
    increasing P\'{e}clet number ($\mathrm{Pe}=10$, $50$, $100$).
    Excursions out of the vorticity-aligned state
    become progressively rarer as $\mathrm{Pe}$ increases. The time step is fixed at $dt=10^{-3}$.}
    \label{fig:langevin_phi}
\end{figure}
 
The analogy with polymer tumbling in turbulent shear is direct.
\citet{chertkov2005polymer} and \citet{turitsyn2007polymer}
have shown that a dumbbell in a mean shear flow with random velocity
gradient fluctuations exhibits aperiodic tumbling, with the
orientation PDF developing algebraic tails around the
shear-preferred direction. In their framework, the Weissenberg
number plays the role of $\mathrm{Pe}$ and the ratio of noise
intensity to mean shear plays the role of $1/\mathcal{R}$.
\citet{gerashchenko2006statistics} have confirmed these predictions
experimentally for single DNA molecules in shear flow. The
Kramers phase-slip statistics derived here are the rigid-particle
analogue of their tumbling-time statistics.
 
We note that the one-dimensional rate
(\ref{eq:kramers_rate}) applies rigorously only when the
equatorial circle attracts, that is when $\lambda_{\perp}<0$ in
(\ref{eq:lamperp}), so that $\phi$ evolves in a true
one-dimensional potential. For the fully coupled system on
$S^{2}$, the deterministic drift is not a pure gradient and the
system lacks detailed balance \citep{brenner1974transport}. The appropriate escape rate then
involves the quasipotential from Freidlin-Wentzell theory
\citep{freidlin1998random}, obtained by solving the associated
Hamilton-Jacobi equation rather than from a simple potential
difference. This is the situation for the prolate
case, to which we now turn.
 
For $\mathcal{K}_{p}<0$ the attractor of the
flow-aligned configuration is the pole, and the relevant escape is
from there. Its two eigenvalues follow from the linearisation
(\ref{eq:lin_pole_gx}),
\begin{equation}\label{eq:lampole}
    \lambda_{\pm}=\frac{\mathcal{K}_{p}\pm\mathcal{G}_1}{2},
    \qquad
    \mathrm{tr}\,\mathbb{J}=\mathcal{K}_{p},
    \qquad
    \det\mathbb{J}=\frac{1-\mathcal{B}^{2}}{4},
\end{equation}
both negative for $\mathcal{K}_{p}<0$. The determinant is
independent of the settling strength, so increasing
$|\mathcal{K}_{p}|$ does not deepen the basin uniformly: it
stiffens one direction, $\lambda_{-}\to\mathcal{K}_{p}$, while
softening the orthogonal one,
$\lambda_{+}\to-(1-\mathcal{B}^{2})/(4|\mathcal{K}_{p}|)$. The
weakly attracting eigendirection turns progressively towards the
great circle normal to gravity, which is the broadside plane, and
its relaxation rate decays as $|\mathcal{K}_{p}|^{-1}$. Noise
therefore drives excursions preferentially along that circle, at a
rate which increases with $|\mathcal{K}_{p}|$, in contrast
with the oblate case where the barrier (\ref{eq:barrier}) grows.
 
To illustrate this, we integrate the Langevin
equations governing the evolution of $\theta$ and $\phi$
numerically for the flow-aligned configuration ($\alpha=\pi/2$,
$\beta=0$) with $\mathcal{B}=0.5$.
Figure~\ref{fig:langevin_phi} shows $\theta(t)$ at three values of
$\mathrm{Pe}$ ($10$, $50$ and $100$) and three values of
$\mathcal{K}_{p}$ ($-1$, $-2$ and $-3$), all above the
bifurcation and at $dt=10^{-3}$. The time series exhibits three
distinct regimes. At low $\mathrm{Pe}$ the noise randomises the
orientation continuously and no separation of time scales is
visible. As $\mathrm{Pe}$ increases the particle dwells near the
vorticity axis for long intervals, punctuated by rapid excursions
in which the director sweeps around the broadside circle, and
these become progressively rarer. At fixed $\mathrm{Pe}$ they become more frequent as $|\mathcal{K}_{p}|$
increases, which is the opposite of what the azimuthal barrier
(\ref{eq:barrier}) would give, since that barrier grows with
$|\mathcal{K}_{p}|$. A plausible mechanism is the softening of
$\lambda_{+}$ in (\ref{eq:lampole}), which turns the weak eigendirection of the
polar attractor towards the broadside circle. Because a quasipotential barrier
need not follow the local curvature, we do not infer the rate from
(\ref{eq:lampole}) but measure it directly. Single realisations such as those
of figure~\ref{fig:langevin_phi} show the morphology of the excursions and
cannot by themselves establish a rate law, so we integrated ensembles of the
It\^{o} system (\ref{eq:lethtphi}) and counted excursions under a stated
criterion: the director is assigned to the state $\mathrm{sgn}(p_{z})$ only
once $|p_{z}|$ exceeds a threshold $p_{c}$, and an excursion is a change of that
state. The two thresholds required to register an event remove the recrossings
that occur while the director sits near the equator, so the count is a genuine
transition count and not a level-crossing count. Table~\ref{tab:excursion}
reports the results for $\mathcal{B}=0.5$ with $p_{c}=0.5$, from $400$ paths of
duration $800$ at $dt=10^{-3}$ with the first $15\%$ discarded.

\begin{table}
  \begin{center}
  \begin{tabular}{lcccccc}
    $|\mathcal{K}_{p}|$ & $\mathrm{Pe}=40$ & $60$ & $80$ & $100$
      & $\Delta U_{\mathrm{eff}}$ & $\Delta U$ from (\ref{eq:barrier})\\[3pt]
    $1$ & $3.48\times10^{-3}$ & $8.97\times10^{-4}$ & $1.88\times10^{-4}$
        & $5.15\times10^{-5}$ & $0.0706\pm0.0024$ & $0.018$\\
        & \small$(947)$ & \small$(244)$ & \small$(51)$ & \small$(14)$ & & \\
    $2$ & $6.39\times10^{-3}$ & $2.42\times10^{-3}$ & $8.86\times10^{-4}$
        & $4.19\times10^{-4}$ & $0.0474\pm0.0013$ & $0.369$\\
        & \small$(1737)$ & \small$(659)$ & \small$(241)$ & \small$(114)$ & & \\
    $3$ & $8.14\times10^{-3}$ & $3.74\times10^{-3}$ & $2.06\times10^{-3}$
        & $1.12\times10^{-3}$ & $0.0341\pm0.0009$ & $0.818$\\
        & \small$(2213)$ & \small$(1018)$ & \small$(559)$ & \small$(305)$ & & \\
  \end{tabular}
  \caption{Broadside-excursion rate per unit strain for a
  prolate spheroid in the flow-aligned configuration
  ($\alpha=\pi/2$, $\beta=0$, $\mathcal{B}=0.5$), from ensembles of
  (\ref{eq:lethtphi}) with the threshold criterion $p_{c}=0.5$. Numbers in
  parentheses are the events observed. $\Delta U_{\mathrm{eff}}$ is the slope
  of a weighted fit of $\ln(\text{rate})$ against $-\mathrm{Pe}$, with
  $R^{2}=0.999$, $0.998$ and $0.995$ for the three rows. The last column is the
  azimuthal barrier (\ref{eq:barrier}) at the same $\mathcal{R}$, which moves in
  the opposite direction.}
  \label{tab:excursion}
  \end{center}
\end{table}

Three conclusions follow. First, $\ln(\text{rate})$ is linear
in $\mathrm{Pe}$ over more than two decades of rate, so the excursions are
noise-activated and possess a well-defined effective barrier. Second, that
barrier decreases as $|\mathcal{K}_{p}|$ grows, from
$\Delta U_{\mathrm{eff}}=0.071$ at $|\mathcal{K}_{p}|=1$ to $0.034$ at
$|\mathcal{K}_{p}|=3$, whereas the azimuthal barrier (\ref{eq:barrier}) rises
over the same range from $0.018$ to $0.818$. The escape route is therefore not
the azimuthal one, and the rate rises with the settling strength rather than
falling with it. Third, $\Delta U_{\mathrm{eff}}$ is an order of magnitude
smaller than (\ref{eq:barrier}) at $|\mathcal{K}_{p}|=2$ and $3$, which is why
the excursions remain observable at P\'{e}clet numbers where azimuthal slips would
be entirely suppressed. Repeating the calculation with $p_{c}=0.7$ gives
$0.0687\pm0.0024$, $0.0441\pm0.0013$ and $0.0324\pm0.0009$, so the fitted
barriers shift by less than $7\%$ while the ordering and the trend with
$|\mathcal{K}_{p}|$ are unchanged. The exponent is therefore a property of the
dynamics and not of the counting criterion, which sets the prefactor. We do not identify $\Delta U_{\mathrm{eff}}$ with the
Freidlin-Wentzell quasipotential, whose computation on $S^{2}$ requires the
solution of a Hamilton-Jacobi equation that we do not attempt; it is the
measured activation energy of the transition, and the two coincide only in the
limit $\mathrm{Pe}\to\infty$.

Each excursion carries
the director from $\theta=0$ to $\theta=\pi$; because
$\boldsymbol{p}$ and $-\boldsymbol{p}$ describe the same physical
orientation, such an event returns the particle to the same
nematic state, and is a large fluctuation about a single attractor
rather than a transition between two distinct equilibria. The
trajectory shown in Appendix~\ref{appbmom} makes this explicit:
the director is concentrated near the $\hat{\boldsymbol{y}}$-$\hat{\boldsymbol{z}}$
plane, which is the broadside plane for gravity along $\hat{\boldsymbol{x}}$, and
traverses it repeatedly. This is consistent with figure~\ref{fig:ffgx}, where
the orientation distribution function is large in the $\hat{\boldsymbol{y}}$-$\hat{\boldsymbol{z}}$
plane; in the presence of noise the particle explores a broad
range of orientations within the broadside plane. The rate of
these events remains exponentially sensitive to $\mathrm{Pe}$, so
the picture of rare, noise-activated reorientation carries over,
but the controlling barrier is the quasipotential on $S^{2}$
rather than (\ref{eq:barrier}); its computation is left for future
work.
 
Near the SNIC bifurcation
($\mathcal{R}\to 1^{+}$) the barrier vanishes as
$\Delta U\simeq(\sqrt{2}/3)(\mathcal{R}-1)^{3/2}$, so that
$\mathrm{Pe}\,\Delta U\to 0$ and the exponential factor in
(\ref{eq:kramers_rate}) tends to unity. The prefactor
$\sqrt{\mathcal{R}^{2}-1}/2\pi$ vanishes on the same approach,
however, and the harmonic approximation underlying
(\ref{eq:kramers_rate}) fails because the well becomes shallow and
strongly anharmonic. Equation (\ref{eq:kramers_rate}) is therefore
not uniformly valid as $\mathcal{R}\to 1^{+}$, and the escape time
in that limit is governed by the slow passage through the
saddle-node ghost rather than by a barrier crossing. The dynamics
there is intermittent: the spheroid alternates between prolonged
dwelling near the equilibrium orientation and rapid rotational
excursions during which it completes a near-$\pi$ phase slip.
Below the bifurcation ($\mathcal{R}<1$), no potential wells exist
and the particle tumbles continuously. Noise then causes phase diffusion. The correct statement is
about the unwrapped phase: $\mathrm{Var}[\phi(t)-\omega_m t]=2D_{\rm eff}t$ with
$D_{\rm eff}=O(\mathrm{Pe}^{-1})$, where $\omega_m$ is the mean angular
frequency. The period of a single cycle is a first-passage random variable with
a stationary distribution at fixed parameters; its variance does not grow with
the observation time, and any statement about it requires a separate
first-passage calculation.
 
For the vorticity-aligned case ($\alpha=0$), the situation is
intermediate: the azimuthal dynamics remains of Jeffery type
(no potential wells in $\phi$), but the settling creates a
deterministic restoring force in $\theta$. The noise in $\theta$
is of the Ornstein-Uhlenbeck type (Gaussian fluctuations around
the deterministic attractor $\theta^{*}$), while the noise in
$\phi$ is of the diffusive type (phase diffusion along the
periodic orbit). This mixed character explains the behaviour
observed in figures~\ref{fig:gz_pzsq}:
$\langle p^{2}_{z}\rangle$ (sensitive to the $\theta$
confinement) decreases monotonically as $\mathrm{Pe}^{-1}$, while
$\langle p^{2}_{x}p^{2}_{y}\rangle$ (sensitive to the azimuthal
distribution) approaches a finite plateau set by the
orbit-averaged Jeffery statistics.
The description above is for a prolate spheroid, whose attractor in this
configuration is the tumbling orbit at $\theta^{*}=\pi/2$. For an oblate
spheroid the attractor is instead the vorticity pole $\theta^{*}=0$, where
$\phi$ is undefined, and the two angles do not exchange roles. Linearising
about the pole in a local Cartesian tangent plane gives eigenvalues
$-\mathcal{K}_{p}\pm\tfrac{i}{2}\sqrt{1-\mathcal{B}^{2}}$, so both tangent
components are restored at rate $\mathcal{K}_{p}$ while rotating together at
$\tfrac{1}{2}\sqrt{1-\mathcal{B}^{2}}$. The fluctuations are then a single
coupled two-dimensional Ornstein-Uhlenbeck process with no freely diffusing
phase.
 
\subsubsection{Experimental implications: inferring noise
intensity from flip statistics}
 
The exponential sensitivity of the Kramers rate
(\ref{eq:kramers_rate}) to the ratio $\Delta U/D$ (where
$D=1/\mathrm{Pe}$) suggests a practical experimental protocol
for characterising the effective rotary diffusivity in
suspensions. If an oblate settling spheroid is
observed in a simple shear
flow at known $\mathcal{B}$ and $\mathcal{K}$ (so that
$\Delta U$ can be computed from (\ref{eq:barrier})), the mean
time $\langle\tau\rangle$ between successive azimuthal phase
slips can be measured from particle tracking data. Since
\begin{equation}\label{eq:lntau}
    \ln\langle\tau\rangle \approx \mathrm{Pe}\,\Delta U
    - \tfrac{1}{2}\ln(\mathcal{R}^{2}-1)
    + \ln(2\pi),
\end{equation}
a plot of
$\ln\langle\tau\rangle+\tfrac{1}{2}\ln(\mathcal{R}^{2}-1)-\ln(2\pi)$
against $\Delta U$, assembled by varying the field strength
$\mathcal{K}$, is a straight line through the origin whose slope
is $\mathrm{Pe}=\gamma/\mathcal{D}_{r}$. Retaining the prefactor
is necessary because $\mathcal{R}$, and hence
$\ln(\mathcal{R}^{2}-1)$, varies with $\mathcal{K}$ along the same
sequence of measurements. This
provides a direct measurement of $\mathcal{D}_{r}$ from
single-particle observations.

How the sequence of measurements is assembled requires care.
Combining $\mathcal{K}=\rho_{f}\tau_{p}^{2}g^{2}/(8\pi\mu\gamma)$ with
$\mathrm{Pe}=\gamma/\mathcal{D}_{r}$ and $W=\tau_{p}g$ gives
\begin{equation}\label{eq:PeKid}
    \mathrm{Pe}\,\mathcal{K}
        =\dfrac{\rho_{f}W^{2}}{8\pi\mu\,\mathcal{D}_{r}},
\end{equation}
in which the shear rate has cancelled. For thermal noise this reduces further,
using $\mathcal{D}_{r}=k_{B}T/(8\pi\mu \ell^{3}Y_{C})$, to
$\mathrm{Pe}\,\mathcal{K}=\rho_{f}W^{2}\ell^{3}Y_{C}/(k_{B}T)$, a property of the
particle and the suspending fluid alone. Varying the shear rate therefore moves
the system along a hyperbola $\mathrm{Pe}\,\mathcal{K}=\mathrm{const}$ rather
than along a line of fixed $\mathrm{Pe}$, and the sequence of measurements must
instead be assembled by varying the body force at fixed $\gamma$, for instance
with an applied electric or magnetic field in place of gravity, or in a
centrifuge. The same identity sets a second constraint. The particle is
displaced by $W/\gamma=8\pi\mu\mathcal{K}/(\rho_{f}W)$ per unit strain, so a
Brownian spheroid of semi-major axis $10\,\mu\mathrm{m}$ and twice the density
of water reaches $\mathcal{K}=O(1)$ only at
$\gamma\approx2\times10^{-3}\,\mathrm{s}^{-1}$, and then falls of order a metre
in a single Jeffery period. Since a phase slip takes many Jeffery periods, the
protocol is not realisable by gravitational settling of a thermal particle. It
becomes accessible when the noise is athermal, so that $\mathcal{D}_{r}$ is set
independently of the Stokes-Einstein relation and $\mathrm{Pe}$ can be lowered
at fixed $\mathcal{K}$, or in a configuration in which the mean settling is
compensated.
 
When noise is purely thermal, $\mathcal{D}_{r}$ is related
to temperature and viscosity through the Stokes-Einstein
relation, and the measurement constitutes a microrheological
probe. When noise has an athermal origin, as in turbulent
suspensions or active fluids containing swimming microorganisms,
the measured $\mathcal{D}_{r}$ reflects the effective noise
intensity of the non-equilibrium environment, which can vastly
exceed the thermal value
\citep{bechinger2016active}. In either case, the settling
spheroid acts as a transducer that converts microscale
orientational fluctuations into a macroscopically observable
flip rate, with the exponential amplification provided by the
Kramers mechanism ensuring high sensitivity to the noise
intensity.
 
We emphasise that this measurement is applicable only in the
regime $\mathcal{R}>1$ (above the SNIC bifurcation), where
well-defined potential barriers exist, and for
oblate spheroids, for which the in-plane equilibria are attracting.
Below the bifurcation, phase slips do not occur, and the relevant
observable is instead the variance of the tumbling period, which
increases linearly with the rotational diffusivity.
For prolate spheroids the analogous observable is
the rate of the broadside excursions described above. It is
likewise exponentially sensitive to $\mathrm{Pe}$, so the same
protocol applies in principle, but the barrier must be obtained
from the quasipotential on $S^{2}$ rather than from
(\ref{eq:barrier}). Moreover, phase slips are
observed only over an intermediate range of P\'{e}clet numbers: at
large $\mathrm{Pe}$ they are strongly suppressed by weak rotational
diffusion, whereas at very small $\mathrm{Pe}$ strong noise
obscures the underlying deterministic dynamics.

\section{Conclusion}\label{sec:conc}
 
We have investigated the orientation dynamics of a settling
spheroidal particle in simple shear flow, combining deterministic
dynamical-systems analysis with a stochastic Fokker-Planck
treatment. The dynamics is governed by the competition between the Jeffery torque from the
background shear and the inertial torque,
with the settling parameter
$\mathcal{K}$ measuring
their relative strength. The phase space is the unit sphere
$S^{2}$, which by the Poincar\'{e}-Bendixson theorem, given finitely
many equilibria, admits only fixed points, periodic orbits, and heteroclinic
connections as possible long-time states, rigorously excluding chaos for the present autonomous system; chaotic orientation dynamics is not precluded in general, and may arise once the effective phase-space dimension is increased, for instance by time-dependent forcing.
 
For the two canonical gravity-vorticity configurations, we have
shown that the azimuthal dynamics reduces to overdamped motion in a
tilted periodic potential with effective amplitude
$\mathcal{R}=\sqrt{\mathcal{B}^{2}
+\mathcal{K}^{2}\mathcal{F}_{p}^{2}}$. A saddle-node bifurcation on an invariant circle (SNIC) occurs at
$\mathcal{R}=1$: below this threshold the spheroid rotates continuously, with a period diverging as $T\sim(1-\mathcal{R})^{-1/2}$, and
above it the azimuthal rotation arrests. The principal result
is not a universal transition from tumbling to a stable fixed point, but a
shape-dependent global classification built on that one reduced bifurcation.
The transverse eigenvalue
$\lambda_{\perp}=-(\mathcal{K}_{p}+\mathcal{G}_{1})/2$ carries the sign of
$-\mathcal{K}_{p}$, since $\mathcal{G}_{1}<|\mathcal{K}_{p}|$ for every
physical spheroid. Oblate spheroids therefore undergo a genuine transition from
an attracting equatorial cycle to an attracting in-plane equilibrium. Prolate
spheroids retain the vorticity axis as the attractor on both sides of
$\mathcal{R}=1$, the new equatorial equilibria being a source and a saddle,
while developing an attracting broadside slow manifold at strong settling. The
same distinction separates the weak-noise mechanisms: azimuthal winding in the
oblate problem, and two-dimensional broadside excursions about a single nematic
attractor in the prolate one. For an oblate spheroid, whose equator attracts, below a
critical value of $\mathcal{K}$ at fixed $\mathcal{B}$, the particle undergoes tumbling motion. Above the critical value, the particle evolves towards a steady equilibrium orientation corresponding to a fixed point, whose nature and stability are examined through linear stability analysis. Further, the rapid-settling analysis reveals that, in the limit $\mathcal{K}\gg 1$, the particle aligns in the broadside plane for a long time, with the resulting polar angle spanning a broad range that depends on the initial orientation. However, when gravity is
parallel to the vorticity axis, the settling parameter does not
enter the azimuthal equation and  a particle exhibits a rotational motion
(tumbling for prolate, log-rolling for oblate spheroids) for all
$\mathcal{K}$.   
 
The stochastic analysis reveals that noise plays qualitatively
different roles depending on whether the settling-induced inertial
torque is present. In the pure Jeffery problem
($\mathcal{K}=0$), noise diffuses over the continuum of orbit
constants, selecting a unique steady-state distribution in the
manner analysed by \citet{leal1971effect} and
\citet{hinch1972effect}. When $\mathcal{K}\neq 0$ and
$\mathcal{R}>1$, the settling-induced potential barriers between
equilibria enable Kramers-type escape events: sudden
$\pi$-rotations (phase slips) or tumbling events. These phase slips are
extreme events governed by large deviation theory, analogous to
the aperiodic tumbling of polymers in turbulent shear. Near the SNIC
bifurcation, the barrier vanishes, the harmonic approximation
underlying the rate expression fails, and the dynamics becomes
intermittent, with long dwells near the equilibrium
orientation punctuated by rapid phase slips. For a prolate spheroid the
same noise produces a different process: the attractor is the vorticity
axis, and the excursions seen in the Langevin time series carry the
director across the broadside plane and back to the same physical
orientation.
 
Asymptotic results in both the small-$\mathrm{Pe}$ and
large-$\mathrm{Pe}$ limits complement the numerical solutions at
intermediate $\mathrm{Pe}$. At low $\mathrm{Pe}$, the
settling and shear contributions to the orientation distribution
function decouple at leading order, entering through orthogonal
spherical harmonics. At
large $\mathrm{Pe}$, the orientation moments exhibit distinct
scalings depending on the attractor type (for $\alpha=0$):
$\langle p_{z}^{2}\rangle\sim\mathrm{Pe}^{-1}$ for periodic
orbits versus approach to a deterministic constant for stable
fixed points. At large $|\mathcal{K}_{p}|$, the orientation distribution becomes increasingly concentrated within the broadside plane.
 
In the present analysis, particle inertia has been neglected.
For sufficiently large or dense particles, inertial effects may
modify the transient dynamics, although experimental evidence
suggests that the equilibrium alignment is primarily governed by
the inertial torque \citep{bhowmick2024inertia}. The interplay
between settling, shear-induced fluid inertia, particle inertia,
and turbulent fluctuations in realistic suspension flows presents
additional complexity that lies beyond the present scope. In
particular, a detailed characterisation of the Kramers escape
statistics near the SNIC bifurcation, including the computation of
the full quasipotential landscape on $S^{2}$ and the instanton
trajectories, represents a natural and promising direction for
future work.
From an experimental perspective, the exponential sensitivity of
the Kramers escape rate to the noise intensity suggests that
flip-time measurements of settling spheroids could serve as a
sensitive probe of the effective rotary diffusivity in turbulent
or active suspensions.

\vspace{3cm}

\textbf{Acknowledgments.} A.R. acknowledges ANRF project ANRF/ARG/2025/010666/ENS for funding. H.M. acknowledges the support from the Prime Minister's Research Fellows (PMRF) scheme at the Ministry of Education, Government of India (project no. SB22230347AMPMRF008746).

\backsection[Declaration of interests]{ The authors report no conflict of interest.}

\backsection[Author ORCIDs]{\\ 
Himanshu Mishra https://orcid.org/0000-0001-6255-2124;\\ 
Anubhab Roy https://orcid.org/0000-0002-0049-2653.
}

\appendix

\section{Phase space topology and the exclusion of chaos}\label{appphase}

The orientation vector $\boldsymbol{p}$ evolves on the unit sphere $S^{2}$, a compact two-dimensional manifold without boundary. By the Poincar\'{e}--Bendixson theorem, the only possible $\omega$-limit sets of a smooth autonomous flow on such a surface are fixed points, periodic orbits, and heteroclinic or homoclinic connections between saddle points \citep{strogatz2015nonlinear}. Consequently, chaotic dynamics is rigorously excluded for all parameter values in the present autonomous system. Chaos may, however, arise in higher-dimensional phase spaces, for example, through the introduction of time-dependent forcing or additional dynamical degrees of freedom.
 
The topology of $S^{2}$ further constrains the fixed-point
structure through the Poincar\'{e}-Hopf index theorem. The Euler
characteristic of $S^{2}$ is $\chi(S^{2})=2$, so the sum of the
indices of all fixed points must equal 2. Nodes and spirals carry
index $+1$, while saddle points carry index $-1$. If the system has $n$
nodes or spirals and $s$ saddles, then $n-s=2$. This global
constraint explains the recurring pattern observed in the phase
portraits of \S\ref{sec:gz}--\S\ref{sec:gx}: pairs of nodes and
saddles always appear together. The physical configuration space is $\mathbb{RP}^{2}$, of
which $S^{2}$ is the antipodal double cover, since $\boldsymbol{p}$ and
$-\boldsymbol{p}$ describe the same director. We work on $S^{2}$ throughout,
where the two poles supply a baseline index $+2$ and each equatorial
source-saddle or node-saddle pair adds zero net index; the corresponding count
on $\mathbb{RP}^{2}$ has Euler characteristic $1$. The distinction matters
whenever equilibria are counted or a $\pi$-slip is described, because a
$\pi$-rotation returns the director to the same physical state. It also means
that intersections of the nullclines drawn at $\theta=0$ or $\pi$ in the
$(\theta,\phi)$ chart are a single point of $S^{2}$, not several, and must be
classified in local Cartesian tangent coordinates as in
(\ref{eq:lin_pole_gx}).

\section{Closed-form solution for the orientation angles}\label{appA}

For gravity in the shear plane the azimuthal equation closes on
itself, as noted in \S\ref{sec:gy}, and integrates in closed form. Writing
$\mathcal{B}=\mathcal{R}\cos\psi$ and
$\mathcal{K}\mathcal{F}_{p}=\mathcal{R}\sin\psi$, and setting
$u=2\phi-\psi$ for $\beta=\pi/2$ or $u=2\phi+\psi$ for $\beta=0$, both
configurations reduce to the single autonomous equation
\begin{equation}\label{eq:ueq}
    \dot u = -1+\mathcal{R}\cos u ,
\end{equation}
whose solution depends on which side of the bifurcation the
parameters lie. Above the threshold
$\mathcal{G}_{1}=\sqrt{\mathcal{R}^{2}-1}$ is real and
\begin{equation}\label{eq:phi_yeq}
    \tan\dfrac{u}{2}
    =\sqrt{\dfrac{\mathcal{R}-1}{\mathcal{R}+1}}\,
    \tanh\!\left[\dfrac{\mathcal{G}_{1}}{2}\left(t-t_{0}\right)\right],
\end{equation}
so $\phi$ approaches its equilibrium monotonically and the time
to reach it diverges as $\mathcal{G}_{1}^{-1}$ as the threshold is approached
from above. For an initial azimuth outside the well the same quadrature gives
$\coth$ in place of $\tanh$, describing approach to the same equilibrium from
the other side. Below the threshold
$\mathcal{G}_{2}=\sqrt{1-\mathcal{R}^{2}}$ is real and
\begin{equation}\label{eq:phixd}
    \tan\dfrac{u}{2}
    =\sqrt{\dfrac{1-\mathcal{R}}{1+\mathcal{R}}}\,
    \tan\!\left[-\dfrac{\mathcal{G}_{2}}{2}\left(t-t_{0}\right)\right],
\end{equation}
a periodic solution in which $\phi$ advances by $\pi$ over
$T=2\pi/\mathcal{G}_{2}$, the period quoted in \S\ref{sec:gy}. In both cases
$t_{0}$ is fixed by the initial azimuth. Setting $\mathcal{K}=0$, so that
$\mathcal{R}=|\mathcal{B}|<1$ and $\psi=0$, recovers the Jeffery solution.
Writing the two branches separately in this way avoids the imaginary radicals
that appear if a single hyperbolic form is used throughout.

The polar angle follows by quadrature. Every
$\theta$-dependence of $\Chi$ enters through $\sin^{2}\theta$, so the polar
equation has the form $\dot\theta=F(\phi)\sin 2\theta$ and
\begin{equation}\label{eq:thquad}
    \tan\theta(t)=\tan\theta_{0}\,
    \exp\!\left[2\!\int_{0}^{t}\!F\bigl(\phi(t')\bigr)\,\mathrm{d}t'\right],
    \qquad
    F(\phi)=\dfrac{\mathcal{B}}{4}\sin 2\phi
    +\dfrac{\mathcal{K}\mathcal{F}_{p}}{2}\cos^{2}(\beta-\phi),
\end{equation}
where the integral may be converted to
$\int F\,\mathrm{d}\phi/\dot\phi$ using (\ref{eq:phi_yeq}) or
(\ref{eq:phixd}). Written out the result is lengthy and is not needed below;
its structure is what matters. Because $\theta$ enters only through
$\sin 2\theta$, the polar angle can cross neither $\theta=0$ nor
$\theta=\pi/2$, and its long-time behaviour is set by the sign of the orbit
average of $F$, which is the slow drift (\ref{eq:slowdrift_gy}) of
\S\ref{sec:gy}. For gravity along the vorticity axis the same construction
closes completely: there $F=(\mathcal{B}/4)\sin 2\phi
-\mathcal{K}\mathcal{F}_{p}/2$ with $\dot\phi$ of Jeffery form, the quadrature
is elementary, and
\begin{equation}\label{eq:thgz_closed}
    \tan\theta(t)=\tan\theta_{0}\,
    e^{-\mathcal{K}\mathcal{F}_{p}t}
    \sqrt{\dfrac{1-\mathcal{B}\cos 2\phi_{0}}
    {1-\mathcal{B}\cos 2\phi(t)}} ,
\end{equation}
an exponential drift at rate $\mathcal{K}\mathcal{F}_{p}$
modulated by a bounded periodic factor, which is the Jeffery orbit constant.

\section{ Numerical convergence, validation, and stochastic dynamics of orientation angles}\label{appbmom}

\begin{figure}
\centering  
\subfigure[]{\includegraphics[width=0.49\linewidth]{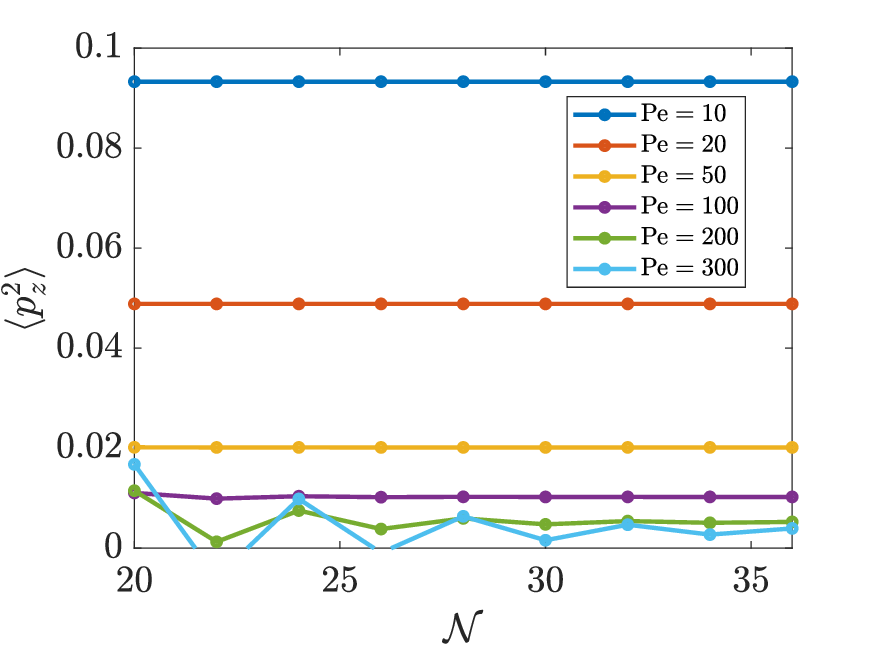}}
\subfigure[]{\includegraphics[width=0.49\linewidth]{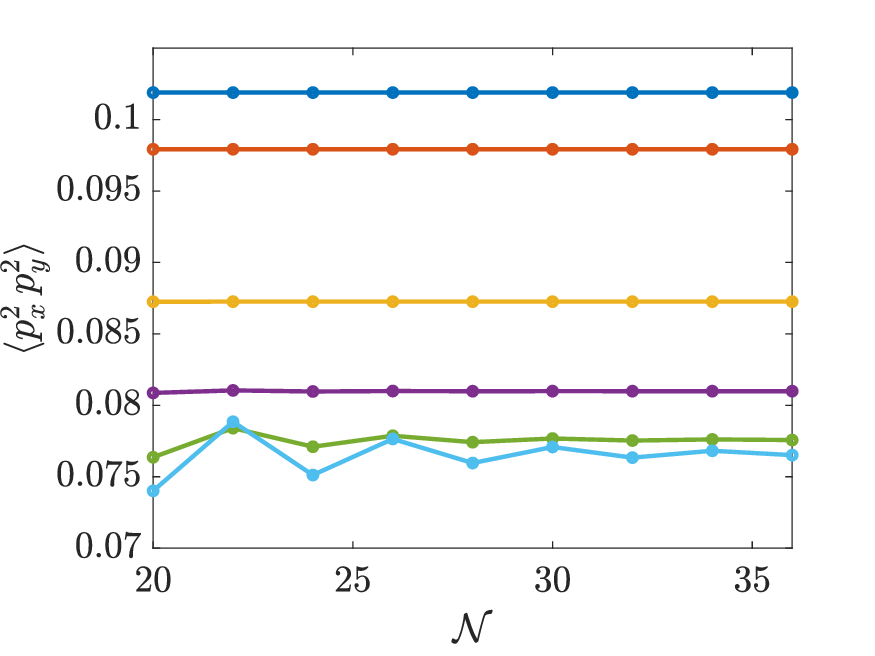}}
\subfigure[]{\includegraphics[width=0.49\linewidth]{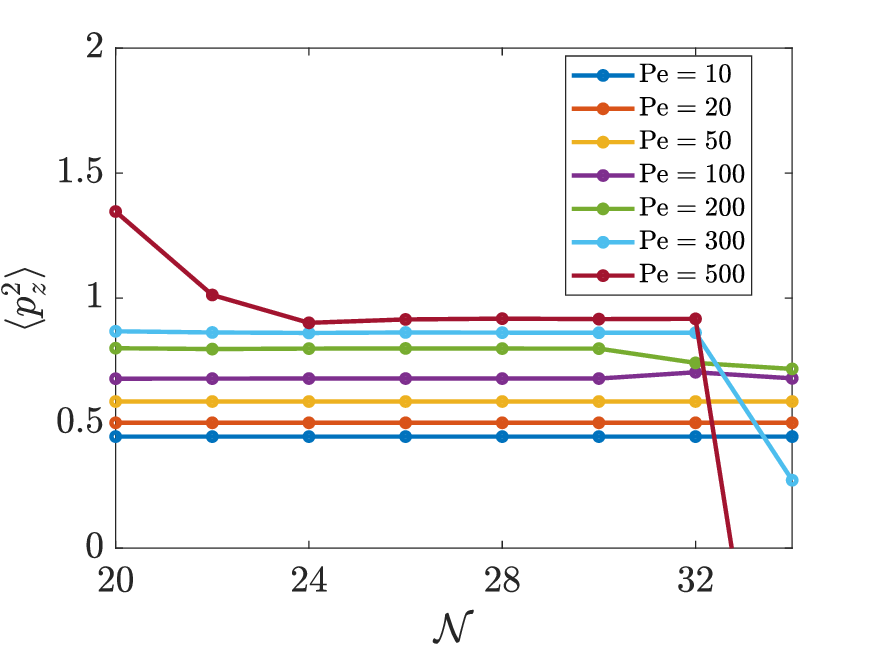}}
\subfigure[]{\includegraphics[width=0.49\linewidth]{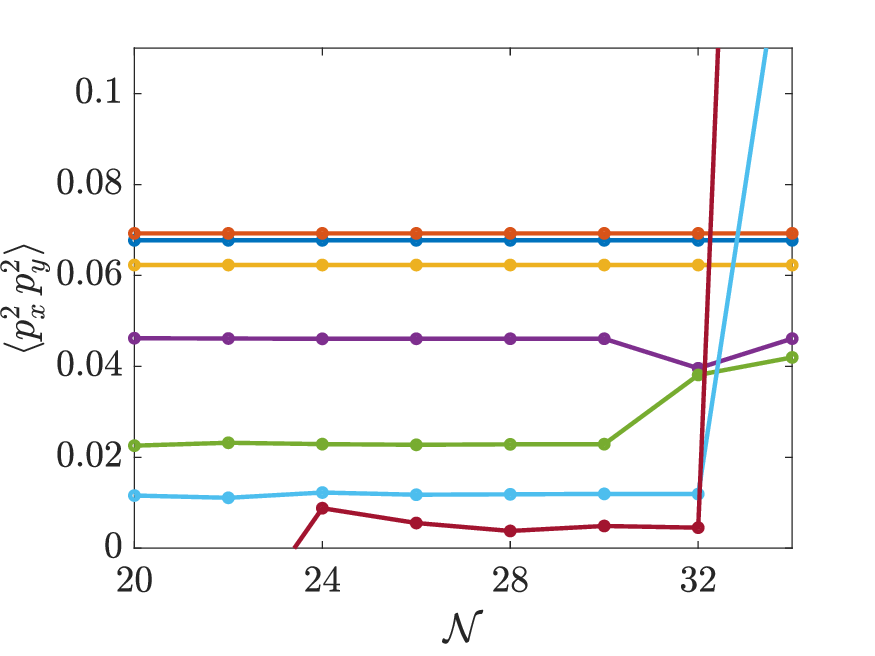}}
\caption{(a) Variation of the orientation moment $\langle p_{z}^{2}\rangle$ versus $\mathcal{N}$ when $\alpha=0$,  (b) variation of $\langle p_{x}^{2}p_{y}^{2}\rangle$ versus $\mathcal{N}$ when $\alpha=0$, (c)  variation of the orientation moment $\langle p_{z}^{2}\rangle$ versus $\mathcal{N}$ when $\alpha=\pi/2$ and $\beta=0$, and (d) variation of $\langle p_{x}^{2}p_{y}^{2}\rangle$ versus $\mathcal{N}$ when $\alpha=\pi/2$ and $\beta=0$. In all plots $\mathcal{B}=0.9$ ($r=4.36$) and $\mathcal{K}_{p}=-1$.  } 
\label{fig:conv_gz_gx}
\end{figure}

In this appendix, we examine the convergence of the numerical solutions for both flow configurations and validate the numerical calculations with the dumbbell model. Figure~\ref{fig:conv_gz_gx} shows the orientation moments as a function of $\mathcal{N}$, for $\mathcal{B}=0.9$ and $\mathcal{K}_{p}=-1$. The first two panels correspond to a case when $\alpha=0$. For the classical case of $\mathcal{K}_{p}=0$, the numerical solution converges at $\mathcal{N}\geq26$ for $\mathrm{Pe}=10^{3}$, consistent with previous studies. In contrast, a finite value of $\mathcal{K}_{p}$ significantly affects the convergence, and the numerical solution becoming numerically unstable and ill-conditioned for $\mathrm{Pe}> 500$. To illustrate this behaviour, we present the orientation moments $\langle p_{z}^{2}\rangle$ and $\langle p_{x}^{2}p_{y}^{2}\rangle$ for various values of $\mathrm{Pe}$. At low $\mathrm{Pe}$, the numerical solution is accurate for all values of $\mathcal{N}$ considered. As $\mathrm{Pe}$ increases, progressively larger values of $\mathcal{N}$ are required to obtain an accurate solution, consistent with the previously reported trends for $\mathcal{K}_{p}=0$. The numerical solution eventually becomes unstable and ill-conditioned at sufficiently large values of $\mathcal{N}$, with this behaviour becoming more pronounced at higher $\mathrm{Pe}$. A similar trend is observed when the gravity vector is aligned with the flow direction (shown in the third and fourth panels).


\begin{figure}
\centering  
\subfigure[]{\includegraphics[width=0.485\linewidth]{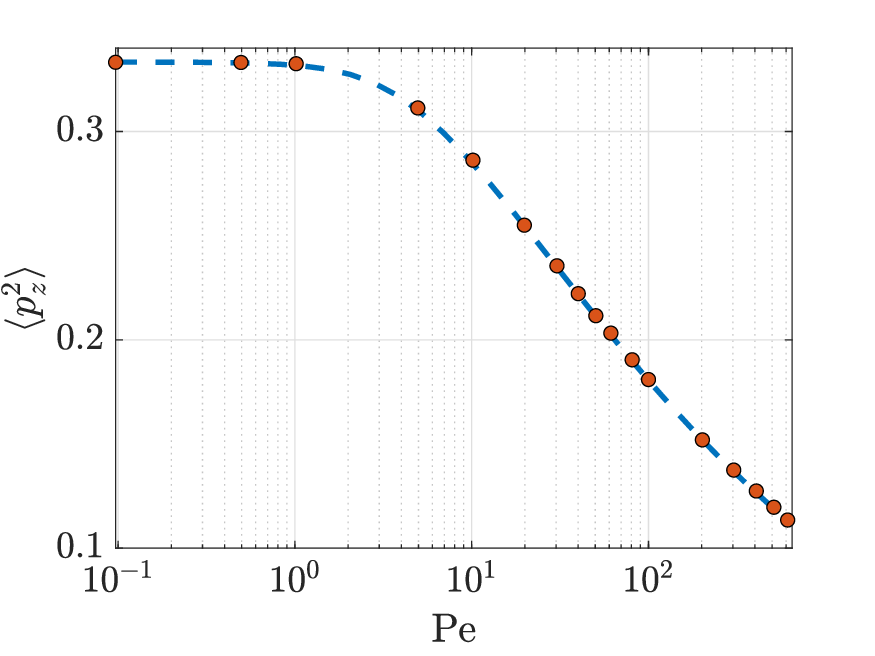}}
\subfigure[]{\includegraphics[width=0.5\linewidth]{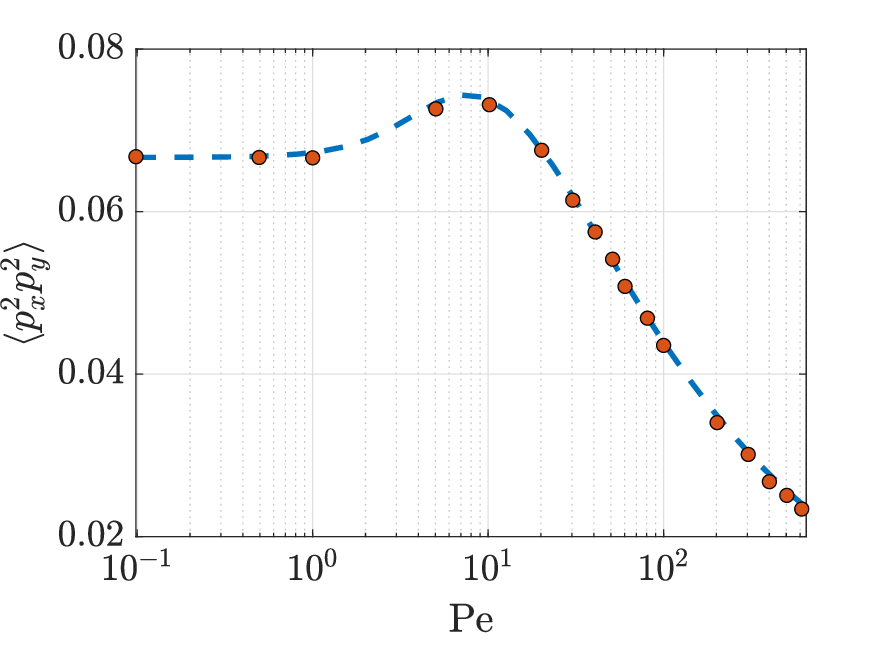}}
\caption{Orientation moments as a function of $\mathrm{Pe}$, for the classical dumbbell model ($\mathcal{B}=1$, $\mathcal{K}_{p}=0$). The blue solid lines represent the present numerical results, while the circles denote the published results of \citet{asokan2002novel}, obtained using the spherical harmonics method. Also, we set $\mathcal{N}=26$.} 
\label{fig:k0_test}
\end{figure} 

Figure~\ref{fig:k0_test} presents the orientation moments as functions of the P\'{e}clet number $\mathrm{Pe}$, for the classical dumbbell model ($\mathcal{B}=1$ and $\mathcal{K}_{p}=0$). The present numerical results are shown by the blue solid lines, while the circles represent the results of \citet{asokan2002novel}, obtained using the spherical harmonics method. The two sets of results are indistinguishable on the scale of the figure, which validates the present numerical implementation against the classical dumbbell problem.

Next, we examine the role of stochastic noise on the evolution of $\theta$ and $\phi$ angles for the case when $\alpha=\pi/2$ and $\beta=0$. Isotropic rotary diffusion on the unit sphere acquires metric terms when written in spherical coordinates, so that the It\^o system corresponding to (\ref{eq:stfp}) is

\begin{subequations}\label{eq:lethtphi}
    \begin{equation}
       d\theta=\left(\dot{\theta}+\dfrac{1}{\mathrm{Pe}}\cot\theta\right)dt+\sqrt{\dfrac{2}{\mathrm{Pe}}}d\mathcal{W}_1,
    \end{equation}
    \begin{equation}
        d\phi=\dot{\phi}\,dt+\dfrac{1}{\sin\theta}\sqrt{\dfrac{2}{\mathrm{Pe}}}\,d\mathcal{W}_2.
    \end{equation}
\end{subequations}

\begin{figure}
    \centering
    \includegraphics[width=0.7\linewidth]{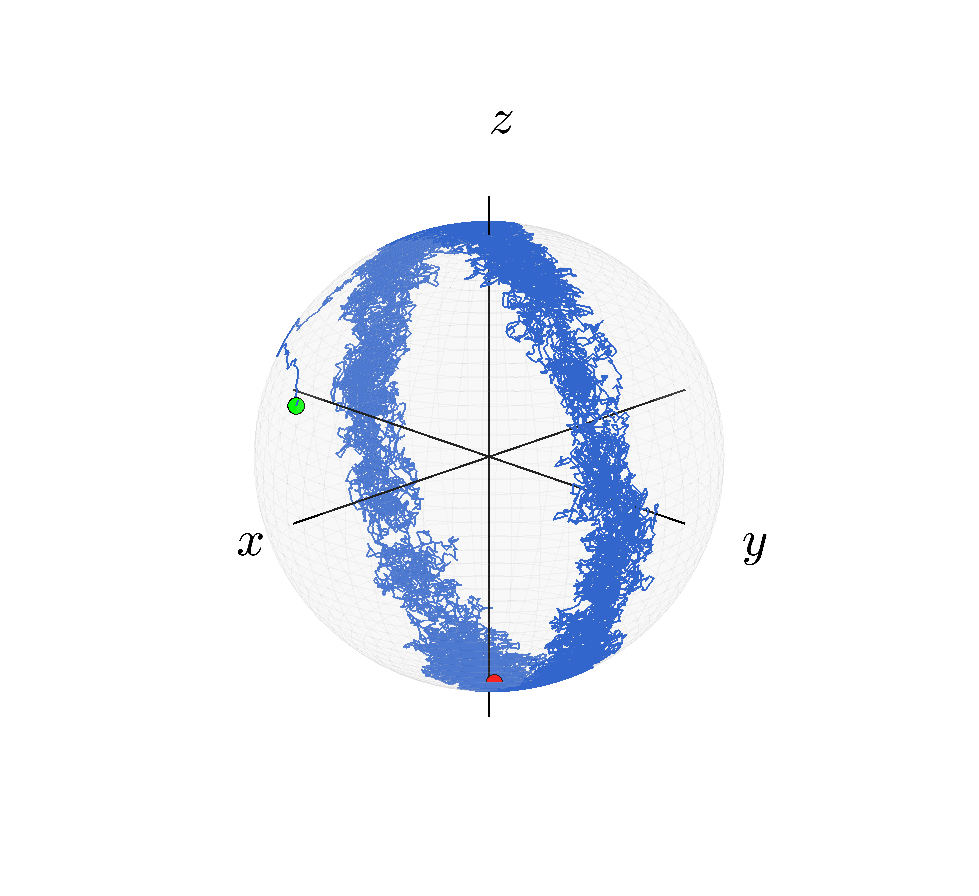}
    \caption{Trajectory of the orientation vector on the unit sphere for the case when
gravity is aligned with the flow direction. Parameters are: $\mathcal{B}=0.5$, $\mathcal{K}_{p}=-3$ and $\mathrm{Pe}=50$. Also, green circle represents the starting point and red circle corresponds to the end of the trajectory.}
    \label{fig:us_gx_le}
\end{figure}

Here $\mathcal{W}_{1}$ and $\mathcal{W}_{2}$ are independent Wiener processes, and $\dot\theta$ and $\dot\phi$ are given by (\ref{eq:gtpeq}). The $\mathcal{D}_{r}\cot\theta$ drift and the $1/\sin\theta$ noise amplitude are the metric terms; they are what make the isotropic distribution the stationary solution of (\ref{eq:lethtphi}) when $\mathcal{B}=\mathcal{K}_{p}=0$, and they dominate near the poles, where the excursions of figures~\ref{fig:langevin_phi} and \ref{fig:us_gx_le} take place. Equivalently, and more convenient numerically, one may integrate the embedded Cartesian form $d\boldsymbol{p}=\dot{\boldsymbol{p}}\,dt+\sqrt{2/\mathrm{Pe}}\,(\mathbb{I}-\boldsymbol{p}\boldsymbol{p})\cdot d\boldsymbol{W}-(2/\mathrm{Pe})\,\boldsymbol{p}\,dt$, which is free of coordinate singularities. We solve the above equations to obtain the trajectory in $\theta$ and $\phi$ and represent it on the unit sphere, as shown in figure~\ref{fig:us_gx_le}. The parameters are $\mathcal{B}=0.5$, $\mathcal{K}_{p}=-3$, and $\mathrm{Pe}=50$. The trajectory is integrated up to $t=400$, where time is scaled by the shear rate $\gamma$, with a time step of $dt=10^{-3}$. The green and red circles denote the initial and final positions, respectively. As shown in figure~\ref{fig:langevin_phi}, $\theta$ exhibits significant tumbling events between successive crossings of $\theta=0$ and $\pi$ in the time series. This behaviour is reflected in the trajectory, which repeatedly transitions between the $\hat{\boldsymbol{z}}$ and $-\hat{\boldsymbol{z}}$ axes within the $\hat{\boldsymbol{y}}$-$\hat{\boldsymbol{z}}$ plane. With isotropic rotary diffusion and the given values of $\mathcal{K}_{p}$ and $\mathcal{B}$, the trajectory is not confined to a plane, but it remains concentrated near the $\hat{\boldsymbol{y}}$-$\hat{\boldsymbol{z}}$ plane, closely resembling the structure of the orientation distribution function shown in figure~\ref{fig:ffgx}.

\bibliographystyle{jfm}
\bibliography{jfm}

\end{document}